 \pdfoutput=1
\documentclass[%
preprint,
 amsmath,amssymb,
 aps,
floatfix,
]{revtex4-1}
\pdfoutput=1
\usepackage{appendix}
\usepackage{amsmath}
\usepackage{graphicx}
\usepackage{epstopdf}
\usepackage{graphics}
\usepackage{epsfig}
\usepackage[byname]{smartref}
\usepackage{hyperref} 
\usepackage{txfonts}
\usepackage{morefloats}
\usepackage{tocloft}
\usepackage{graphicx}
\usepackage{dcolumn}
\usepackage{bm}
\usepackage{hyperref}

\begin{document}

\preprint{V1}

\title{Spin Waves in Ferromagnetic Dots 2D Honeycomb Lattice
Stripes}

\author{Maher Ahmed}
\affiliation{Department of Physics and Astronomy, University of Western Ontario, London ON N6A 3K7, Canada}
\affiliation{Physics Department, Faculty of Science, Ain Shams University, Abbsai, Cairo, Egypt}
\email{mahmed62@uwo.ca}
%



\begin{abstract}
In this work, the spin wave calculations were carried out using the
Heisenberg Hamiltonian to study the allowed spin waves of zigzag
and armchair edged stripes for ferromagnetic nanodots arrayed in a 2D
honeycomb lattice \cite{Selim2011}. The Hamiltonian is used to construct the $\mathbf{E}$
matrix which encodes the exchange flow of magnons in the stripes. It is found
that the allowed spin wave modes are the eigenvalues of the $\mathbf{E}$
matrix and therefore it is used to study the effects of the stripe width,
edge exchange, the edge uniaxial anisotropy, and impurities on the allowed
spin waves of stripes. The obtained results almost coincide with the results
of graphene nanoribbons described by tight binding Hamiltonian for electronic
excitations. Therefore, we suggest the fabrication of the magnetic
counterpart to graphene as a new technology in the field of spintronic
devices and magnetic applications.
\end{abstract}

\pacs{Valid PACS appear here}
\maketitle

\def\baselinestretch{1.66}

\section{Introduction}
The superior physical properties of graphene which lead to its promising applications in technology are mainly attributed to both its crystal structure as 2D honeycomb lattice and its short range interactions. While graphene itself is not a strongly magnetic material, many experimental and theoretical works have been done for related magnetic properties and proposed designs of graphene-based spintronic device \cite{Yazyev2010,Mecklenburg2011}. Graphene is formed due to the nature bonding stability of carbon atoms, but there is no natural atomic elements able to form a stable planar ferromagnetic 2D honeycomb lattice. In this case,  however, ferromagnetic nanodots can be used as magnetic artificial atoms \cite{E.O.KamenetskiiMar2003} with the ability to design the requested magnetic properties in the same way as the quantum dots nanostructures are used as artificial atoms with ability to design tunable electronic properties not found in naturally existence atomic elements \cite{Rontani2011,Astafiev2007,Bratschitsch2006,Kastner1993}. A question of interest here that arises to us from the physics and the technology point of view and from advances in the material science fabrication techniques is wherever are can fabricate a ferromagnetic dots 2D honeycomb lattice stripes as already done for similar magnetic structures \cite{malkinski:7325,PhysRevB.73.052411,Stepanova2011}. In this case, the ferromagnetic dots 2D honeycomb lattice stripes will share both crystal structure and short range interaction with graphene. The expected physical properties of that ferromagnetic dots 2D honeycomb stripes might lead to a new technology especially in the field of spintronic devices. A theoretical study is therefore needed to predict the similarity and the difference between magnetic and electronic short range interaction in the 2D honeycomb lattice.

Many studies have been conducted on the thermodynamic properties of hexagonal and honeycomb atomic spin lattices \cite{Richter2004,Darendelioglu1995,PhysRevB.45.9834,Reger1989}. Many of these  studies are for 3D systems reveal low-dimensional
magnetic behavior with predominantly antiferromagnetic behavior like $\beta{}$-Cu$_{2}$V$_{2}$O$_{7}$ \cite{Tsirlin2010}, or  hexagonal spin lattices, with ferromagnetic and antiferromagnetic interaction \cite{PhysRevB.49.9679}. Some studies have been conducted on the spin wave excitations in ferromagnetic nanostructures array \cite{Nguyen2007,Nguyen2009} but there are no known studies for ferromagnetic nanodots 2D honeycomb lattices stripes which are not Bravais lattices.

In this work, we will study the spin waves of zigzag and armchair stripes of ferromagnetic nanodots 2D honeycomb lattice with the total Hamiltonian~\eqref{dhimaltonian12}, which was used in the study of spin waves of 2D square lattice in \cite{Ahmed}. The new results will give us more understanding of lattice geometry effect on the physical properties of 2D materials.

\section{Theoretical model}
\begin{figure}[h]
\centering
\includegraphics[scale=0.2]{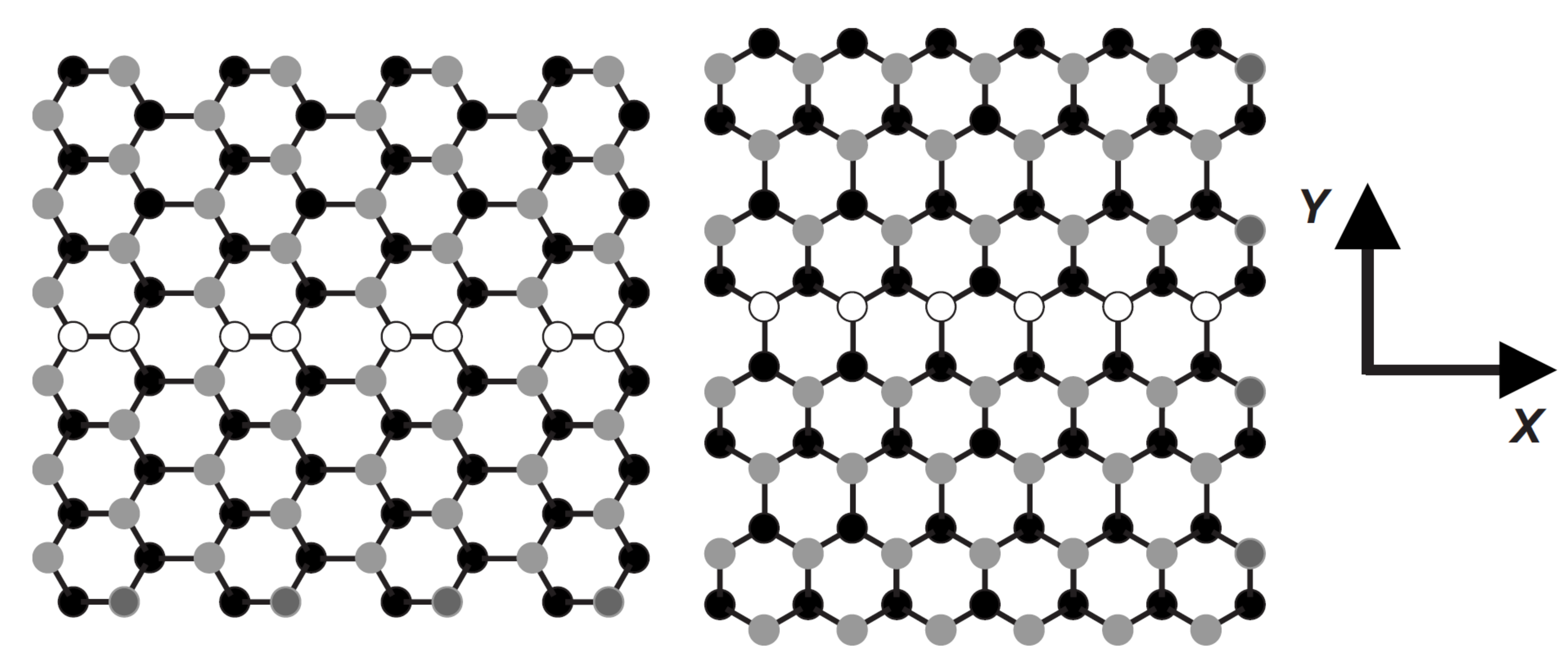}
\caption{Armchair (left) and zigzag (right)  2D Heisenberg ferromagnetic dots honeycomb stripes in $xy$-plane,  where black (gray) dots are the sublattice
A(B) with a line of impurities (white dots) in the middle of
the sheet, and with average spin alignment in $z$ direction. The stripes are finite in $y$ direction with $N$ rows
($n = 1,\cdots,N$) and they are infinite in the $x$ direction.  Figure taken from \cite{rim1}.}\label{fig:graphenelattice3}
\end{figure}

The systems under study here are 2D Heisenberg ferromagnetic stripes (or
nanoribbons) formed from arrays of dots in the $xy$-plane with armchair and zigzag edges. We assume a honeycomb (graphene like) lattice with
crystallographic description given in \cite{Neto1}.
The average spin alignment of the magnetic sites is in the $z$
direction, which is also the direction of the applied magnetic field. The
nanoribbon is of finite width in the $y$ direction with $N$  rows
(labeled as $n = 1,\cdots,N$) and it is infinite in $x$ direction ($-\infty
\Leftrightarrow \infty$)(see Figure
\ref{fig:graphenelattice3}).

The total Hamiltonian of the system is given by
\begin{equation}\label{dhimaltonian12} \hat{H}_{\textbf{Total}}=-\frac{1}{2}
\sum_{i,j} J_{i,j}\mathbf{S}_i\cdot\mathbf{S_j} -g\mu_B H_0 \sum_i S_i^Z
-\sum_i D_i (S_i^Z)^2,
\end{equation}where
the first term is the Heisenberg nearest-neighbor exchange term, the second
term is the Zeeman energy term due to an applied field $H_0$, and the third
term is the uniaxial anisotropy term. The summations over $i$ and $j$ run
over all the sites where $i$ and $j$ always belong to different sublattice A(B). This is because because in the honeycomb
lattice, the nearest neighbors of the $A$
sites are always $B$ sites and vice versa. The nearest neighbor exchange $J_{ij}$ has a
constant ``bulk" value $J$ when either $i$ and $j$ are in the interior of the
nanoribbons, and another constant value $J_e$  when $i$ and $j$ are both at
the edge of the nanoribbon (i.e., in row $n=1$ or $n=N$). Similarly, for the
site-dependent uniaxial anisotropy term $D_i$, we assume that it has a
constant value $D$ when the site $i$ is inside the nanoribbon, and it is
equal to $D_e$ for sites at the edge of the nanoribbon.

To calculate the spin waves for this system at low temperatures $T\ll T_{c}$
where the spins are nearly aligned such that $S^z\simeq S$ for each spin, we
use the Holstein-Primakoff (HP) transformation and follow the procedures of \cite{Selim2011}
to express the total Hamiltonian in terms of boson operators for the two sublattices A and B.
We arrive to the expression
\begin{equation}\label{df}
\hat{H}_{\textbf{Total}}= E_0+\hat{H}_s
\end{equation} where the constant term $E_0$ is the energy of the ground state
for the ferromagnetic system given by 
\begin{equation}\label{fgh}
 E_0=
 S^2\left(-\frac{1}{2}\sum_{i,j} J_{i,j}  -\sum_i D_i\right)-g\mu_B H_0 \sum_i S,
\end{equation}
and the operator
term $\hat{H}_s$ has the following form
\begin{eqnarray}
H_s&=&-\frac{1}{2} S \sum_{i,j} J_{i,j}  \left(a_{i}b^{\dag}_{j}+a^{\dag}_{i}b_{j}- b^{\dag}_{j}b_{j}- a^{\dag}_{i}a_{i} \right)\label{hmalrr3} \\
&&+ \sum_i \left[g\mu_B H_0+(2S-1) D_k\right]a^{\dag}_{i}a_{i}+ \sum_j \left[g\mu_B H_0+(2S-1) D_k\right]b^{\dag}_{j}b_{j}, \nonumber
\end{eqnarray}
where $a^{\dag}_{i}$ ($a_{i}$) and $b^{\dag}_{j}$ ($b_{j}$) are the creation and the annihilation boson
operators for sublattices A and B  respectively.

Since the nanoribbon extends to $\pm\infty$ in the $x$ direction, we may
introduce a 1D Fourier transform to wavevector $q_x$ along the $x$ direction
for the boson operators $a^{\dag}_{i}$ ($a_{i}$) and $b^{\dag}_{j}$ ($b_{j}$) as follows:
\begin{align}
b_{j}(x)&=\frac{1}{\sqrt{N_0}} \sum_{n} b_{n}(q_x) e^{-i \mathbf{q}_x\cdot \mathbf{r}_j},  \hspace{20pt}
b_{j}^\dag(x)=\frac{1}{\sqrt{N_0}} \sum_{n} b^\dag_{n}(q_x) e^{i \mathbf{q}_x\cdot \mathbf{r}_j},\label{far3}\\
a_{i}(x)&=\frac{1}{\sqrt{N_0}} \sum_{n} a_{n}(q_x) e^{-i \mathbf{q}_x\cdot \mathbf{r}_i},  \hspace{20pt}
a_{i}^\dag(x)=\frac{1}{\sqrt{N_0}} \sum_{n} a^\dag_{n}(q_x) e^{i \mathbf{q}_x\cdot \mathbf{r}_i}. \nonumber
\end{align}
Here $N_0$ is the (macroscopically large) number of spin sites in any row, $\mathbf{q_x}$
is a wavevector in the first Brillouin zone of the reciprocal lattice and both
$\mathbf{r}_i$ and $\mathbf{r}_j$ is the position vectors of any magnetic sites $i$ and $j$.
The new operators obey the following commutation relations:
\begin{equation}\label{comu3}
\left[a_{n}(q_x),a^\dag_{n}(q'_x)\right]=\delta_{q_xq'_x},\hspace{30pt}\left[b_{n}(q_x),b^\dag_{n}(q'_x)\right]=\delta_{q_xq'_x}.
\end{equation}

Also, we define the exchange sum:
\begin{equation}\label{exsum4}
\gamma(q_x) = \frac{1}{2} S\sum_{\nu} J_{i,j}  e^{-i\mathbf{q}_x \cdot (\mathbf{r}_i-\mathbf{r}_j)}.
\end{equation}
The sum for the exchange terms $J_{i,j}$ is taken to be over all $\nu$ nearest neighbors
in the lattice which depends on the edge configuration as zigzag or armchair for the stripe.     For the armchair configuration, the exchange sum gives the following amplitude factors $ \gamma_{nn'}(q_x)$
\begin{eqnarray}\label{amchairaf}
   \gamma_{nn'}(q_x)=\frac{1}{2} SJ\left[\exp(iq_xa)\delta_{n',n}+\exp\left(i\frac{1}{2}q_xa\right)\delta_{n',n\pm1}\right],
\end{eqnarray}
while for the zigzag case it gives
\begin{eqnarray}\label{zigzagaf}
  \gamma_{nn'}(q_x)=\frac{1}{2} SJ\left[2\cos\left(\frac{\sqrt{3}}{2}q_xa\right)\delta_{n',n\pm1}+\delta_{n',n\mp1}\right].
\end{eqnarray}
The $\pm$ sign depends on the sublattice
since the sites line alternates from A and B.

Substituting Equations~\eqref{far3} and \eqref{exsum4} in Equation~\eqref{hmalrr3}, and
rewriting the summation over nearest neighbors sites, we get
\begin{eqnarray}
   H_s&=&\sum_{q_x,nn'}  \left \{\alpha \left(a^\dag_{n} a_{n'}+ b^\dag_{n} b_{n'} \right)+\left( \gamma(q_x) a_{n}  b^\dag_{n'}+ \gamma(-q_x) a^\dag_{n}  b_{n'} \right) \right\}. \label{shp3hs}
\end{eqnarray}
The first terms count the elementary excitations on each
sublattice, while the second term describes the coupling between
the sublattices, and $\alpha$ is defined by
\begin{equation}\label{alpha}
\alpha = \left(
g\mu_B H_0+(2S-1) D_{n}\right)\delta_{nn'}.
\end{equation}

In order to diagonalize $\hat{H}_s$ and obtain the spin wave frequencies, we
may consider the time evolution of the creation and the annihilation
operators $a^{\dag}_{i}$ ($a_{i}$) and $b^{\dag}_{j}$ ($b_{j}$), as calculated in the Heisenberg picture. The equations of motion (using
the units with $\hbar=1$) for the annihilation operators $a_{i}$($b_j$) are as follows
\cite{Bes2007,Liboff,Kantorovich2004,Roessler2009,HenrikBruus2004}:
\begin{eqnarray}
  \frac{d a_{n}}{dt} &=&i[H,a_{n}] \nonumber \\
     &=&i\sum_{q_x,nn'}-\alpha a_{n'}-\gamma(-q_x) b_{n'} \label{ch3em1}
\end{eqnarray}
and
\begin{eqnarray}
  \frac{d b_{n}}{dt} &=&i[H,b_{n}]  \nonumber\\
     &=&i\sum_{q_x,nn'}-\alpha b_{n'}-\gamma(q_x) a_{n'} \label{ch3em2}
\end{eqnarray}
where the commutation relations between $b^\dag_i$ and $b_j$ in
\begin{equation}\label{crp}
\left[b_i,b^\dag_j\right]=\delta_{ij},\hspace{20pt}  \left[b^\dag_j,b_i\right]=-\delta_{ij},\hspace{20pt} \left[b_i,b_j\right]=\left[b^\dag_j,b^\dag_i\right]= 0.
\end{equation}
was used and similar commutation relations between $a^\dag_i$ and $a_j$, as well as the operator identity
$[AB,C]=A[B,C]+[A,C]B$.

The dispersion relations of the spin waves (i.e., energy or frequency versus wavevector)
can now be obtained by solving the above operator equations of motion. The spin
wave energy can be expressed in terms of the spin wave frequency using the relation
$E = \hbar\omega$, and assuming that spin wave modes behave like $\exp[-i\omega(q_x)t]$. We get the following sets of coupled equations:
\begin{eqnarray}
\omega(q_x) a_{n}  &=&\sum_{q_x,nn'}\alpha a_{n'}+\gamma(-q_x) b_{n'}  \\
& &  \nonumber\\
\omega(q_x) b_{n}  &=&\sum_{q_x,nn'}\gamma(q_x) a_{n'}+\alpha b_{n'}
\end{eqnarray}
The above equations can be written in matrix form as following
\begin{eqnarray}
 \omega(q_x)\left[%
\begin{array}{c}
a_{n}   \\
b_{n}   \\
\end{array}%
\right]  &=& \left[%
\begin{array}{cc}
  \alpha I_N  & T(q_x)  \\
  T^*(q_x) & \alpha I_N  \\
\end{array}%
\right] \left[%
\begin{array}{c}
a_{n}   \\
b_{n}   \\
\end{array}%
\right], \label{eq3eqing}
\end{eqnarray}
where the solution of this matrix equation is given by the
condition
\begin{eqnarray} \label{graphenee}
\det \left[%
\begin{array}{cc}
 -(\omega(q_x) -\alpha) I_N  & T(q_x)  \\
  T^{*}(q_x) & -(\omega(q_x)-\alpha) I_N  \\
\end{array}%
\right]=0.
\end{eqnarray}
Here, $T(q_x)$ is the exchange matrix, which depends on the orientation of the ribbon, and $\omega(q_x)$ are the energies of the spin wave modes. The matrix $T(q_x)$ is given by
\begin{equation}\label{}
\left(
  \begin{array}{ccccc}
   \varepsilon & \beta   &     0 & 0 & \cdots \\
    \beta   & \varepsilon & \gamma& 0 & \cdots \\
    0       & \gamma  &\varepsilon&  \beta    &\cdots \\
    0       &      0  & \beta & \varepsilon & \cdots \\
     \vdots & \vdots& \vdots & \vdots & \ddots \\
  \end{array}
\right),
\end{equation}
where the parameters $\varepsilon$, $\gamma$, and $\beta$ depend on the stripe edge geometry
and are given in Table \ref{tabexchn}.

Equation~\eqref{graphenee} is very similar to the one obtained for graphene ribbons in reference~\cite{rim1}, where the tight binding model was used. Where the next nearest-neighbor hopping term is neglected, the only essential  difference between the Heisenberg model and the tight binding model results is the existence of the $\alpha$ term in the Heisenberg model. This extra term is in the diagonal of the Hamiltonian matrix which is shifting the total spin waves energy by amount related to the in-site Zeeman energy term and the uniaxial anisotropy energy term. This similarity between graphene and ferromagnetic stripes shows that Heisenberg and tight binding model are closely-related models with the nearest neighbor interactions represented by $t_{ij}$ (Nearest neighbor hopping) and $J_{ij}$.

\begin{table}[h]
\caption{Nearest neighbor exchange matrix elements for 2D magnetic honeycomb lattice}\label{tabexchn}
\begin{tabular}{lcl}
  \hline\hline
    Parameter     &   Zigzag                  &   Armchair                 \\\hline
   \hspace{50pt}  &  \hspace{50pt}            &  \hspace{50pt}             \\
   $\varepsilon$  &         0                 &$\frac{SJ}{2}e^{-iq_xa}$   \\[10pt]
   $\beta$        &$SJ \cos(\sqrt{3}q_x a/2)$&$\frac{SJ}{2}e^{iq_xa/2}$  \\[10pt]
   $\gamma$       & $\frac{SJ}{2}$           &$ \frac{SJ}{2}e^{iq_xa/2}$  \\
 \hline
\end{tabular}
  \centering
\end{table}
\section{Numerical calculations}\label{ncref}
The dispersion relations for the above 2D Heisenberg ferromagnetic dots honeycomb stripes are obtained numerically as the eigenvalues \cite{algebra,RefWorks:27} for the matrix Equation \eqref{eq3eqing}. The first step for solving this eigenvalue problem, for given value of the wavevector $q_x$, is constructing the matrix
\begin{equation}\label{ematrix}
\mathbf{E}=\left[%
\begin{array}{cc}
  \alpha I_N  & T(q_x)  \\
  T^*(q_x) & \alpha I_N  \\
\end{array}%
\right],
\end{equation}
which is $2N\times 2N$ since both $I_N$  and $T(q_x)$ are $N\times N$ for the number $N$ of rows in the stripe.

First, the matrix $\alpha I_N$ is independent of the value of the wavevector $q_x$ and it is simply constructed using the material properties of the stripe for evaluating $\alpha$ values by Equation \eqref{alpha}, and since $\alpha$ is real the matrix $\alpha I_N$ is also real. Second, the matrices $T(q_x)$ and $T^*(q_x)$ depend on the value of the wavevector $q_x$, the material properties of the stripe $SJ$, and the stripe edge geometry as zigzag or armchair (see Table \ref{tabexchn}). For the zigzag case the element of the exchange matrix $T(q_x)$ are real (see Table \ref{tabexchn}). Consequently $T(q_x)=T^*(q_x)$, and therefore the matrix $\mathbf{E}$ is real too. A standard procedure, following reference \cite{RefWorks:27}, to obtain the eigenvalues for a real matrix is:

First, balance the real matrix  $\mathbf{E}$ by using similarity transformations in order to have comparable norms for
corresponding rows and columns of the matrix, which then
reducing the sensitivity of the eigenvalues to rounding errors. It is done here using the subroutine \verb"balanc" \cite{RefWorks:27}.

Second, reduce the matrix $\mathbf{E}$ to a matrix that has zeros everywhere below
the diagonal except for the first subdiagonal row, i.e., to upper Hessenberg form. It is done here using the subroutine \verb"elmhes" \cite{RefWorks:27}.

Third, find all eigenvalues of the matrix $\mathbf{E}$ in the upper Hessenberg form. It is
done here using the subroutine \verb"hqr" \cite{RefWorks:27}.

Forth, sort the obtained eigenvalues of the matrix $\mathbf{E}$ (done here using the subroutine \verb"piksrt"  \cite{RefWorks:27}) and plot the dispersion relations for the given stripe.

For the armchair case, the element of the exchange matrix $T(q_x)$ are complex (see Table \ref{tabexchn}), and $T(q_x)$ is  Hermitian conjugate to $T^*(q_x)$, so consequently $\mathbf{E}$ is a Hermitian matrix. One way to obtain the eigenvalues of  Hermitian complex  matrix like $\mathbf{E}$ is to convert it to an equivalent real matrix  \cite{RefWorks:27}, and then use the above standard procedures to obtain the eigenvalues for real matrix.

The conversion to a real matrix is done as follows: First the Hermitian complex  matrix $\mathbf{E}$ can be written as real and imaginary parts
\begin{equation}\label{1}
\mathbf{E}=\mathbf{Re}(q_x)+i\mathbf{Im}(q_x)
\end{equation}
where $\mathbf{Re}(q_x)$ and $\mathbf{Im}(q_x)$ are $2N\times2N$ real matrixes, using the above representation of  $\mathbf{E}$ in the Equation \eqref{eq3eqing}, we get the following $2N\times2N$ complex eigenvalue problem
\begin{eqnarray}
 \omega(q_x)\left (%
\mathbf{u}_n+i\mathbf{v}_n\right)  &=& \left(\mathbf{Re}(q_x)+i\mathbf{Im}(q_x)\right)\cdot \left (%
\mathbf{u}_n+i\mathbf{v}_n\right) \label{eq3eqing2}
\end{eqnarray}
where $\mathbf{u}_n$ and $\mathbf{v}_n$ represent the operators column vector. The above Equation \eqref{eq3eqing2} is equivalent to solving the following $4N\times4N$ real eigenvalue problem
\begin{eqnarray}
 \omega(q_x)\left[%
\begin{array}{c}
\mathbf{u}   \\
\mathbf{v}  \\
\end{array}%
\right]  &=& \left[%
\begin{array}{cc}
  \mathbf{Re}(q_x) &-\mathbf{Im}(q_x) \\
  \mathbf{Im}(q_x) & \mathbf{Re}(q_x)\\
\end{array}%
\right] \left[%
\begin{array}{c}
\mathbf{u}   \\
\mathbf{v} \\
\end{array}%
\right]. \label{eq3eqing32}
\end{eqnarray}

\subsection{Introducing effects of edges and  impurities}\label{egedgsf}
The study of edge effects on the 2D magnetic stripes properties are very important. They are introduced in this model numerically by using edges material properties in the elements $(1,1)$ and $(N,N)$ in $\alpha I_N$ matrix, and the elements $(1,1), (1,2)$ and $(2,1)$ for lower edge and the elements $(N,N), (N,N-1)$ and $(N-1,N)$ for upper edge in the matrixes $T(q_x)$ and $T^*(q_x)$.

The pure 2D magnetic stripes offer very interesting dispersion relations, but to be suitable for technological devices applications one needs to engineer its properties. One way to do so is the introduction of magnetic impurities lines substitutionally into the stripe materials, which is very similar to suggested  graphene materials engineering \cite{rim1}. The effects of one or two lines of impurities at any chosen rows numbers $n_0$ and $n'_0$ in the stripe are introduced in this model numerically by using magnetic impurities material properties in the elements $(n_0,n_0)$ and $(n'_0,n'_0)$ in $\alpha I_N$ matrix, and in the elements of the matrixes $T(q_x)$ and $T^*(q_x)$ that express the interaction of the impurity line with itself in the diagonal element $(n_0,n_0)$ for first line and $(n'_0,n'_0)$ for second line, the interaction of the impurity line with line before it in the stripe in the elements $(n_0-1,n_0),(n_0,n_0-1)$ for first line and $(n'_0-1,n'_0),(n'_0,n'_0-1)$ for second line, and the interaction of the impurity line with line after it in the stripe in the elements $(n_0,n_0+1),(n_0+1,n_0)$ for first line and  $(n'_0,n'_0+1),(n'_0+1,n'_0)$ for second line.

\section{Results}
To compare our results for ferromagnetic dots stripes using the Heisenberg Hamiltonian with those of graphene nanoribbons using the tight-binding Hamiltonian, where the nearest neighbor (NN) interactions are represented by $t_{ij}$ and $J_{ij}$, we choose our stripes sizes, scaling our result to be dimensional less quantities, and choosing physical parameters matched that ones used in reference \cite{rim1} for graphene.

\subsection{Zigzag stripes results}
\begin{figure}[hp!]
\centering
\includegraphics[scale=1]{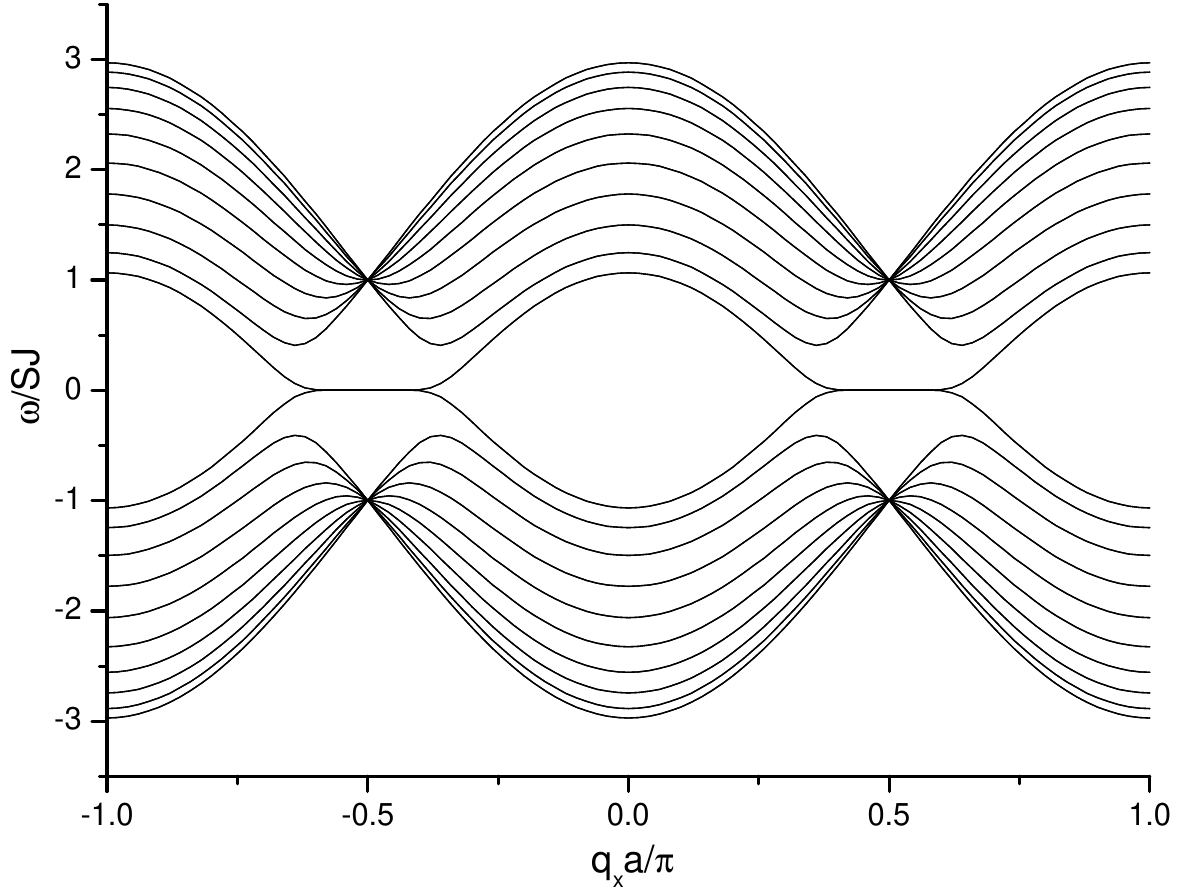}
\caption{Spin waves dispersion for zigzag 2D Heisenberg ferromagnetic dots honeycomb stripes
with $N=20$, $J=J_e=1$, $D=D_e=0$ and $\alpha=0.01$.} \label{fig:mgzf1}
\vspace{10pt}
\centering
\includegraphics[scale=1]{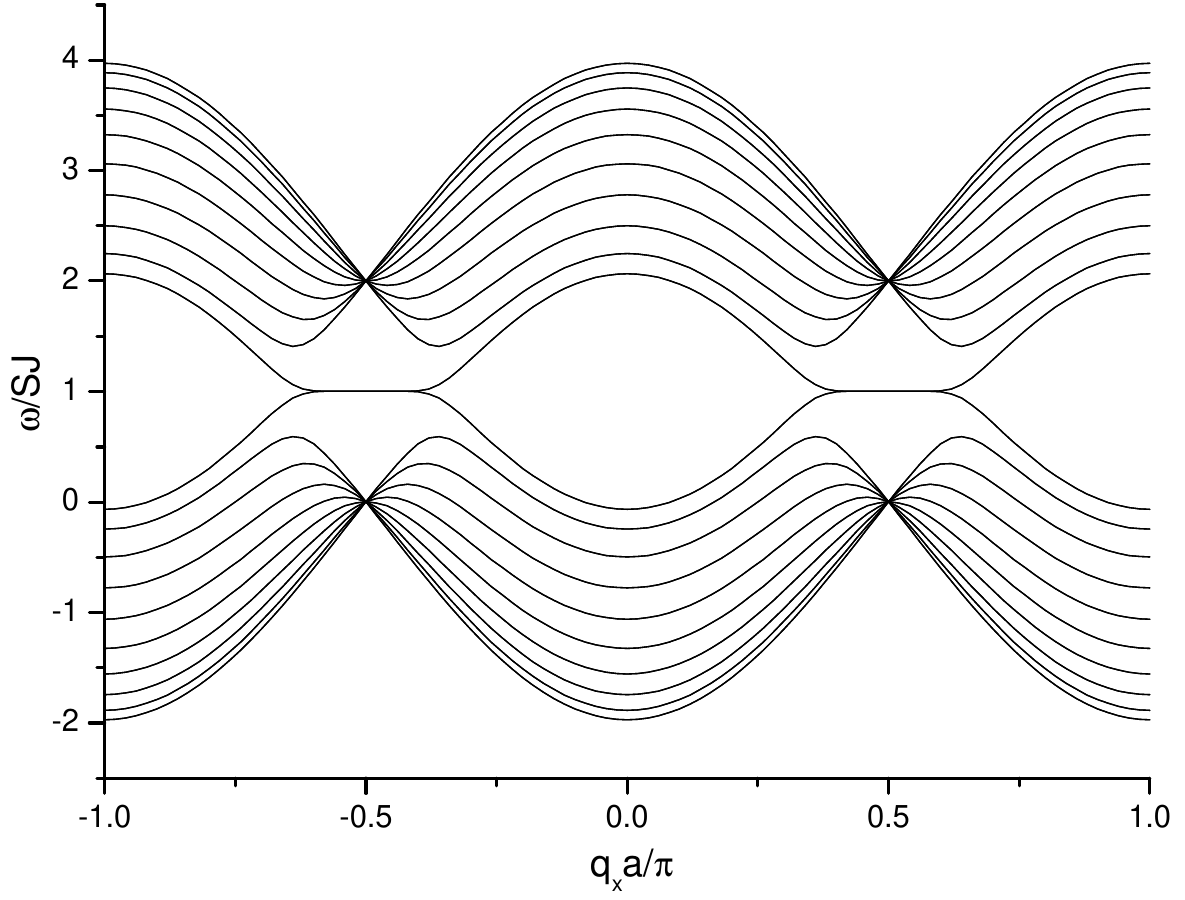}
\caption{Spin waves dispersion for zigzag 2D Heisenberg ferromagnetic dots honeycomb stripes
with $N=20$, $J=J_e=1$, $D=D_e=1.0$ and $\alpha=1.01$.} \label{fig:mgzf2}
\end{figure}

\begin{figure}[ht!]
\centering
\includegraphics[scale=1]{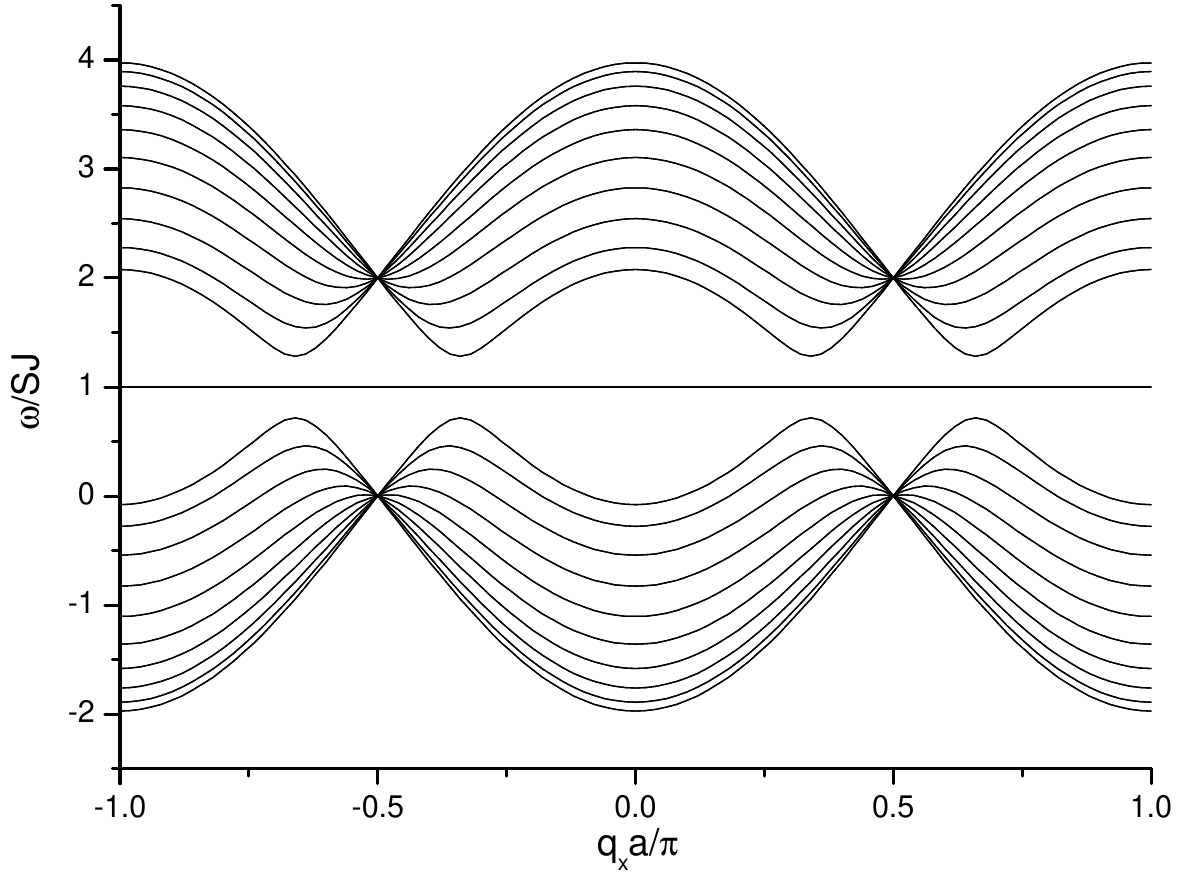}
\caption{Spin waves dispersion for zigzag 2D Heisenberg ferromagnetic dots honeycomb stripes
with $N=21$, $J=J_e=1$, $D=D_e=1$ and $\alpha=1.01$.} \label{fig:mgzf5}
\vspace{20pt}
\end{figure}

Figure \ref{fig:mgzf1} shows the dispersion relation for zigzag 2D Heisenberg ferromagnetic dots honeycomb stripe
with 20 lines without impurities, where the nearest neighbor exchange $J_{ij}$ has a constant
value $J$ through all the stripe including the stripe edges. The same goes for uniaxial anisotropy term $D_i$, which was chosen here to be
zero such that the  $\alpha$ is small and equal to $0.01$. In this case, the Fermi level is $0.01$, and the obtained dispersion relation is very near  to the obtained dispersion relation zigzag graphene ribbon with same size \cite{rim1}, as  uniaxial anisotropy term $D_i$ increase to $1$ the $\alpha$ increase to $1.01$. In this case, the Fermi level is $1.01$, and all the dispersion curves shifted (see Figure \ref{fig:mgzf2}) as we discussed before about $\alpha$  effect.
The figures \ref{fig:mgzf2} and \ref{fig:mgzf5} show that all modes have degeneracy of degree two which reflects the symmetry between the parallel rows of the two sublattices $A$ and $B$ in case of zigzag stripes. Also, the Figure  \ref{fig:mgzf2} shows the famous localized edge states at Fermi level around $q_x=\pm0.5\pi/a$ for even zigzag graphene ribbons \cite{Tao2011,JPSJ.65.1920,PhysRevB.54.17954,rim1}.  The dispersion relation for even rows stripes is different from odd row (see Figure \ref{fig:mgzf5}) stripes especially for edges localized states at Fermi level. This is due to the fact that edges states depends on the probability of exchange between a site in the edge to an interior site. The two edges sites
for even stripes have coordination number equal to 2 (i.e. each site at edges is connected to two interior sites of the stripe, where the stripe begins with sublattice A and ends with sublattice B (see Figure \ref{fig:graphenelattice3})). While the situation is different for the case of odd stripes, as one edge sites have coordination number equal to 2, while the other edge sites have coordination number equal to 1, as the odd stripe begins with sublattice A and ends with sublattice A (see Figure \ref{fig:graphenelattice3}). Which increases the localized edge states and extent it to fill the whole Brillouin zone for odd stripes.

\begin{figure}[hp!]
\centering
\includegraphics[scale=1]{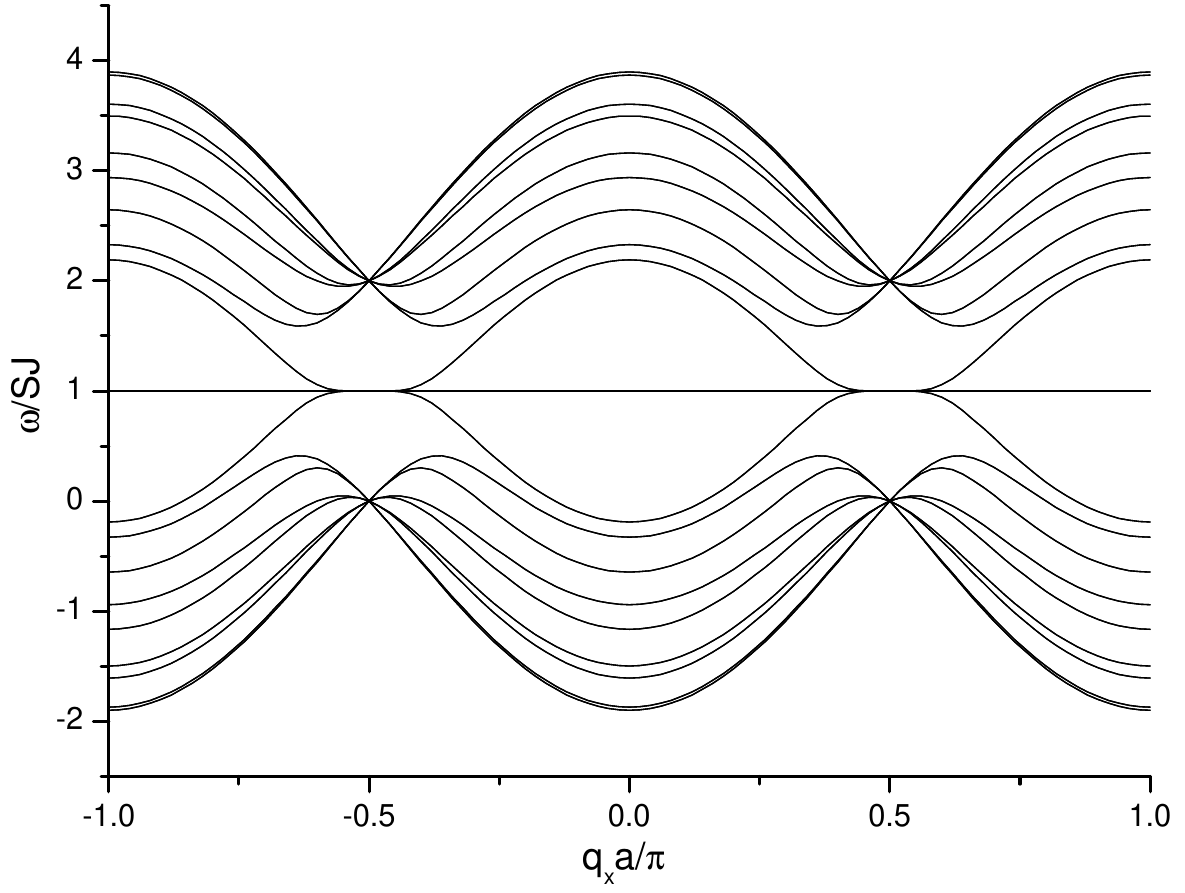}
\caption{Spin waves dispersion for zigzag 2D Heisenberg ferromagnetic dots honeycomb stripes with an impurity line at line number 11, where $N=20$, $J=J_e=1$, $J_I=0.0J$ $D=D_e=D_I=1.0$  and $\alpha=1.01$.} \label{fig:mgzf3}
\vspace{10pt}
\centering
\includegraphics[scale=1]{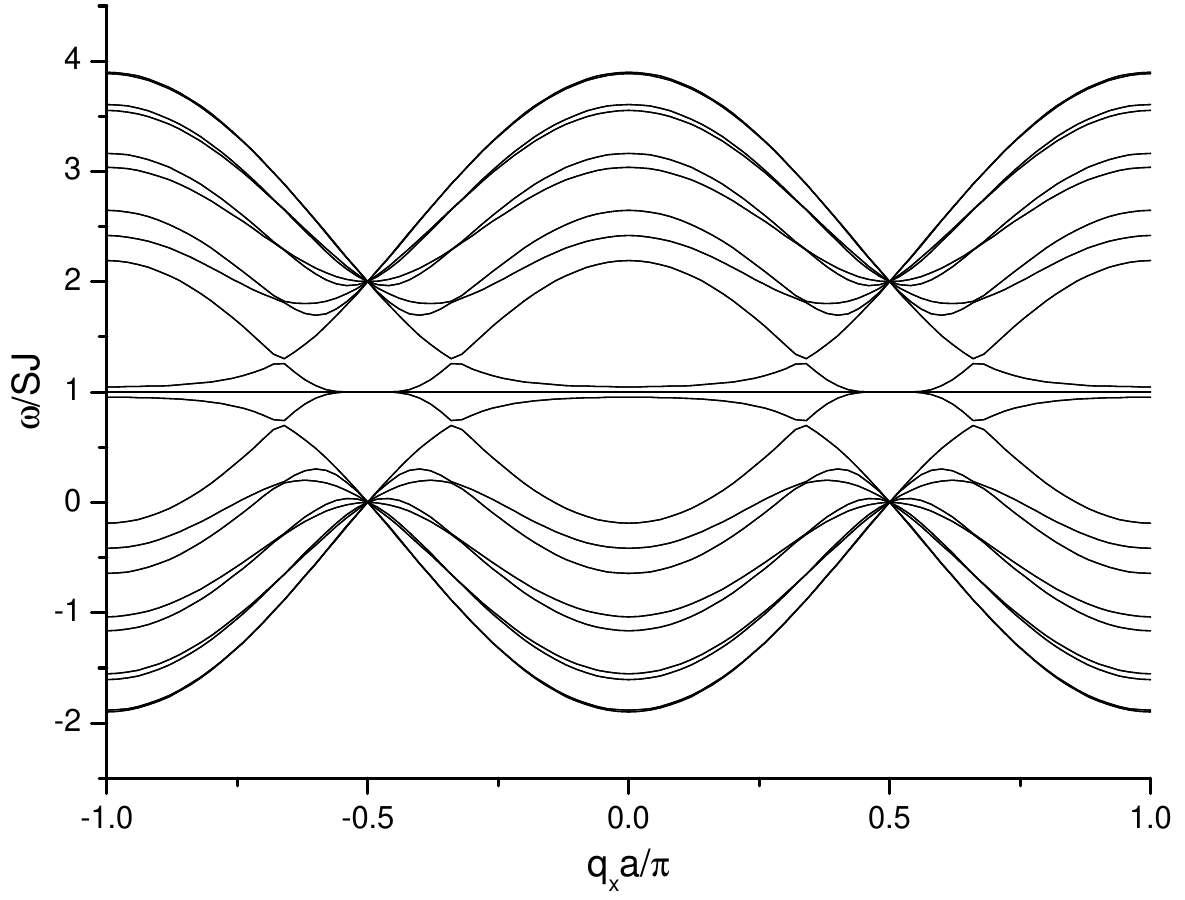}
\caption{Spin waves dispersion for zigzag 2D Heisenberg ferromagnetic dots honeycomb stripes with an impurity line at line number 11, where $N=21$, $J=J_e=1$, $J_I=0.0J$ $D=D_e=D_I=1.0$  and $\alpha=1.01$.} \label{fig:mgzf6}
\end{figure}

Figures \ref{fig:mgzf3} and \ref{fig:mgzf6} show the modified dispersion relations due to the effect of introducing substitutional a magnetic impurities line at row 11 of the zigzag stripes with 20 and 21 lines. The new dispersions show exactly the same behavior seen in the same case for zigzag graphene \cite{rim1}, but shifted in the case of magnetic stripes due to the effect of $\alpha$. The introducing of the impurities line has the effect of splitting the stripe to two interacted stripes with different sizes. In case of 20 line stripe the new stripes are 10 lines and 9 lines, in case of 21 line stripe the new two stripes each 10 lines. The strength of the interaction between the two sub stripes depends on the value of the impurities exchange value $J_I$, the figures show case when $J_I=0$. In this case, the expanded edge localized states in Fermi level are appear. Figure \ref{fig:mgzf6} shows the dispersions relation result for impurities line at row 11 of 20 lines zigzag stripe as a superposition of dispersion relations for even stripe (here 10 lines) with odd stripe (here 9 lines), while Figure \ref{fig:mgzf3}  shows
 the dispersions relation result for impurities line at row 11 of 21 lines zigzag stripe as a superposition of dispersion relations for two even stripe (here 10 lines) but shifted since the two stripes begin with different sublattice one $A$ and the other $B$ \cite{rim1}.

\begin{figure}[hp!]
\centering
\includegraphics[scale=1]{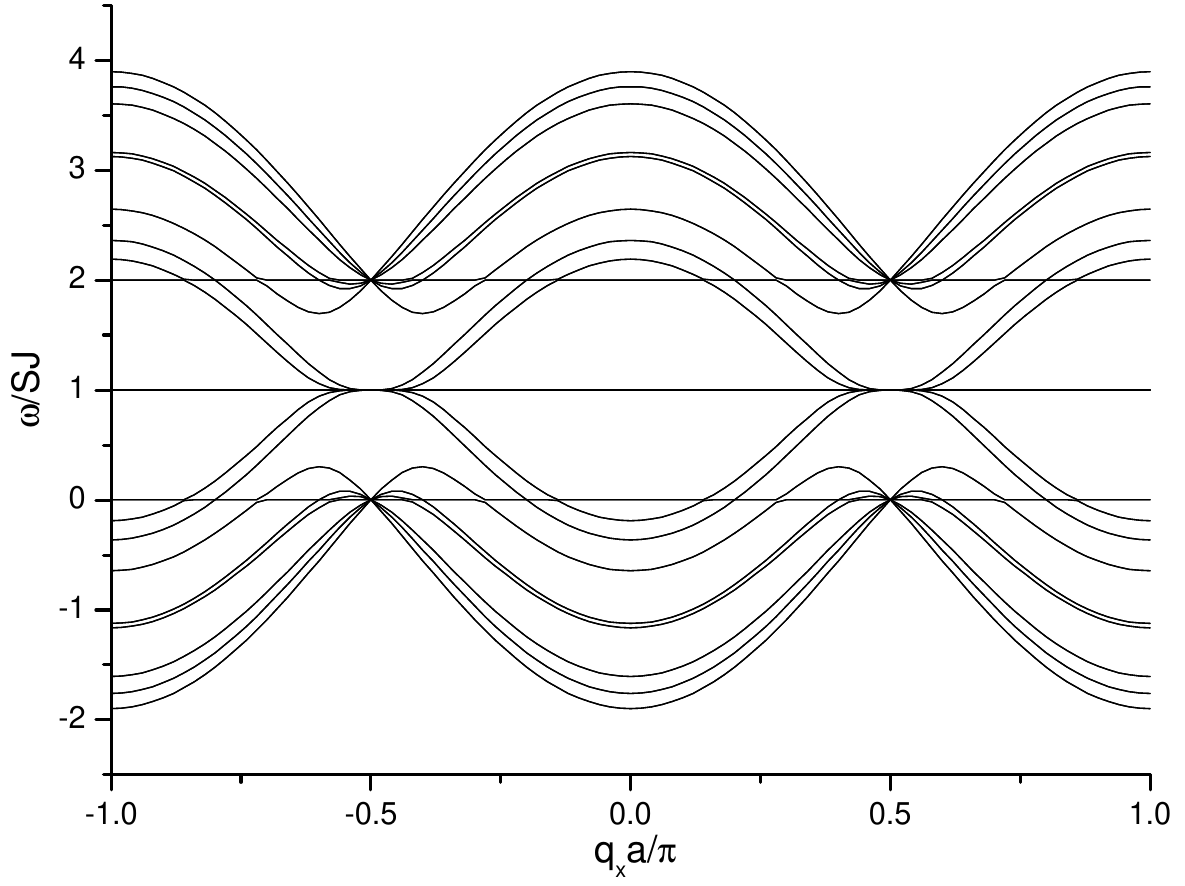}
\caption{Spin waves dispersion for zigzag 2D Heisenberg ferromagnetic dots honeycomb stripes with two impurities lines at line number 11 and line number 14, where $N=20$, $J=J_e=1$, $J_I=0.0J$, $J_{II}=0.0J$, $D=D_e=D_I=1.0$  and $\alpha=1.01$.} \label{fig:mgzf4}
\vspace{10pt}
\centering
\includegraphics[scale=1]{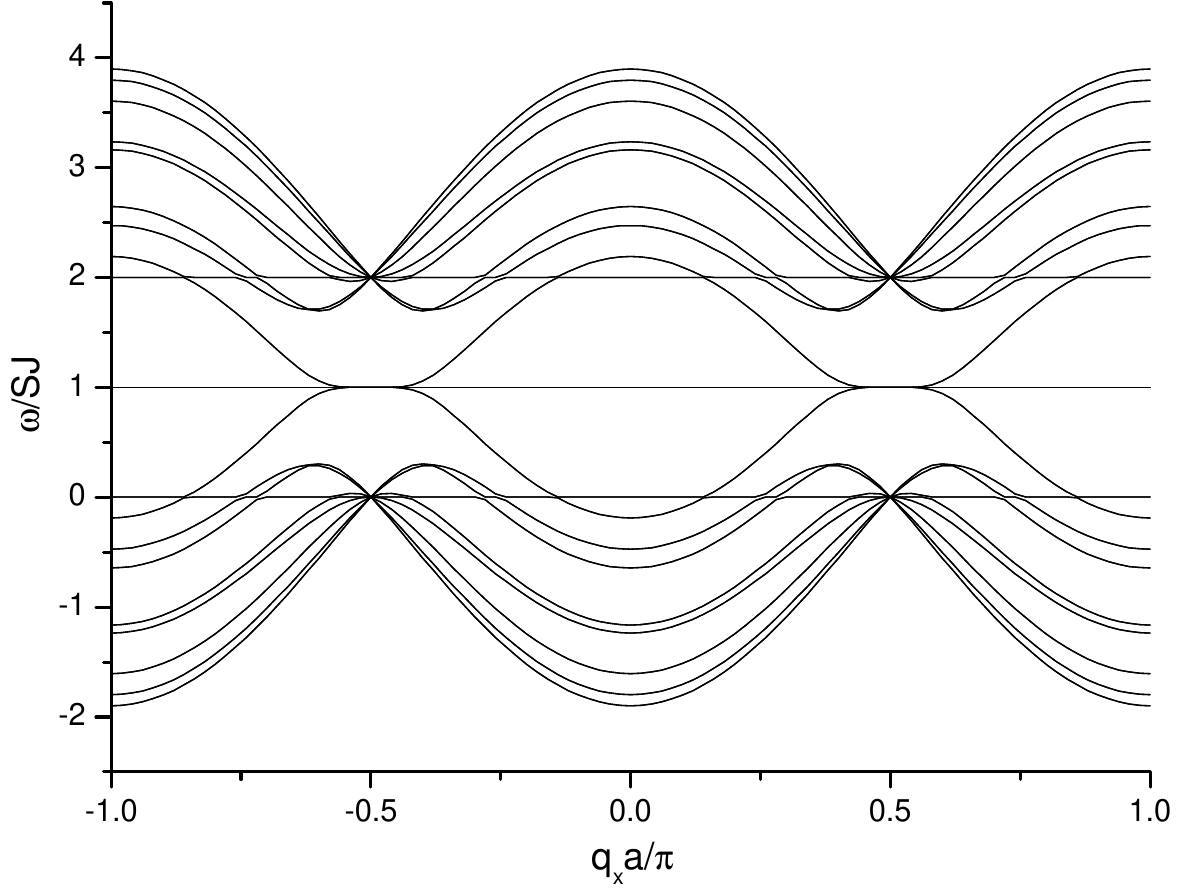}
\caption{Spin waves dispersion for zigzag 2D Heisenberg ferromagnetic dots honeycomb stripes with two impurities lines at line number 11 and line number 14, where $N=21$, $J=J_e=1$, $J_I=0.0J$, $J_{II}=0.0J$, $D=D_e=D_I=1.0$  and $\alpha=1.01$.} \label{fig:mgzf7}
\end{figure}

Figures \ref{fig:mgzf4} and \ref{fig:mgzf7} show the modified dispersion relations due to the effect of introducing magnetic impurities lines at rows 11 and 14 of the zigzag stripes with 20 and 21 lines. Again, the introduction of impurity lines has the effect of splitting the stripe to three interacted stripes with different sizes; in case of 20 line stripe the new stripes are 10 lines, 2 lines and 6 lines, in case of 21 line stripe the new two stripes  10 lines, 2 lines and 7 lines. The existence of stripe with 2 lines between zero exchange lines creates accumulation sites for magnons which then create two flat localized states: one in conduction band and the other in valance band. If the stripe size increases, the accumulation decreases and the localized states are removed.

We see from the results above that zigzag type stripes are gapless even when impurities are introduced. To study the effects of edges and impurities on their dispersions relations, some parameters are needed to reflect the important change in their dispersions relations due those effects. For the applications of zigzag graphene nanoribbons it has been found that the localized states at Fermi level are very important \cite{Bing,Neto1}.  Nakada et al.\cite{PhysRevB.54.17954}, used the density of states at Fermi level and at center band to study size effect on zigzag graphene ribbons. Here, we use the following two parameters to study the effects of edges and impurities on zigzag stripes: The first parameter is the relative density of states near the Fermi level (RDSFL), which computationally is calculated by finding the total number of points in the dispersion relations between  $\alpha-0.0002$ and $\alpha+0.0002$, which is relative measure for the density of states at Fermi level. The second
 is the relative density of states of center band (RDSCB), which computationally is calculated by the total number of points in the dispersion relations between $\alpha-1.0002$ and $\alpha+1.0002$, which is relative measure for the density of states at center band.

\subsubsection{The effect of zigzag stripe width on RDSFL and RDSCB}
To use the RDSFL and RDSCB as parameters to study the effects of edges and impurities on zigzag stripes dispersions relations, we need to make a calibration for those two parameters. We do that by studying the effect of zigzag stripes width on RDSFL and RDSCB, where edge and impurities effects are not taken in the account.

\begin{figure}[ht!]
\centering
\includegraphics[scale=1]{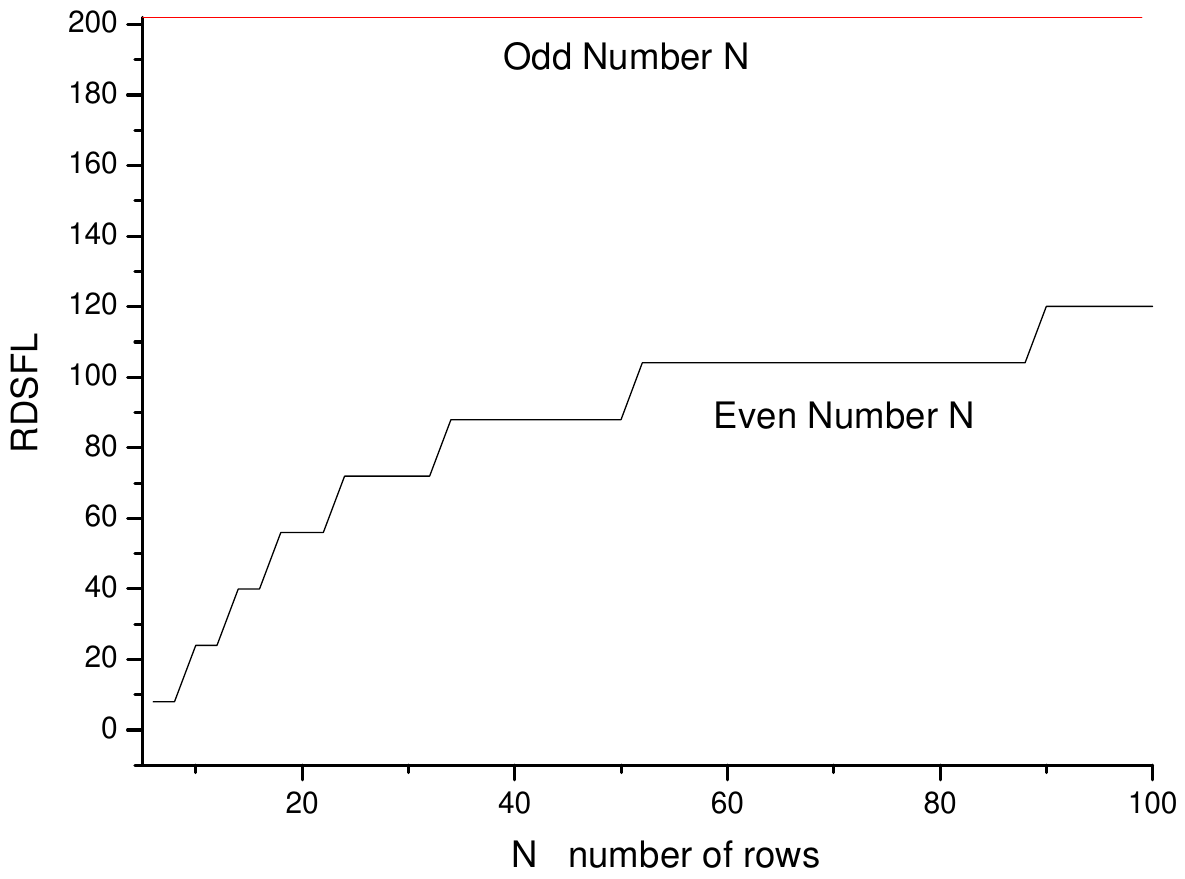}
\caption{The effect of zigzag stripe width on the relative density of states near Fermi level (RDSFL)} \label{fig:mgfw1}
\end{figure}
Figure \ref{fig:mgfw1} shows the effect of zigzag stripe width on RDSFL, which is dependent on the stripe width parity as odd or even. In the case of odd stripe width, the RDSFL is constant and independent on the stripe width. This is understood from the large contributions of the localized edge states that extend over the entire Brillouin zone for edge sites with 1 coordination number. For even stripe width, the situation is different.  The RDSFL is dependent on the stripe width, it is increase stepwise, where the step width increases with increasing stripe size. This could be explained by the fact that the probability of tunneling (or the diffusion length) for magnons between the two edges is high for small even width stripes and decreases with increasing even width stripes. This result for even width stipes is close to the results of Nakada et al.\cite{PhysRevB.54.17954} for graphene zigzag ribbon, with the caveat that their zigzag line numbering for width always gave them eve zigzag stripes with our convention.

\begin{figure}[h]
\centering
\includegraphics[scale=1]{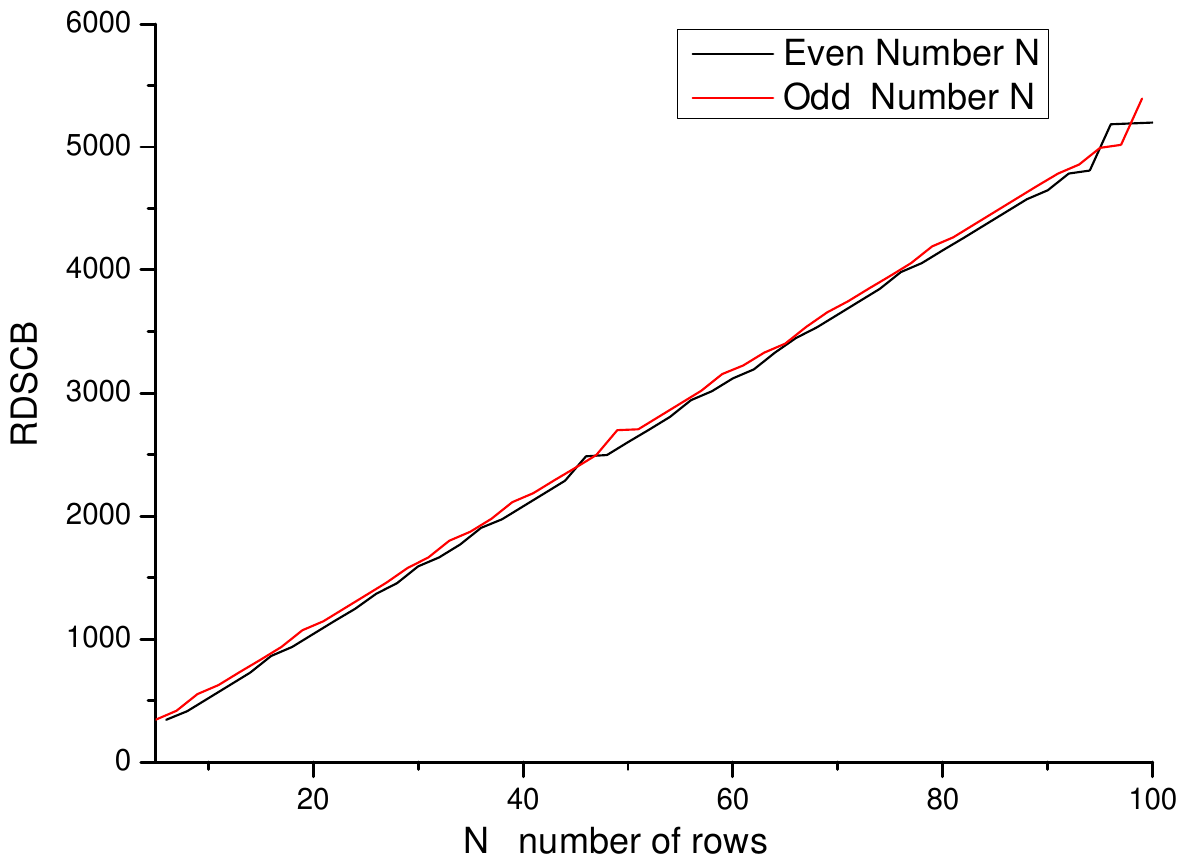}
\caption{The effect of zigzag stripe width on the relative density of states of center band (RDSCB)} \label{fig:mgfw2}
\end{figure}
While Figure \ref{fig:mgfw2} shows the effect of zigzag stripe width on RDSCB. RDSCB is weakly dependent on stripe width parity as odd or even (i.e. the odd RDSCB is almost equal to the even RDSCB, save for a slight increase). This is understood since RDSCB includes RDSFL in its computation for both even and odd stripe width, while the center band almost the same for even and odd stripes.  The RDSCB linearly increases with the stripe width for both even and odd width.

\subsubsection{The effects of edge uniaxial anisotropy on zigzag stripe RDSFL and RDSCB}
Figures \ref{fig:mgfw1} and \ref{fig:mgfw2} show RDSFL and RDSCB when the edge uniaxial anisotropy $D_e$ is equal to the stripe interior uniaxial anisotropy $D$. As $D_e$ becomes unequal to $D$, the RDSFL becomes zero for both even and odd stripes. This is understood since the edge localized states are a reflection of the decreasing probability of exchange interaction between the edge sites and interior sites, which in case of $D_e =D$ come from different coordination number for edge sites. At the same time, insite energy is equal between the edge and interior sites. As $D_e$ becomes unequal to $D$ a difference in insite energy between the edge and interior sites  is created. This difference in insite energy increases the probability of exchange interaction between the edge and interior sites which removes the localized states at the edge of zigzag stripes.
\begin{figure}[ht!]
\centering
\includegraphics[scale=1]{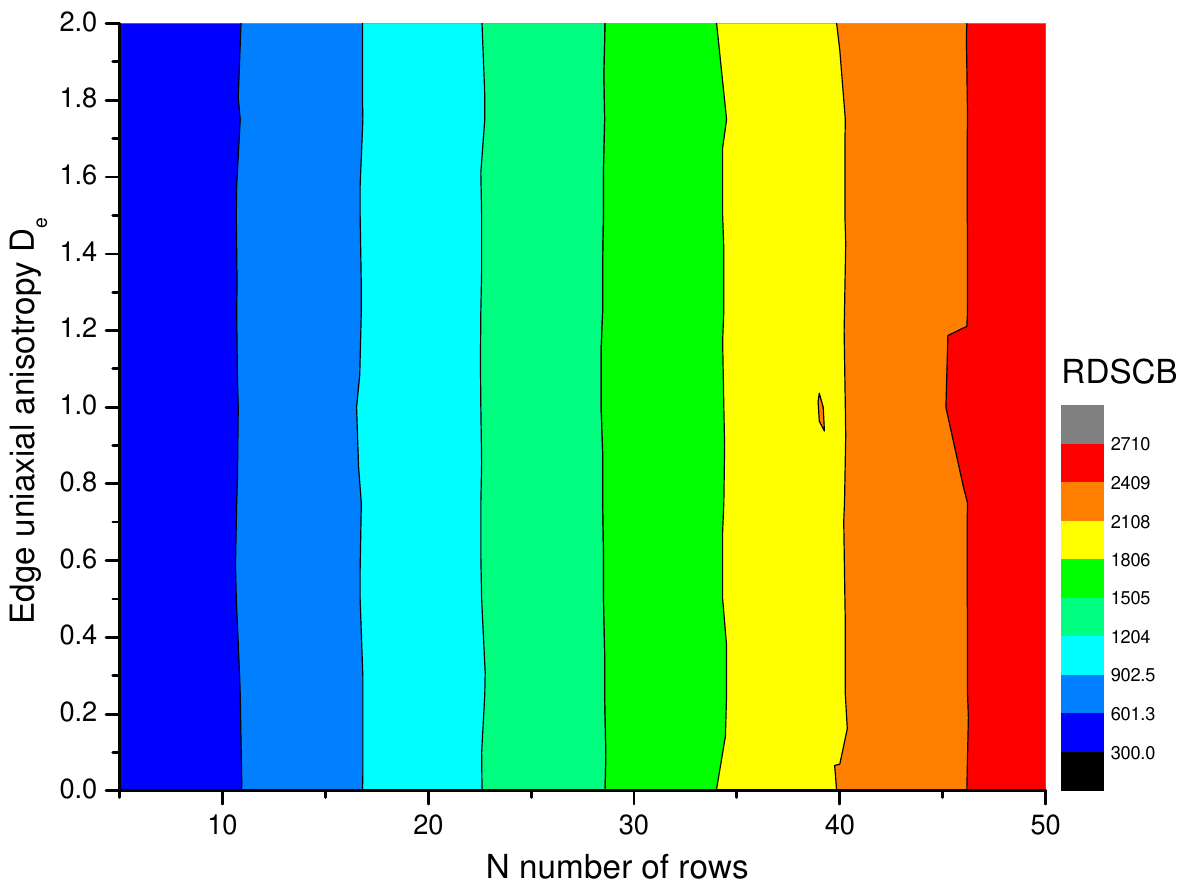}
\caption{The effects of edge uniaxial anisotropy and zigzag stripe width on its RDSCB} \label{fig:mgfedged2}
\end{figure}

While RDSFL becomes zero as $D_e\neq D$, Figure \ref{fig:mgfedged2} shows the color contour plot for the effects of edge uniaxial anisotropy and zigzag stripe width on its RDSCB. The figure shows that RDSCB is independent of the change of edge uniaxial anisotropy which is reflected in parallel colored stripes which is clear since the edge localized states change to center band states, which keep the RDSCB is nearly constant. The figure also shows that RDSCB increases with increasing the stripe width. This conclusion agrees with the result in Figure \ref{fig:mgfw2}.

\subsubsection{The effects of edge exchange on zigzag stripe RDSFL and RDSCB}
\begin{figure}[hp!]
\centering
\includegraphics[scale=1]{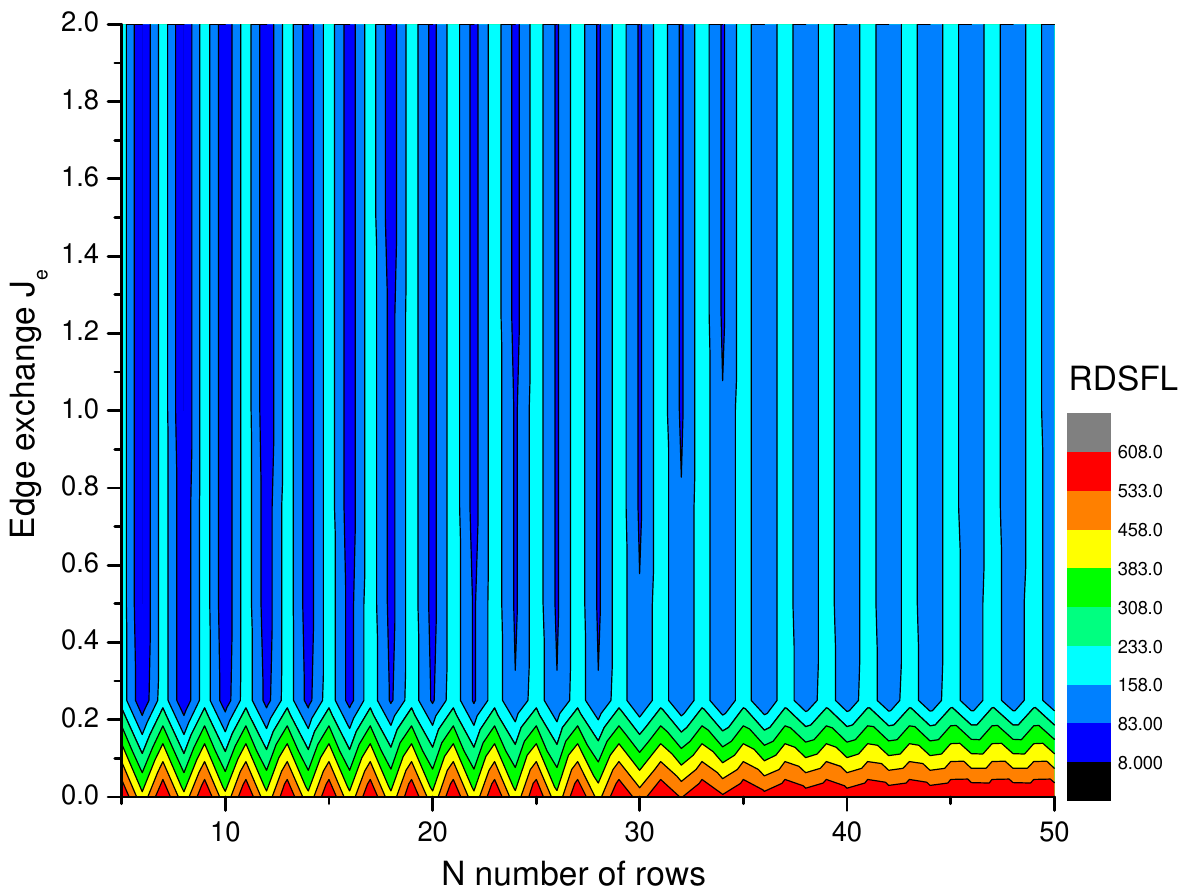}
\caption{The effects of edge exchange and zigzag stripe width on its RDSFL} \label{fig:mgfedge1}
\vspace{20pt}
\centering
\includegraphics[scale=1]{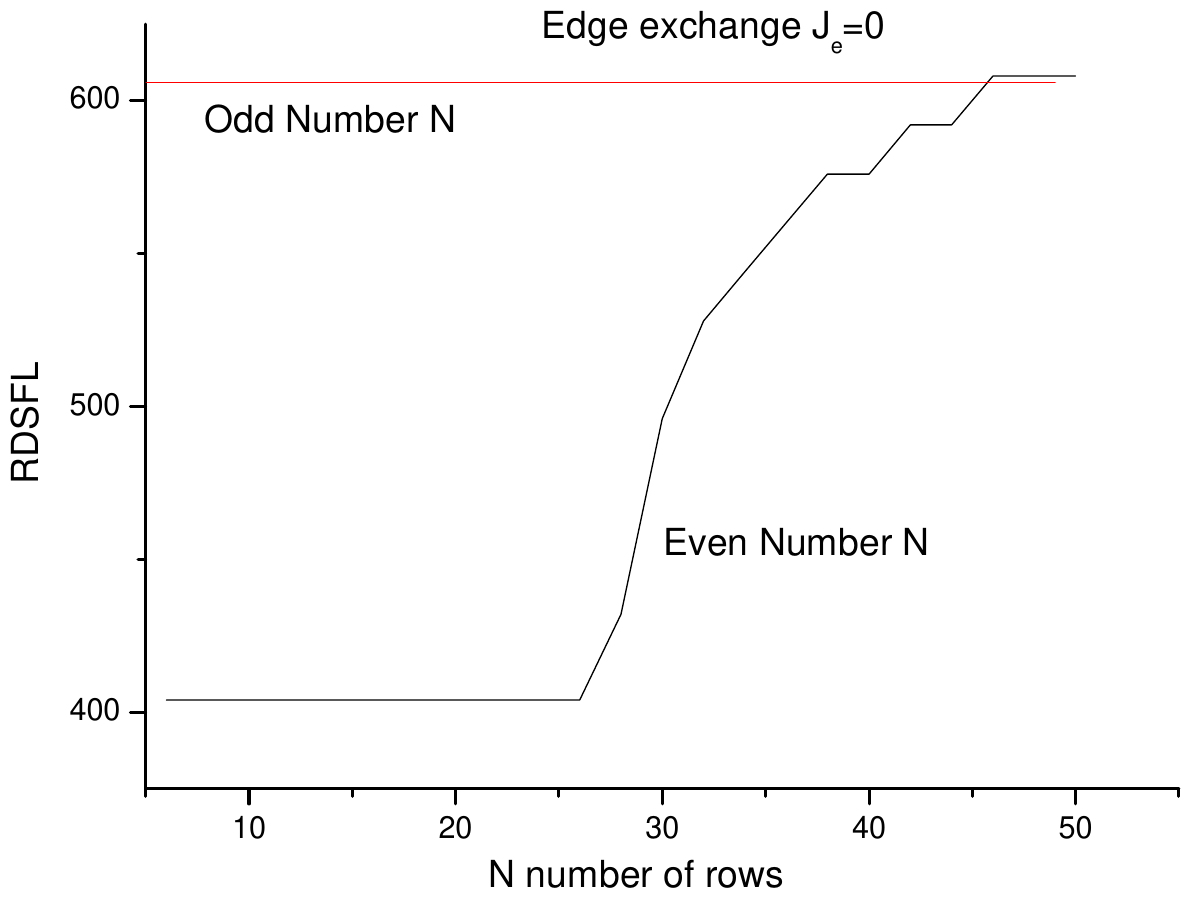}
\caption{The effect of edge exchange at $J_e=0$ and zigzag stripe width on its RDSFL} \label{fig:mgfedge2}
\end{figure}
Figure \ref{fig:mgfedge1} shows the color contour plot for the effects of edge exchange and zigzag stripe width on its RDSFL. In these types of plots, the line with edge exchange equal to interior exchange, i.e. $J_e=1$, represents the calibration line of the figure. Here, this line is no more than the Figure \ref{fig:mgfw1}. It is clear that RDSFL is nearly independent of the value of edge exchange. The main change on the RDSFL is when $J_e=0$. Figure \ref{fig:mgfedge2} shows the effect of edge exchange at $J_e=0$ and zigzag stripe width on its RDSFL, which by comparing it with Figure \ref{fig:mgfw1} shows the large increase in RDSFL values for both odd and even stripes width as the effect of edge exchange at $J_e=0$. The behavior of odd width stripes does not change with edge exchange at $J_e=0$. The behavior of even width stripes does change with edge exchange at $J_e=0$, first in the RDSFL become constant for small even width stripes at about $N=28$ the RDSFL has fast increase to become just above the RDSFL odd width stripes RDSFL, which show that the probability of tunneling of edge localized states depend on both the even stripe width and the edge exchange.

\begin{figure}[h!]
\centering
\includegraphics[scale=1]{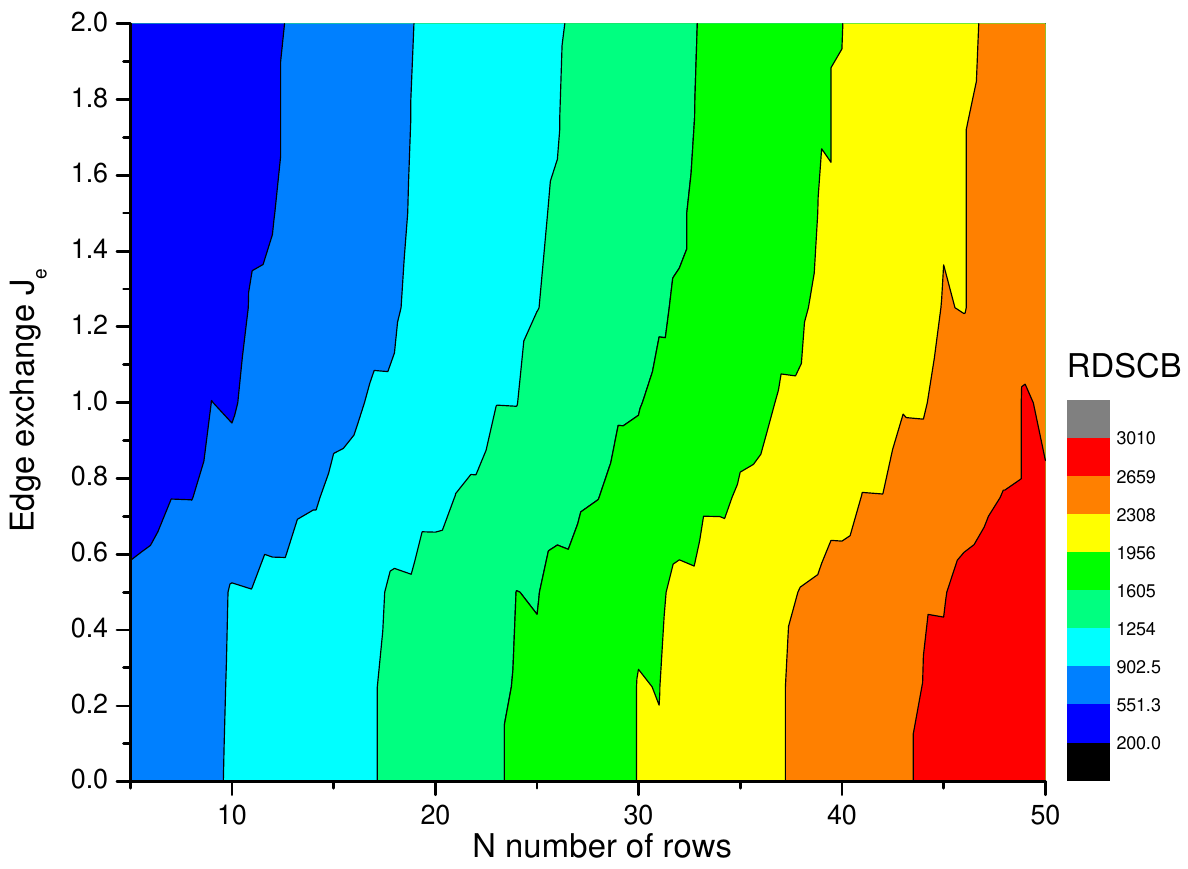}
\caption{The effects of edge exchange and zigzag stripe width on its RDSCB} \label{fig:mgfedge3}
\end{figure}

Figure \ref{fig:mgfedge3} shows the color contour plot for the effects of edge exchange and zigzag stripe width on its RDSCB.  It is clear from the figure that RDSCB decreases with an increase of the exchange. This decrease is particularly large in edge exchange range from 0.5 to 1.0 which shown as a curvature in the colored RDSCB stripes, while the RDSCB increases when increasing the stripe width. Out of edge exchange range 0.5 to 1.0 the RDSCB depends mainly on stripes width which is reflected in parallel colored stripes.

\subsubsection{The effects of impurities on zigzag stripe RDSFL and RDSCB}
The study of magnetic impurities effects on zigzag stripe is important for expected applications. In this section the results are represented  for
the effects of introducing one and two lines of magnetic impurities on zigzag 20 and 21 width stripes on their RDSFL and RDSCB.

There are two parameters for the impurities
that engineering the energy band for magnetic  zigzag  stripes: The first one is the strength of magnetic interaction represented by line of impurity exchange $J_I$  between the impurities line and the stripe materials \cite{rim1}, which here take the range of values from 0 to 2 in the units of stripe materials magnetic exchange $J$. The second parameter is the impurities line position, which can take the value from second to one line before the stripe end, the line position is alternative between even position number in sublattice B and odd position number in sublattice A (see Figure \ref{fig:graphenelattice3}).

\begin{figure}[hp]
\centering
\includegraphics[scale=1]{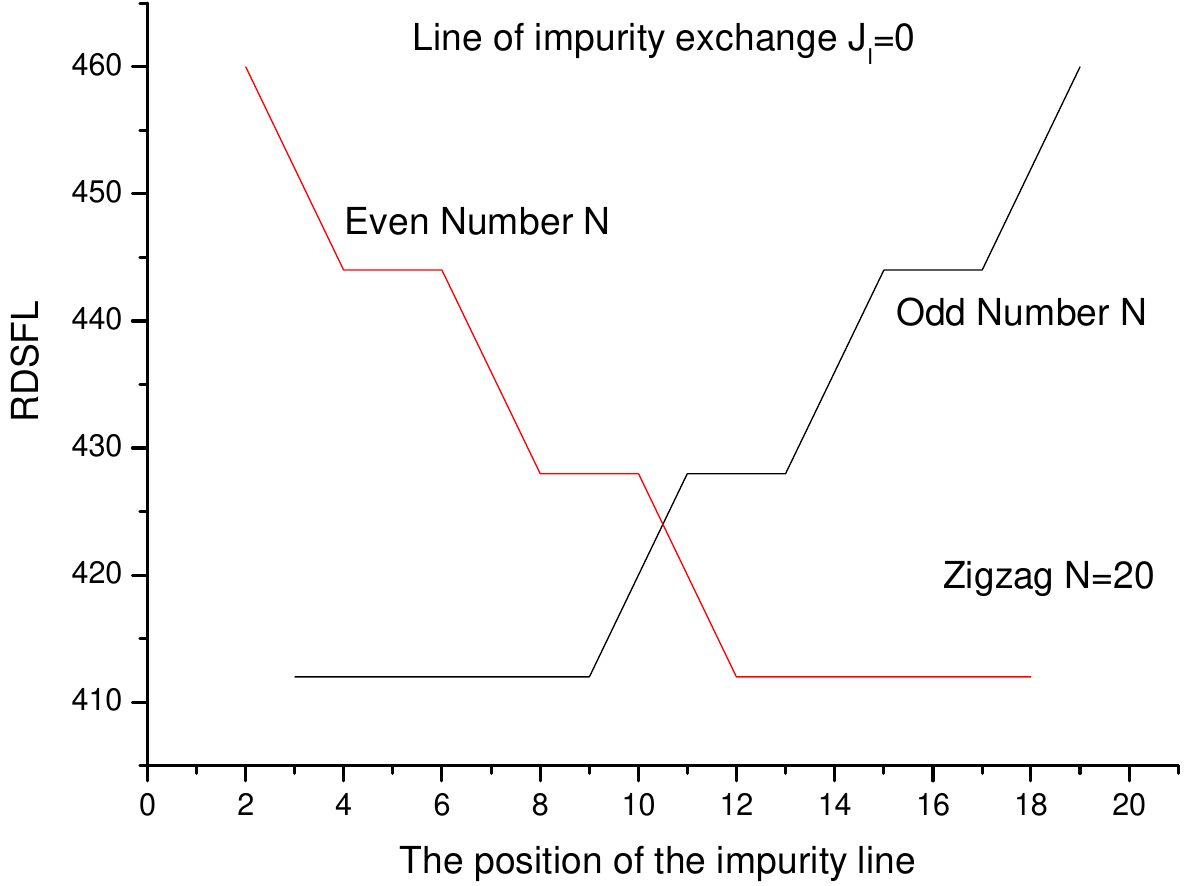}
\caption{The effect of one line of impurities position with impurities exchange $J_I=0$  for  $N=20$ zigzag stripe on its RDSFL} \label{fig:mgf20i1}
\vspace{20pt}
\centering
\includegraphics[scale=1]{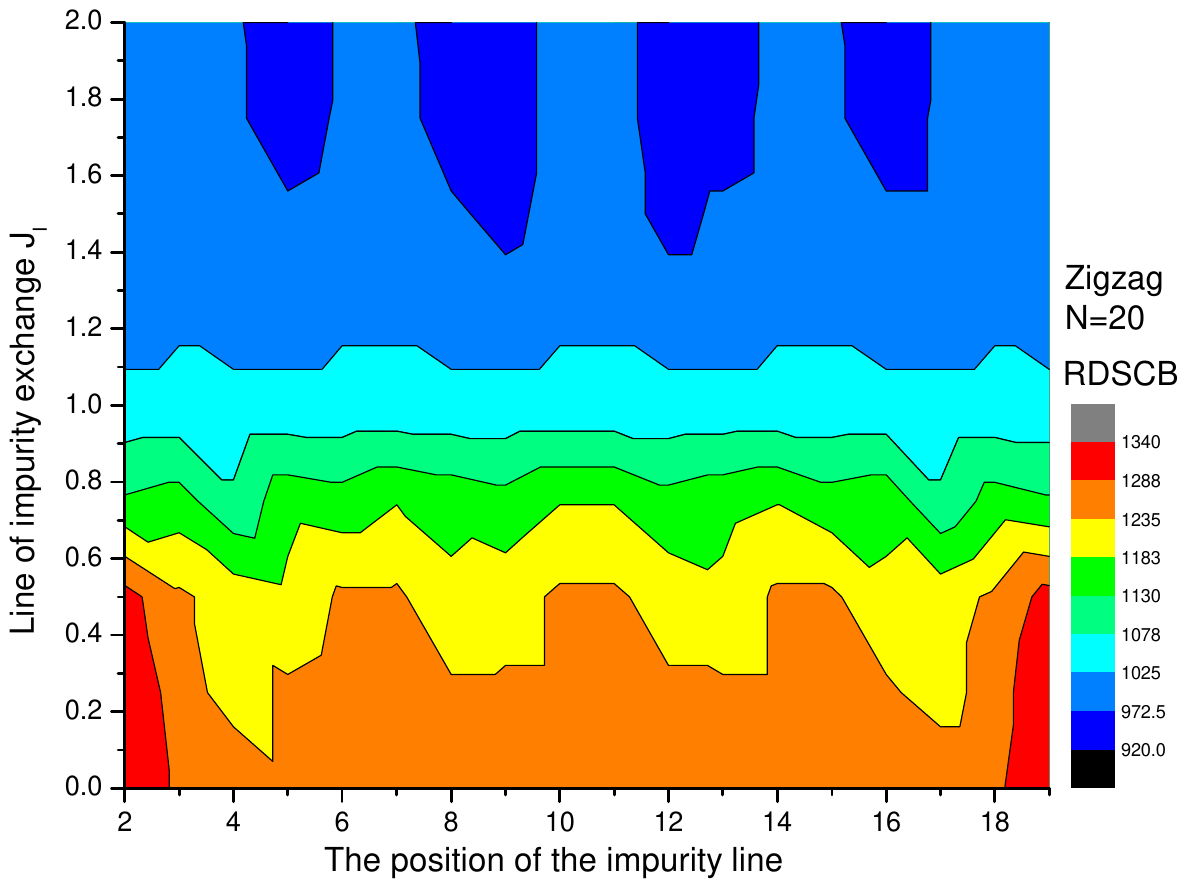}
\caption{The effects of one line of impurities position and impurities exchange  for  $N=20$ zigzag stripe on its RDSCB} \label{fig:mgf20i2}
\end{figure}

The first case to be shown here is the introduction of one line of magnetic impurities on zigzag 20 width stripe. The results for the effects of the position of one impurities line with impurities exchange $J_I$ from 0 to 2 for  $N=20$ zigzag stripe show that RDSFL is nearly independent of the position of impurities line and it is impurities exchange value except when $J_I=0$ which is similar to the case of edge exchange effect.

Figure \ref{fig:mgf20i1} shows the effect of the impurities line position for impurities exchange at $J_I=0$ on RDSFL of zigzag 20 width stripe. RDSFL is dependent on the parity of impurities line position. The RDSFL for even positions of impurities line begins high and decreases stepwise and beginning from position 12 become constant until position 18. The RDSFL for odd positions of impurities line have opposite behaviors; it begins small constant and beginning from position 11 begin to increases stepwise.

Since edge localized states at Fermi level depend mainly on the edge geometry and the width of zigzag stripes, we expected that above behavior is related to the geometries and the widths of zigzag sub stripes and their interaction.  The geometries of zigzag sub stripes as follows for even positions the impurities line is in the sublattice B and subdivide the stripe to one odd stripe type A (i.e. begin and end with sublattice A) and one even stripe begin with sublattice A and end with sublattice B. As the even position increases, the odd stripe A increases and the even stripe decreases. It is clear that odd stripe contribution is higher in edge localized. This is because of its edge has 1 coordination number, while the even stripe has both edges with 2 coordination number.

For odd positions the impurities line is in the sublattice A and subdivide the stripe to one even stripe begin with sublattice A and end with sublattice B and odd stripe type B (i.e. begin and end with sublattice B). As the odd position increase the even stripe increases, and the odd stripe decreases. It is therefore clear that odd stripe contribution is higher in edge localized. This occurs when one of its edge has 1 coordination number, while the even stripe has both edges with 2 coordination number.

Figure \ref{fig:mgf20i2} shows the effects of one line of impurities position and impurities exchange for $N=20$ zigzag stripe on its RDSCB, beginning from impurities exchange with value $1.0$ which means no impurities. As the impurities exchange increases, the RDSCB decreases. As impurities exchange reach 1.4 the RDSCB decreases more at the line of impurities positions 9 and 12. As impurities exchange from 1.6 to 2, the RDSCB decreases more at the line of impurities positions 5, 8, 9, 12, 13 and 16. As impurities exchange decreases than 1.0, the RDSCB increases as impurities exchange reach 0.4 the RDSCB increases more at the line of impurities positions 3, 6, 7, 10, 11, 14, and 15. As impurities exchange from 0.4 to 0.0, the RDSCB has peak values for impurities line positions 2 and 19.

\begin{figure}[hp]
\centering
\includegraphics[scale=1]{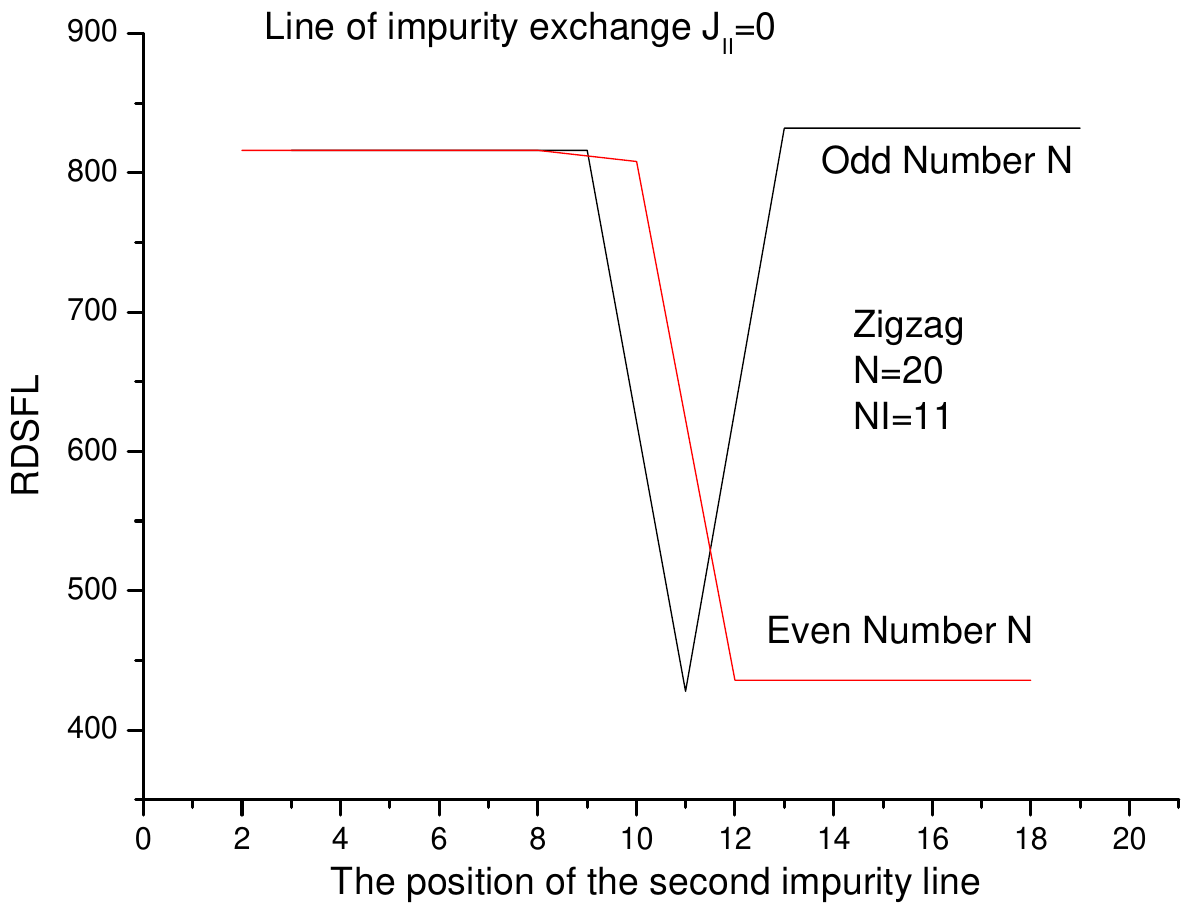}
\caption{The effect of second line of impurities position with impurities exchange $J_{II}=0$ for $N=20$ zigzag stripe with one line of impurities @ $N=11$ with impurities exchange $J_I=0$ on its RDSFL} \label{fig:mgf20ii1}
\vspace{20pt}
\centering
\includegraphics[scale=1]{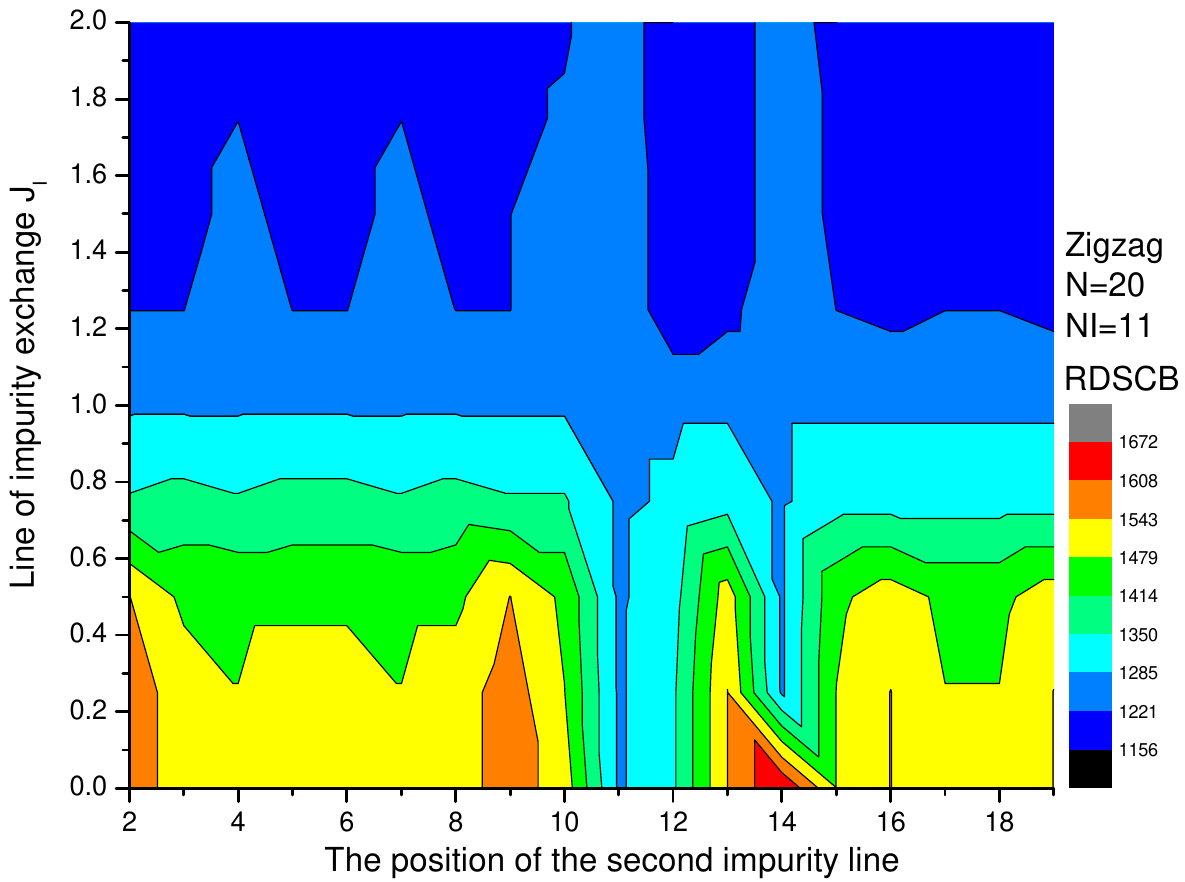}
\caption{The effect of second line of impurities position and impurities exchange for $N=20$ zigzag stripe  with one line of impurities @ $N=11$ with impurities exchange $J_I=0$  on its RDSCB} \label{fig:mgf20ii2}
\end{figure}

The addition of second impurities line to the zigzag stripe increases the possibility to tune the magnetic properties of the stripes to suite more  expected technological devices applications. Figures \ref{fig:mgf20ii1} and \ref{fig:mgf20ii1} show the effects of second line of impurities position and its impurities exchange on RDSFL and RDSCB of $N=20$ zigzag stripe.

The results for the effects of second impurities line position with impurities exchange $J_{II}$ from 0 to 2 for $N=20$ zigzag stripe with first line of impurities at position 11 with impurities exchange $J_{I}=0$  show that RDSFL is nearly independent on the position of second impurities line and its impurities exchange value except when $J_{II}=0$ which is similar to the cases of one impurities line and edge exchange effect.

Figure \ref{fig:mgf20ii1} shows the effect of second  impurities  line position with impurities exchange $J_{II}=0$ for $N=20$ zigzag stripe with one line of impurities at $N=11$ with impurities exchange $J_I=0$ on its RDSFL. The addition of second line increases the RDSFL more than one line of impurities and the behavior change since the new sub stirpes edges geometries and their interaction change. The RDSFL becomes nearly constant and independent on the second line position parity until the first impurities line position. The second line superimposed on the first line left only the effect of first line on the RDSFL. After this position the even position number switch to low nearly constant value while odd position number switch back to high nearly constant value.

Figure \ref{fig:mgf20ii2} shows the effect of second impurities line position with impurities exchange $J_{II}=0$ for $N=20$ zigzag stripe with one line of impurities at $N=11$ with impurities exchange $J_I=0$ on its RDSCB, beginning from impurities exchange with value $1.0$ which means no impurities. As impurities exchange increases, the RDSCB decreases and as impurities exchange reach 1.4, the RDSCB decreases more at certain second impurities line positions. As impurities exchange from 1.8 to 2, the RDSCB decreases more at most second impurities line positions.  As impurities exchange decreases than 1.0, the RDSCB increases gradually in most positions.  As second impurities line positions 2 and 9 there are especial increase in RDSCB at impurities exchange from 0.4 to 0.0.

At first impurities line position, the second line superimposed on the first line left only the effect of first line on the RDSCB. The second impurities line position 14  have especial RDSCB behavior due to the existence of stripe with 2 lines
between two zero exchange lines creating accumulation sites for magnons which create
two localized states: one in conduction band and the other in valance band displayed as apeak RDSCB at second impurities exchange equal to $0$ as the impurities exchange increases the accumulation sites decreased very fast to remove the two localized states and drop RDSCB to the value of second impurities exchange equal to one.

\begin{figure}[hp]
\centering
\includegraphics[scale=1]{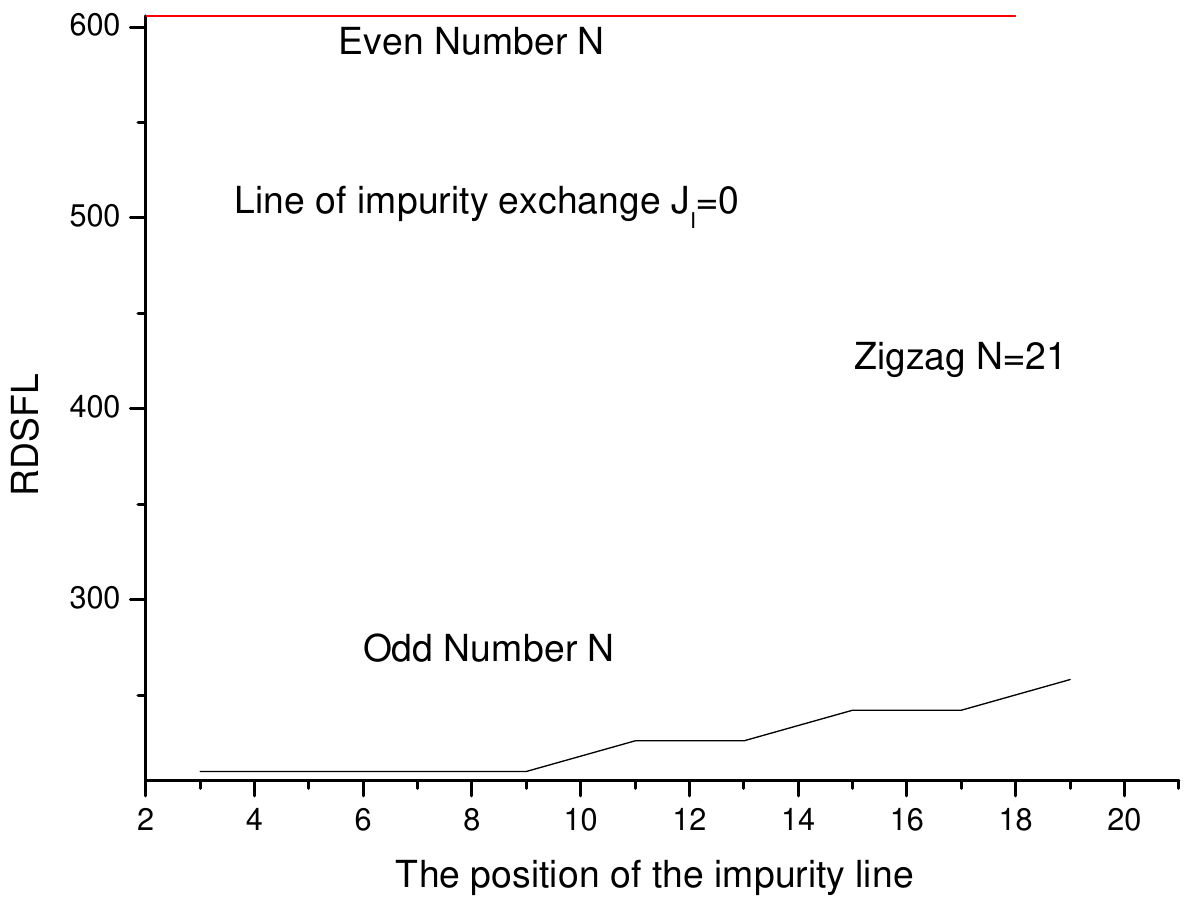}
\caption{The effect of one line of impurities position with impurities exchange $J_I=0$  for  $N=21$ zigzag stripe on its RDSFL} \label{fig:mgf21i1}
\vspace{20pt}
\centering
\includegraphics[scale=1]{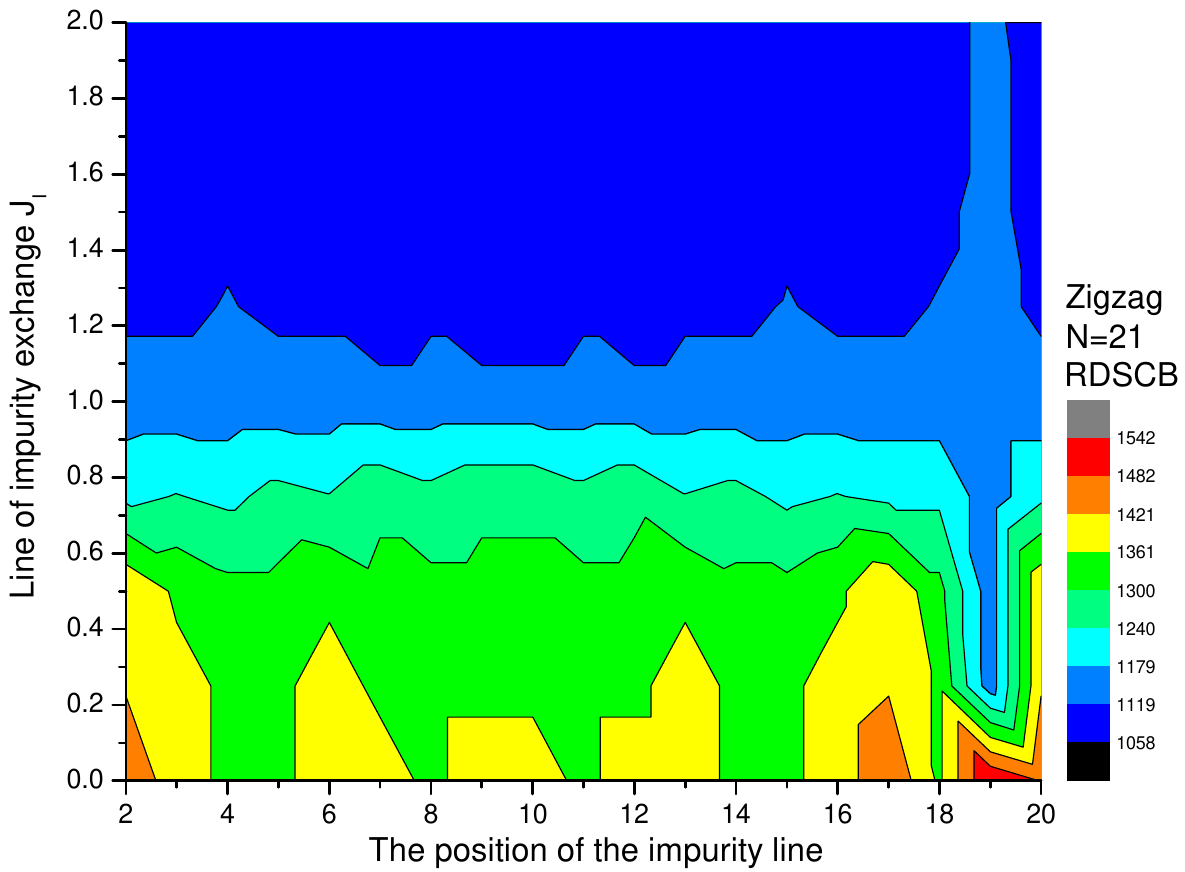}
\caption{The effect of one line of impurities position and impurities exchange  for  $N=21$ zigzag stripe on its RDSCB} \label{fig:mgf21i2}
\end{figure}

The results for the effects of one impurities line position with impurities exchange $J_{I}$ from 0 to 2 for $N=21$ zigzag stripe show that RDSFL is nearly independent on the position of the impurities line and its impurities exchange value except when $J_{I}=0$ which is similar to the cases of one impurities line and edge exchange effect.

Figure \ref{fig:mgf21i1} shows the effect of the impurities line position for impurities exchange at $J_I=0$ on RDSFL of zigzag 21 width stripe. RDSFL is dependent on the parity of impurities line position. The RDSFL for even positions of impurities line is high and constant independent on even lines positions which is very similar to general zigzag odd stripes behavior. While the RDSFL for odd positions of impurities line have revers behavior it begins small constant and beginning from position 9 begin to increases slowly stepwise which is similar to general zigzag even stripes behavior.

The above behavior is related to the geometries and the widths of zigzag sub stripes and their interaction in this case. The geometries of zigzag sub stripes as follow for even positions the impurities line is in the sublattice B and subdivide the stripe to two odd stripe type A (i.e. begin and end with sublattice A).  As the even position increases the first odd stripe A increases and the other odd stripe A decreases. It is clear that both odd stripe contribution is equal in edge localized states since one of their edges has 1 coordination number, which result very similar behavior to general zigzag odd stripes.

While for odd positions the impurities line is in the sublattice A and subdivide the stripe to one even stripe begin with sublattice A and end with sublattice B and even stripe too but begin with sublattice B and end with sublattice A. As the odd position increases first even stripe increases and the other even stripe decreases, the interaction between the two even stripes gives very similar behavior to general zigzag even stripes.

Figure \ref{fig:mgf21i2} shows the effects of one line of impurities position and impurities exchange for $N=21$ zigzag stripe on its RDSCB, beginning from impurities exchange with value $1.0$ which mean no impurities, as impurities exchange increases the RDSCB decreases as impurities exchange reach 1.2 the RDSCB decreases more at all lines of impurities positions except at position 19.

As impurities exchange decreases than 1.0, the RDSCB increases gradually in most position.  At impurities line positions 2 and 17 there are especial increases in RDSCB at impurities exchange from 0.2 to 0.0. Impurities line position 19 have special RDSCB behavior due to the existence of stripe with 2 lines between one zero exchange line and the edge creating accumulation sites for magnons. This creates two localized states: one in conduction band and the other in valance band displayed as apeak RDSCB at impurities exchange equal to $0$ as the impurities exchange increases the accumulation sites decreased very fast to remove the two localized states and drop RDSCB to the value of second impurities exchange equal to one.

\begin{figure}[hp]
\centering
\includegraphics[scale=1]{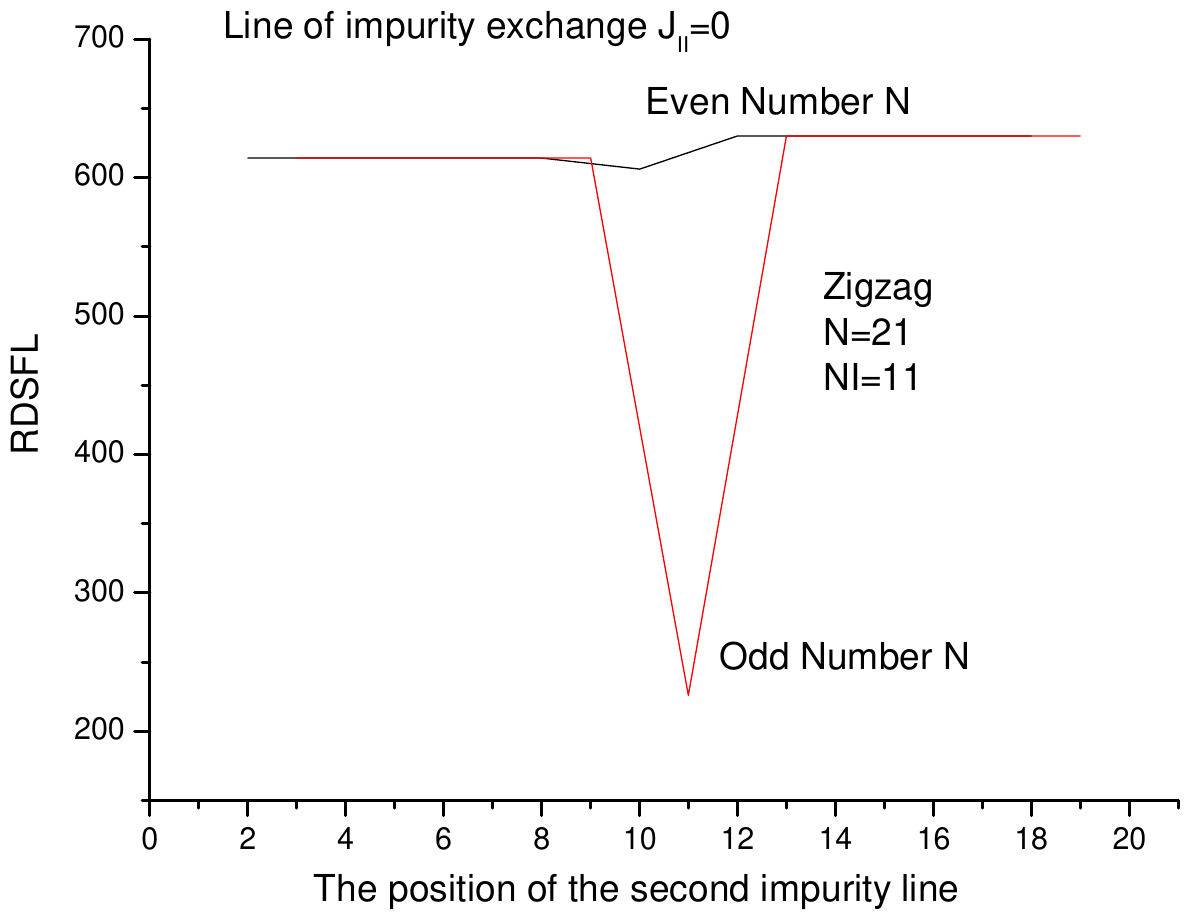}
\caption{The effect of second line of impurities position with impurities exchange $J_{II}=0$ for $N=21$ zigzag stripe with one line of impurities @ $N=11$ with impurities exchange $J_I=0$ on its RDSFL} \label{fig:mgf21ii1}
\vspace{20pt}
\centering
\includegraphics[scale=1]{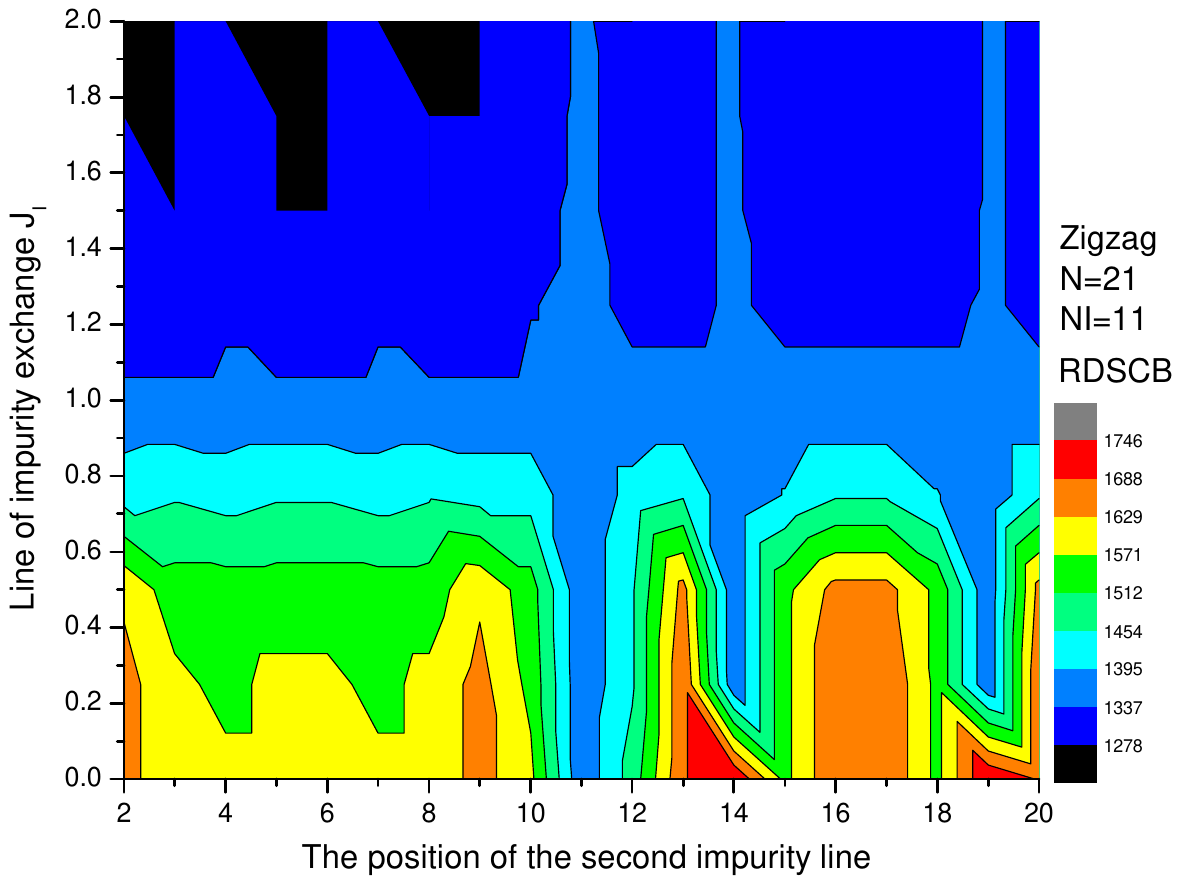}
\caption{The effect of second line of impurities position and impurities exchange for $N=20$ zigzag stripe  with one line of impurities @ $N=11$ with impurities exchange $J_I=0$  on its RDSCB} \label{fig:mgf21ii2}
\end{figure}

The results for the effects of second impurities line position with impurities exchange $J_{II}$ from 0 to 2 for $N=21$ zigzag stripe with first line of impurities at position 11 with impurities exchange $J_{I}=0$  show that RDSFL is nearly independent on the position of second impurities line and its impurities exchange value except when $J_{II}=0$ which is similar to many cases  before.

Figure \ref{fig:mgf21ii1} shows the effect of second  impurities  line position with impurities exchange $J_{II}=0$ for $N=21$ zigzag stripe with one line of impurities at $N=11$ with impurities exchange $J_I=0$ on its RDSFL. The addition of second line increase the RDSFL more than one line of impurities and the behavior change since the new sub stirpes edges geometries and their interaction change. The RDSFL become nearly constant and independent on the second line position parity until the first impurities line position, and the second line superimposed on the first line left only the effect of first line on the RDSFL. After this position the even position number keep nearly constant value while odd position number switch back to high nearly constant value.

Figure \ref{fig:mgf21ii2} shows the effect of second impurities line position with impurities exchange $J_{II}$ for $N=21$ zigzag stripe with one line of impurities at $N=11$ with impurities exchange $J_I=0$ on its RDSCB, beginning from impurities exchange with value $1.0$ which means no impurities. As impurities exchange increases the RDSCB decreases as impurities exchange reach 1.6 the RDSCB decreases more at certain second impurities line positions.   As impurities exchange decreases than 1.0, the RDSCB increases gradually in most positions.  At second impurities line positions 2, 9, 16, and 17 there are special increases in RDSCB as impurities exchange from 0.4 to 0.0.

At first impurities line position, the second line superimposed on the first line left only the effect of first line on the RDSCB. The second impurities line positions 14 and 19 have special RDSCB behavior due to the existence of stripe with 2 lines
between two zero exchange lines creating accumulation sites for magnons. This creates
two localized states: one in conduction band and the other in valance band displayed as a peak RDSCB at second impurities exchange equal to $0$ as the impurities exchange increases the accumulation sites decreased very fast to remove the two localized states and drop RDSCB to the value of second impurities exchange equal to one.

\subsection{Armchair stripes results}

\begin{figure}[hp]
\centering
\begin{tabular}{cc}
\includegraphics[scale=.7]{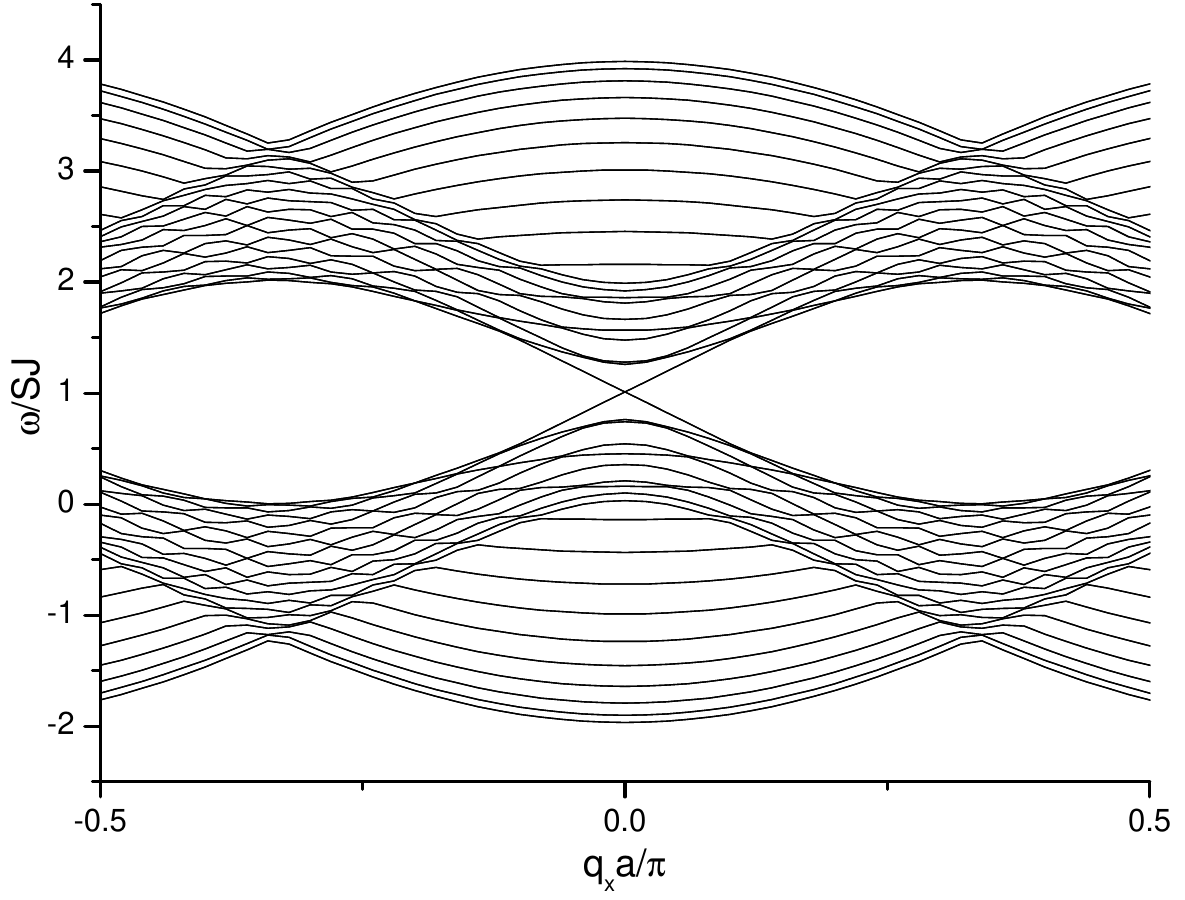}
 \\
\includegraphics[scale=.7]{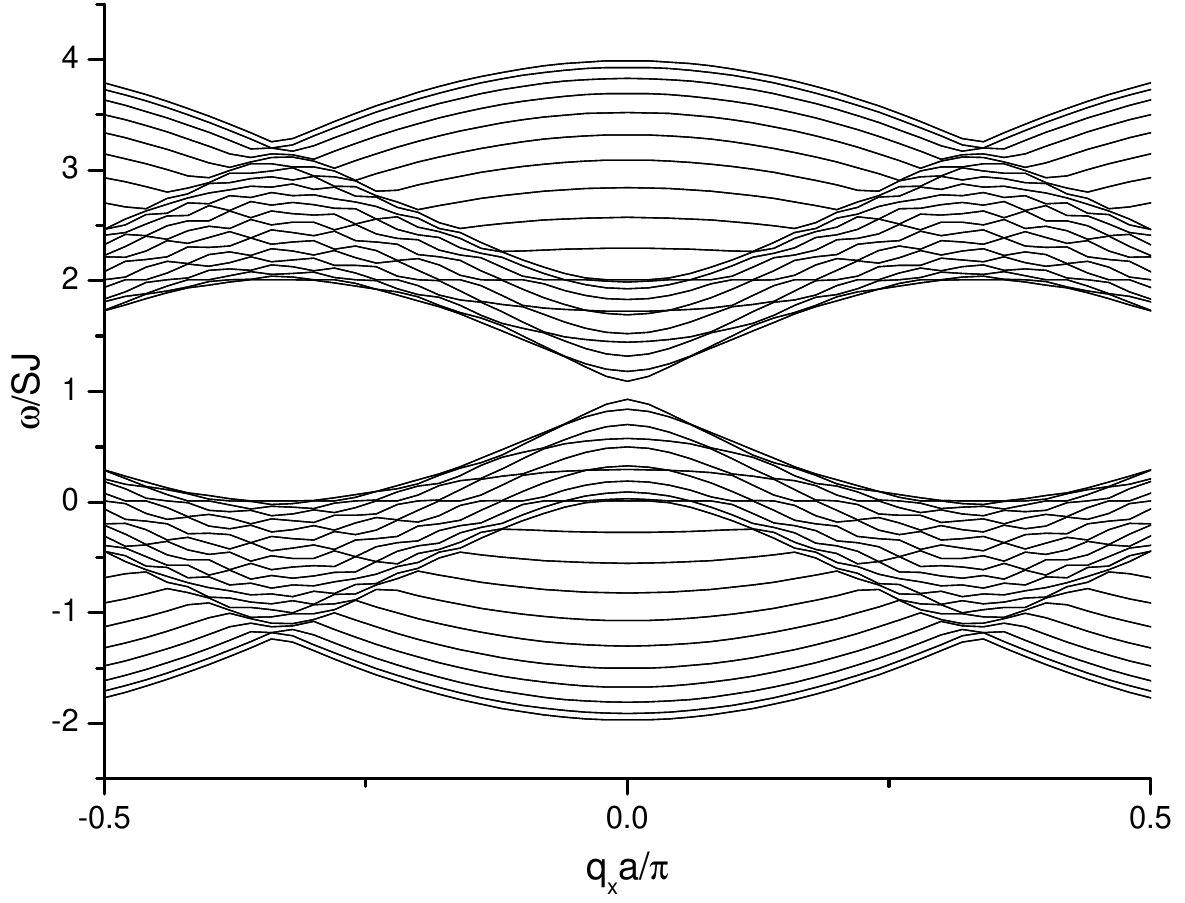} \\
\includegraphics[scale=.7]{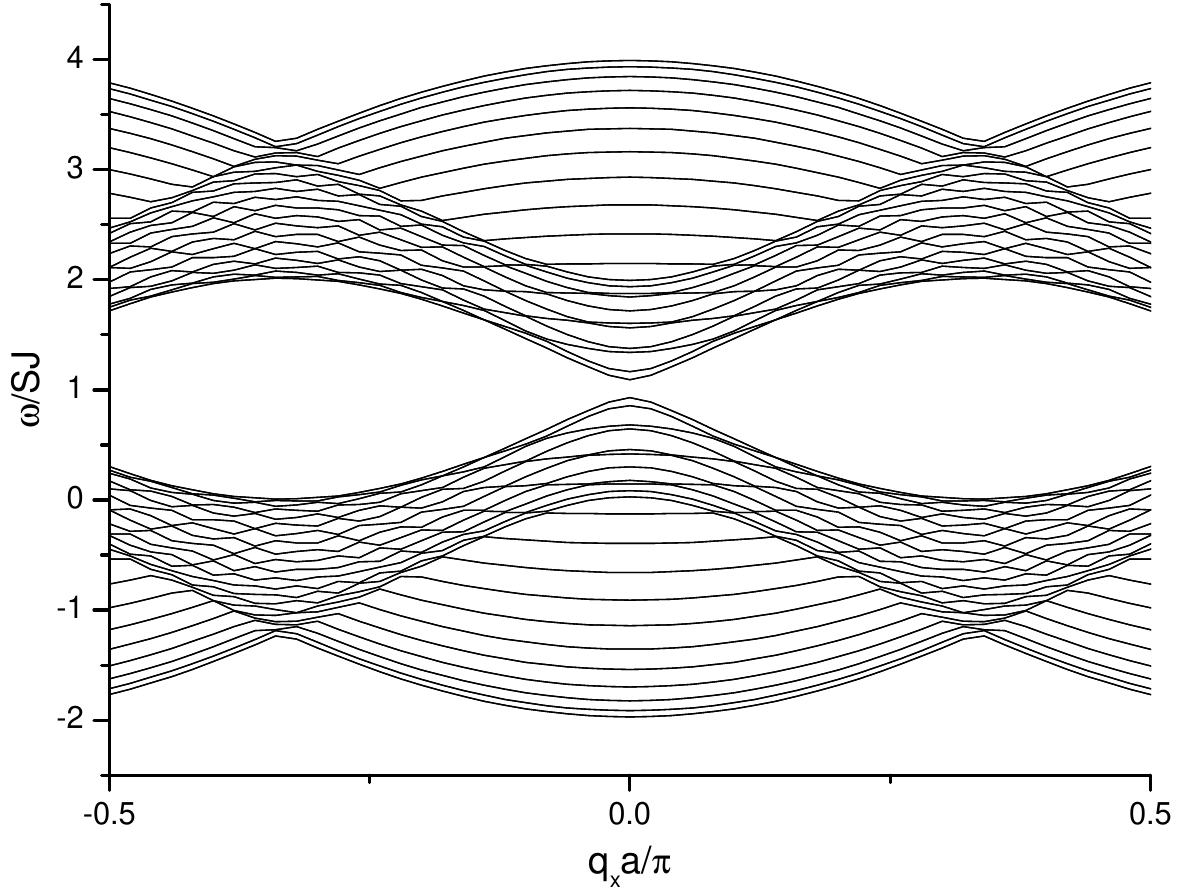}
\end{tabular}
\caption{Spin waves dispersion for armchair 2D Heisenberg ferromagnetic honeycomb stripes
with  $J=J_e=1.0$, $D=D_{e}=1.0$,  and $\alpha=1.01$ for $ N=20,21$ and 22 respectively.}\label{figarmcair1}
\end{figure}

Figures \ref{figarmcair1} show the dispersion relations for three armchair stripes with widths
20, 21, and 22. Where the nearest neighbor exchange $J_{ij}$ has a constant
value $J$ through all the stripe including the stripe edges, the same for uniaxial anisotropy term $D_i$, the obtained dispersion relation is very near to the obtained dispersion relation for armchair graphene ribbons with same size \cite{rim1}, only the dispersion curves here are shifted due to $\alpha$  effect.

It is clear that the shape of the dispersion relations for armchair stipes depends on the
stripe width. In general, the minimum of conduction band and the maximum of valence band are located at
$q_x=0$ for each stripe. In 20 lines stripe they touch each other at the Dirac point, while for 21 and 22 lines stripes they have two different types band gaps. This behavior is a famous behavior seen in graphene armchair ribbons \cite{PhysRevB.75.165414,rim1}. This repeated pattern of the dispersion relations for armchair stripes can be described mathematically as periodic function in the number of
lines as $3i$ and $3i+1$ for stripe with band gaps  while $3i+2$ for gap less stripes where $i=1,2,3,\cdots$. The origin of this behavior is understood as consequence of topologically ladder system nature for armchair geometry, i.e. here cyclic chains with interchain hopping  \cite{PhysRevB.54.17954,PhysRevB.75.165414,PhysRevB.73.045432,JPSJ.65.1920,PhysRevB.49.8901,PhysRevB.46.3159,H.Hosoya1990}.

In armchair geometry the sites from sublattice A are in the same line with sites from sublattice B, which is not the case in zigzag stripes. This removes the symmetry between adjusted lines and eliminates the degeneracy in armchair stripes without impurities \cite{rim1}. There is no localized edge states show up in armchair without impurities as in case of graphene \cite{JPSJ.65.1920,PhysRevB.54.17954,rim1}.

\begin{figure}[hp]
\centering
\begin{tabular}{cc}
\includegraphics[scale=.7]{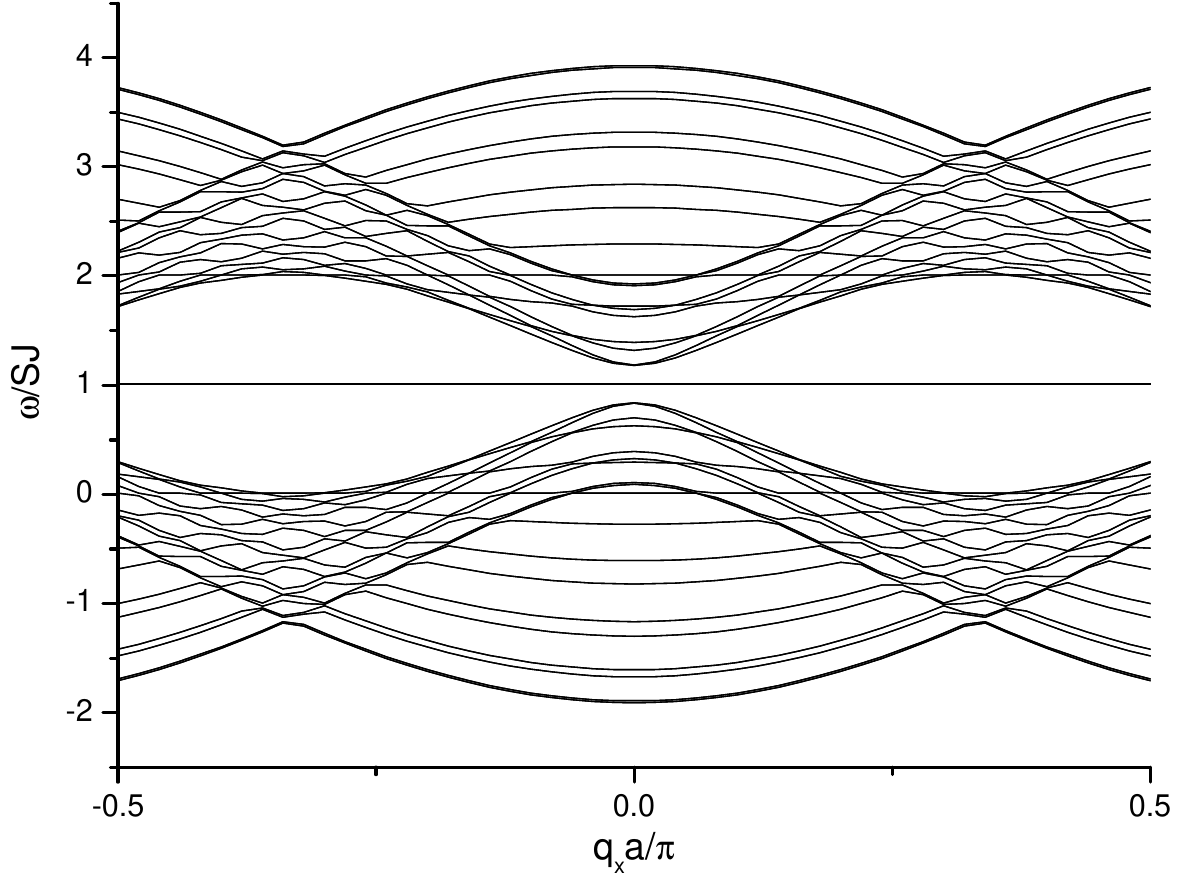}
 \\
\includegraphics[scale=.7]{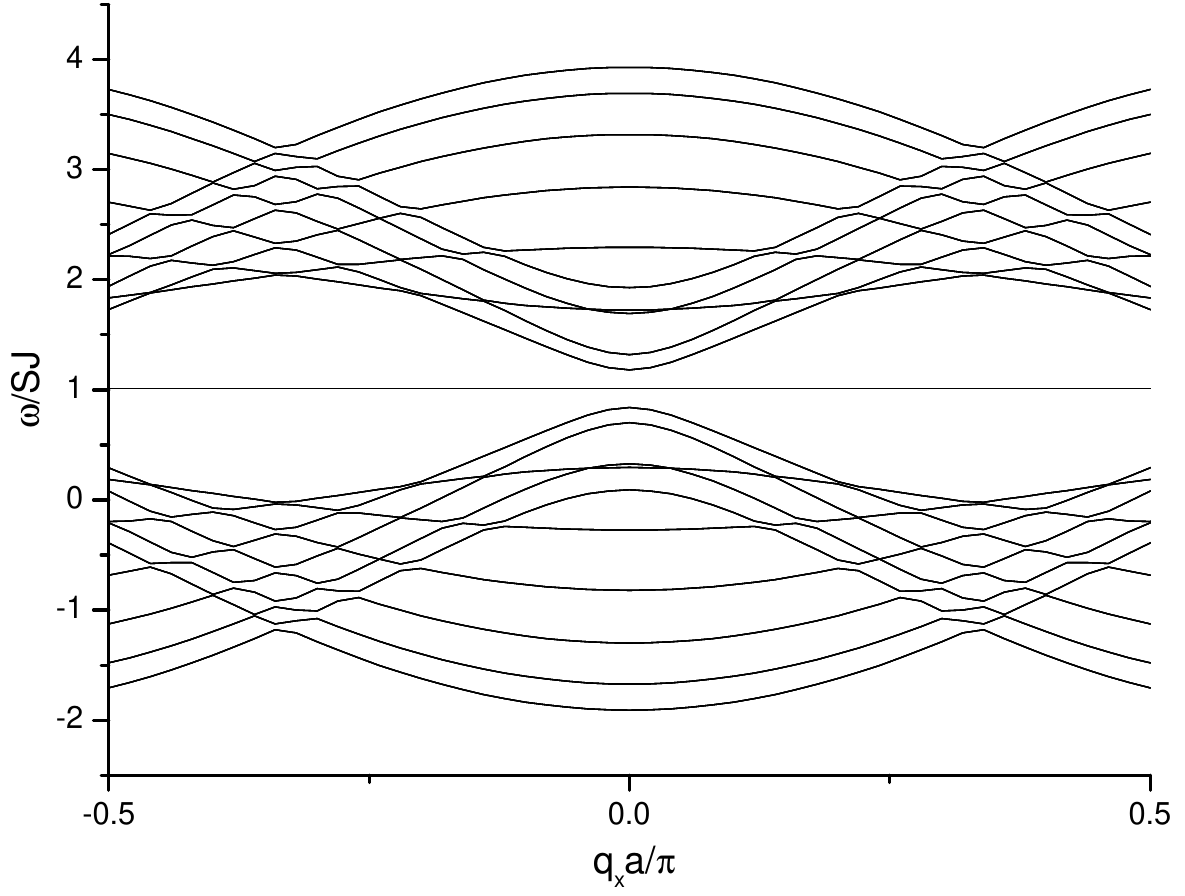} \label{fig:mgfarm2i}\\
\includegraphics[scale=.7]{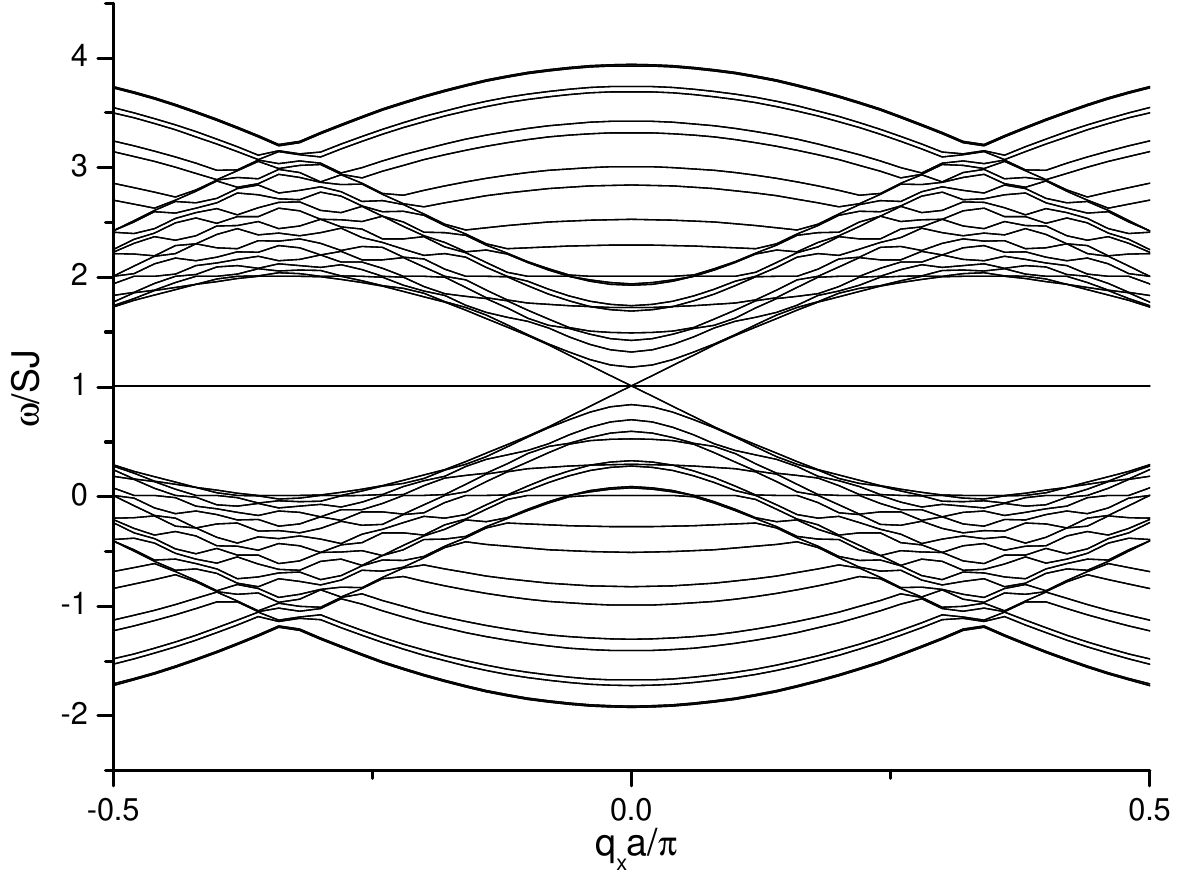}\label{fig:mgfarm3i}
\end{tabular}
\caption{Spin waves dispersion for armchair 2D Heisenberg ferromagnetic honeycomb stripes
with an impurity line at line number 11, where $J=J_e=1$, $J_I=0.0J$ $D=D_e=D_I=1.0$  and $\alpha=1.01$  for $ N=20,21$ and 22 respectively.}\label{fig:mgfarm1i}
\end{figure}

Figures \ref{fig:mgfarm1i} show the modified dispersion relations due to the effect of
introducing substitutional a magnetic impurities line at row 11 of the armchair stripes
with 20, 21, and 22 lines.  Again as in the case of magnetic zigzag stripe, the new dispersions with impurities line for armchair magnetic stripes show exactly the same behavior seen in
the same case for armchair graphene ribbons \cite{rim1}, but shifted in the case of magnetic stripes due the $\alpha$ effect.
The introduction of the impurities line have the effect as the case of zigzag stripes which is splitting the stripe to two interacted stripes with different sizes. In case of 20 line stripe the new stripes are 10 lines and 9 lines, in case of 21 line stripe the new two stripes each 10 lines which lead to completely degenerate dispersion, and in case of 22 line stripe the new two stripes new stripes are 10 lines and 11 lines. The strength of the interaction between the two sub stripes depends on the value of the impurities exchange value $J_I$. The figures show case when $J_I=0$, in this case the expanded edge localized states in Fermi level are appear as the case of zigzag stripes. Those localized states are understood as accumulation sites for magnons in the interface created by the tunneling between the two substripes through the impurities line, as the distance between the two substripes increase the localized states density decreases until the total dispersion for the system show non-interacted individual dispersions for the two substripes without any localized states, while the situation is different in the case of zigzag stripes where there are intrinsic edge localized states beside the one due to the impurities line tunneling interface.

\begin{figure}[hp]
\centering
\begin{tabular}{cc}
\includegraphics[scale=.7]{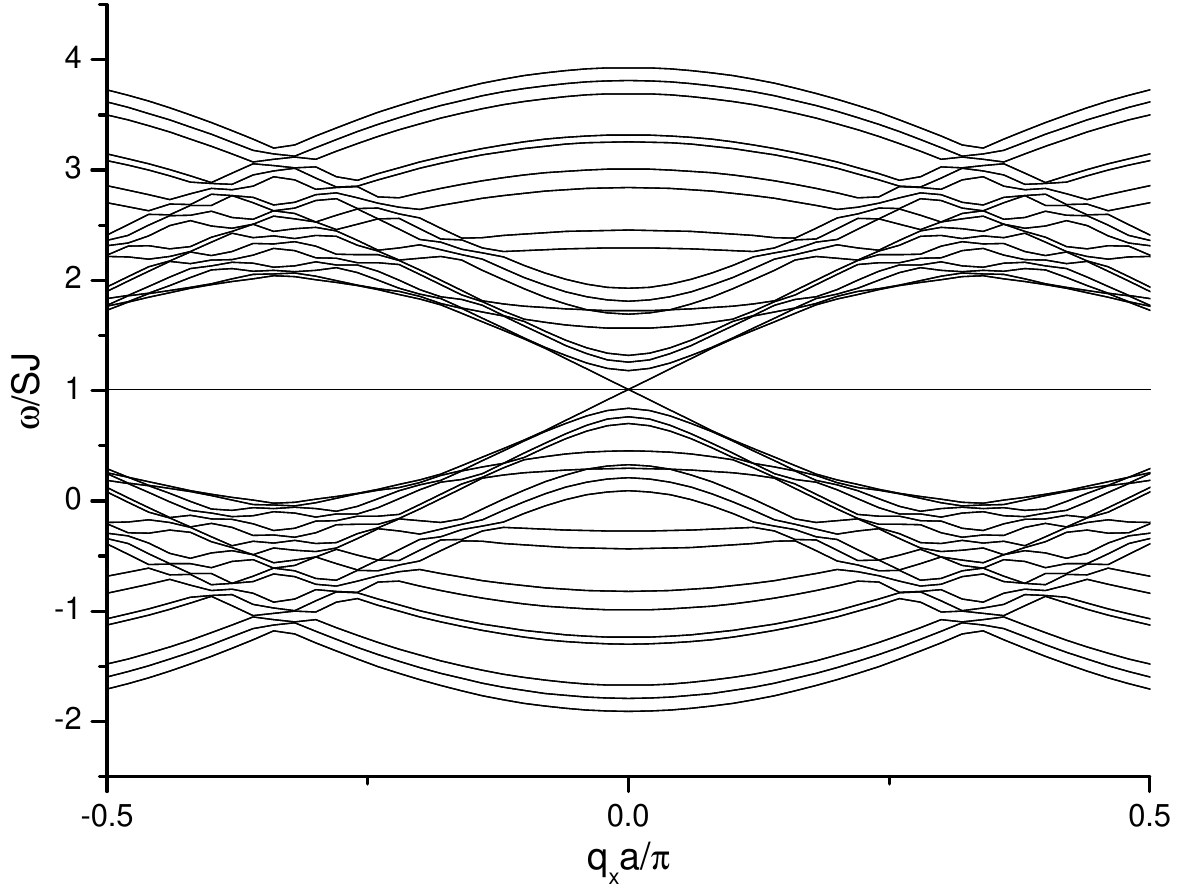}
 \\
\includegraphics[scale=.7]{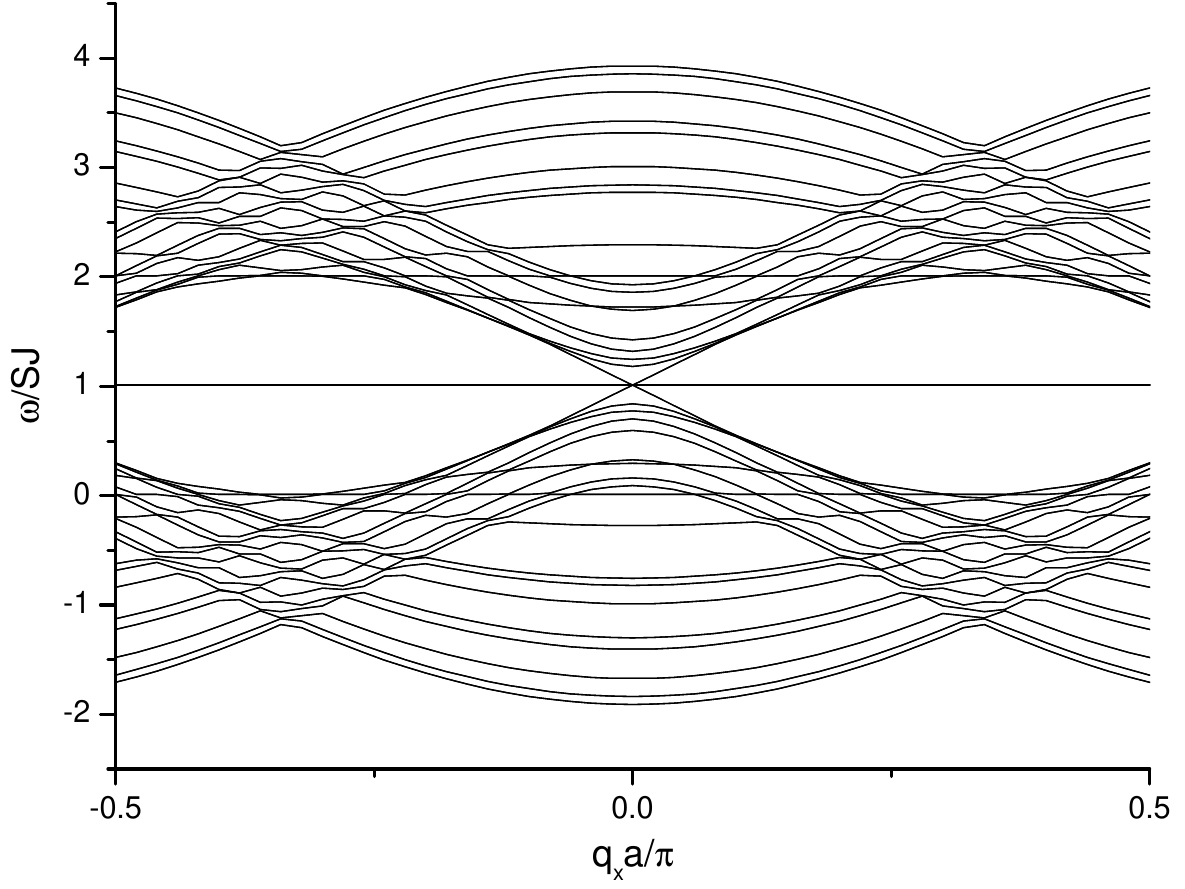} \label{fig:mgfarm2ii}\\
\includegraphics[scale=.7]{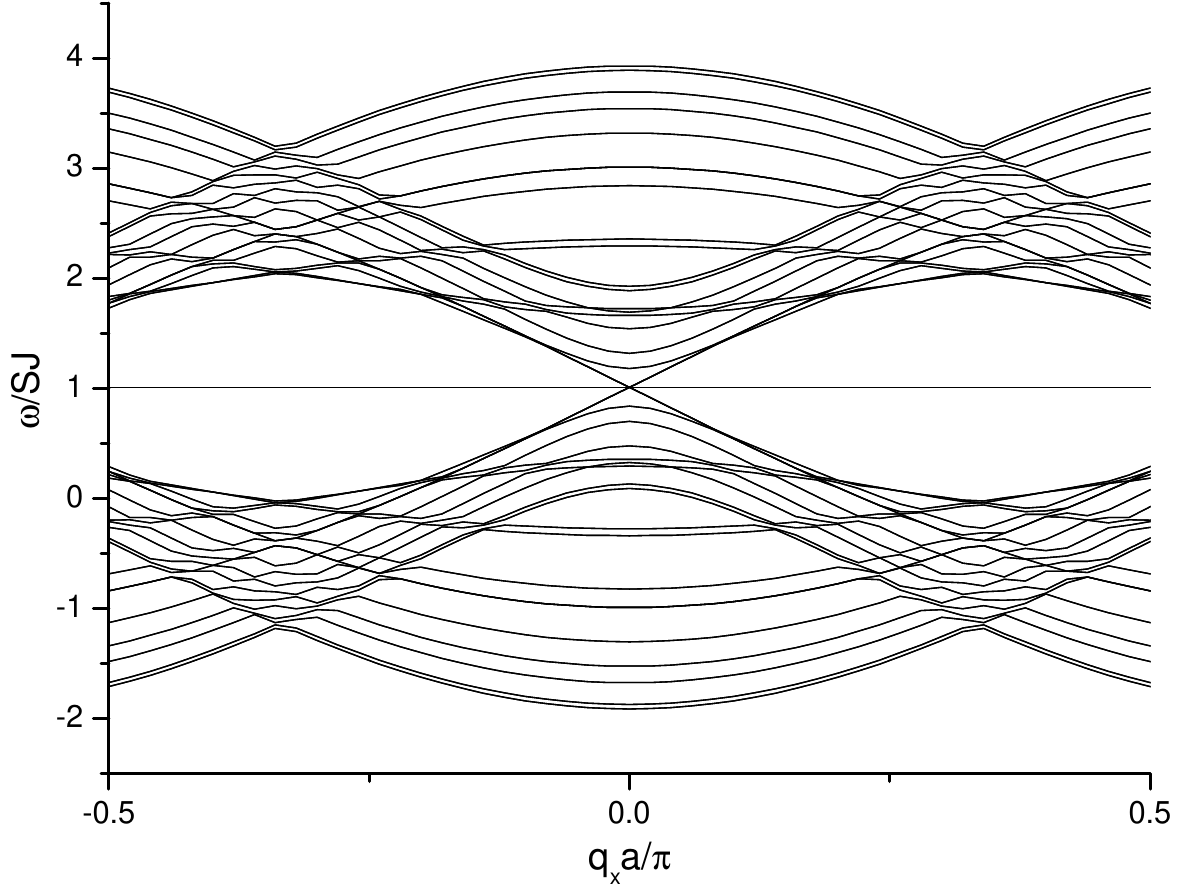}\label{fig:mgfarm3ii}
\end{tabular}
\caption{Spin waves dispersion for armchair 2D Heisenberg ferromagnetic honeycomb stripes
with an impurity lines at line number 11 and line number 14, where $J=J_e=1$, $J_I=J_{II}=0.0J$ $D=D_e=D_I=1.0$  and $\alpha=1.01$  for $ N=20,21$ and 22 respectively.}\label{fig:mgfarm1ii}
\end{figure}

Figures \ref{fig:mgfarm1ii} show the modified dispersion relations due to the effect of introducing substitutional a magnetic impurities lines at rows 11 and 14 of the armchair stripes with 20, 21 and 22 lines. Again, the introducing of the impurities lines have the effect of splitting the stripe to three interacted stripes with different sizes, in case of 20 line stripe the new stripes are 10 lines, 2 lines and 6 lines, in case of 21 line stripe the new two stripes  10 lines, 2 lines and 7 lines, and in case of 22 line stripe the new two stripes  10 lines, 2 lines and 8 lines. The existence of stripe with 2 line which is stripe of type 3i+2 where i=0, force the band gapless feature in the three stripe. The absence of spacial phenomenon like intrinsic localized edge states and accumulation two line stripe in armchair stripes reflect the importance of stripe topology as armchair or zigzag in the follow of the nearest neighbors exchange inside the stripe which will be studied later in this work.

As we see from the results above, that armchair type stripes have band gaps while there is no intrinsic localized edge state, and for the applications of similar armchair graphene nanoribbons the energy band gap is very important \cite{Bing,Neto1,rim1}. Therefore RDSCB and the change in the band gap are the good parameters for armchair stripes to study the effects of edges and impurities on their dispersions relations.

\subsubsection{The effect of armchair stripe width on its RDSCB}

\begin{figure}[h]
\centering
\includegraphics[scale=1]{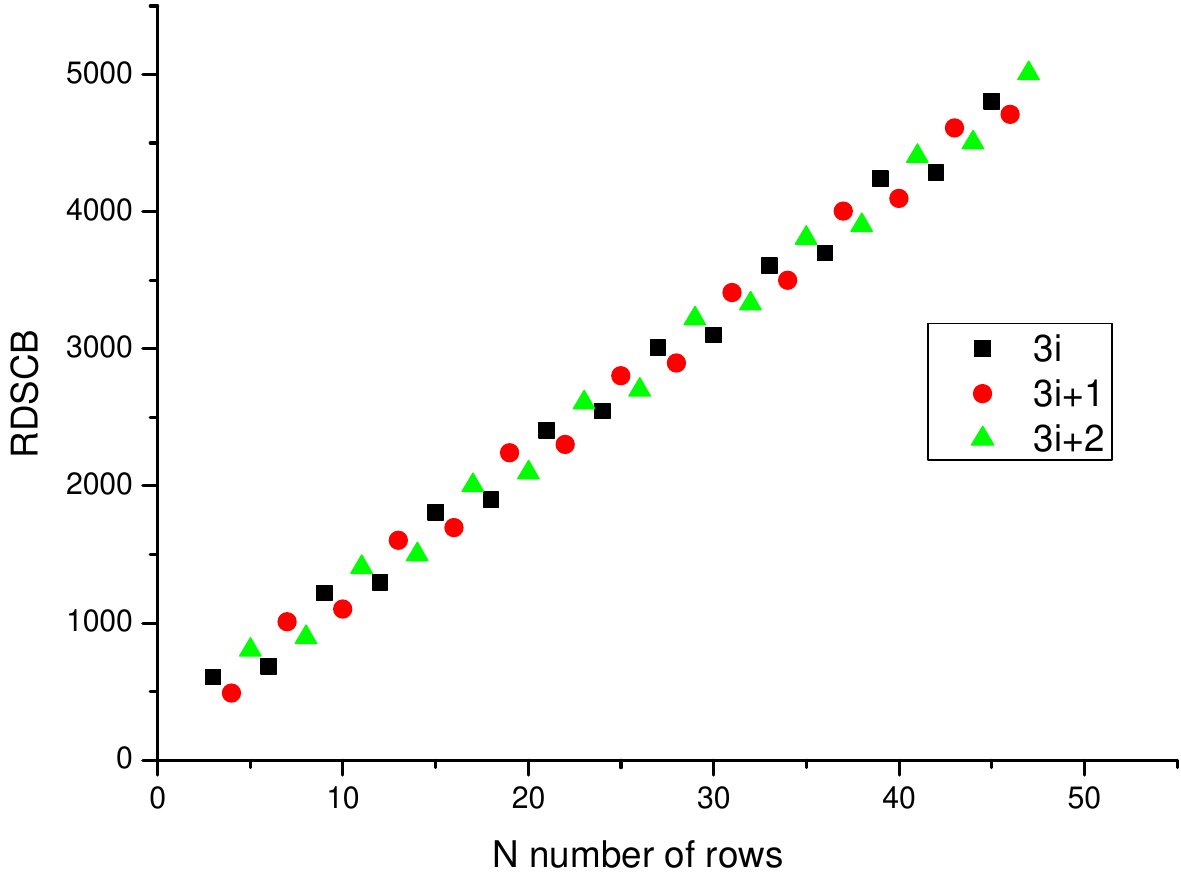}
\caption{The effect of armchair stripe width for the three armchair types on the relative density of states of center
band (RDSCB)} \label{fig:mgfarmwidth}
\end{figure}

Figure \ref{fig:mgfarmwidth} shows the effect of armchair stripe width on its RDSCB. The RDSCB is highly dependent on the armchair stripe width type as 3i, 3i+1, and 3i+2. There is a repeated pattern between the stripe width types and their RDSCB as follow the sequence of first step is 3i+1, 3i, 3i+2, 3i+1, and 3i+2 which then repeated as the width increases. Overall, the average value of the RDSCB linearly increases with increasing stripe width.

\subsubsection{The effects of edge uniaxial anisotropy on armchair stripe RDSCB}

\begin{figure}[h]
\centering
\includegraphics[scale=1]{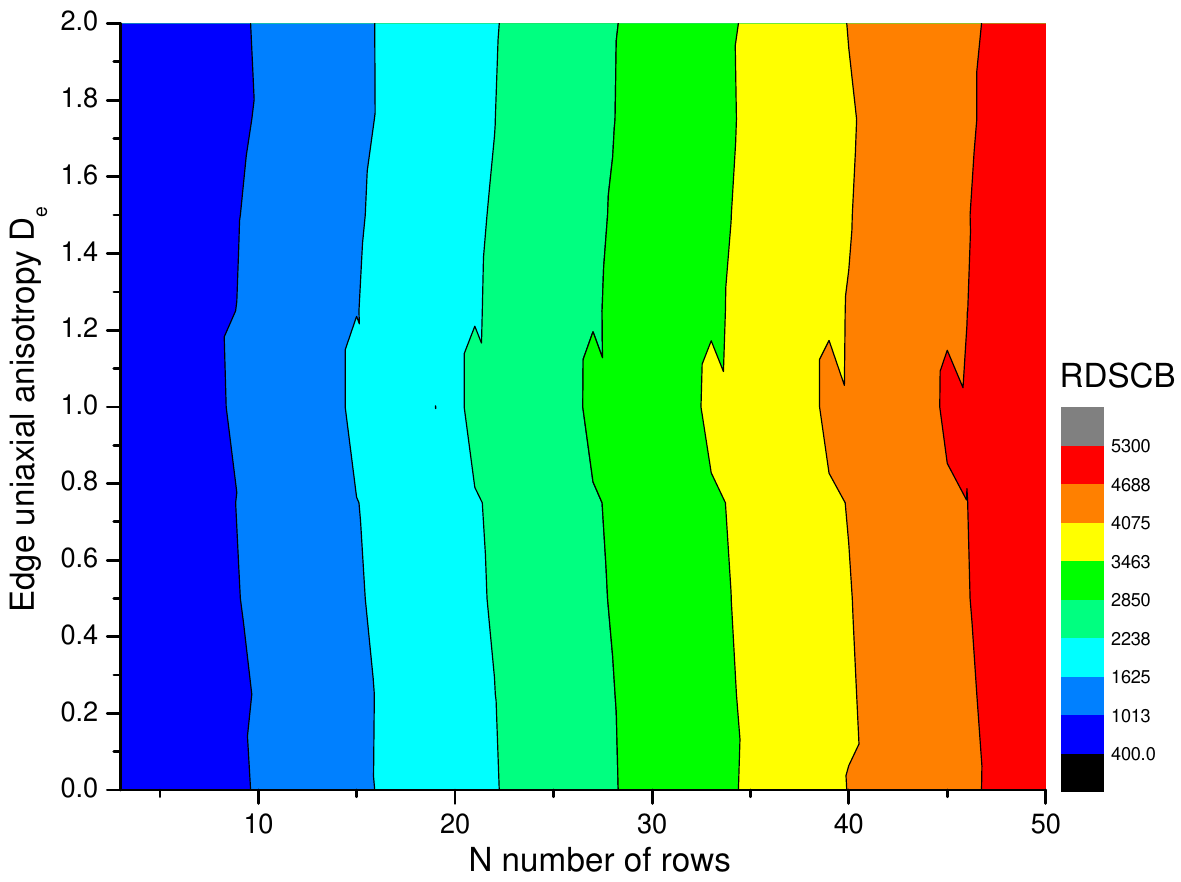}
\caption{The effects of edge uniaxial anisotropy and armchair stripe width on its
RDSCB} \label{fig:mgfarmedgei}
\end{figure}

Figure \ref{fig:mgfarmedgei} shows the color contour plot for the effects of edge uniaxial anisotropy and armchair stripe width on its
RDSCB. The Figure shows that RDSCB is nearly independent on the change of edge uniaxial anisotropy
which as the case of zigzag stripe, is reflected in parallel colored stripes, except at $D_e = D$  where there are certain stripe widths with higher value RDCSB than surrounding numbers, which is first time to be seen in armchair stripes. Those numbers are 9, 15, 21, 27, 33, 39, and 45 which of type $3i$ for which $i$ odd primary number 3, 5, 7, 9, 11, 13 and 15. They are also clear in Figure \ref{fig:mgfarmwidth} as the most outer $3i$ stripes widths (black squares) which show that at those lengths the system wave function has high density of states at the center band \cite{PhysRevB.75.165414}. In general, changing the edges insite energy breaks the symmetry of the dispersion relation and move slightly the Fermi Level which also seen in graphene nanoribbon \cite{PhysRevB.73.045432}, but nearly has no effect on the RDSCB value. It is clear that RDSCB is increasing with increasing the stripe width which agree
with result in figure \ref{fig:mgfarmwidth}.

\subsubsection{The effects of edge exchange on armchair stripe RDSCB}

\begin{figure}[h]
\centering
\includegraphics[scale=1]{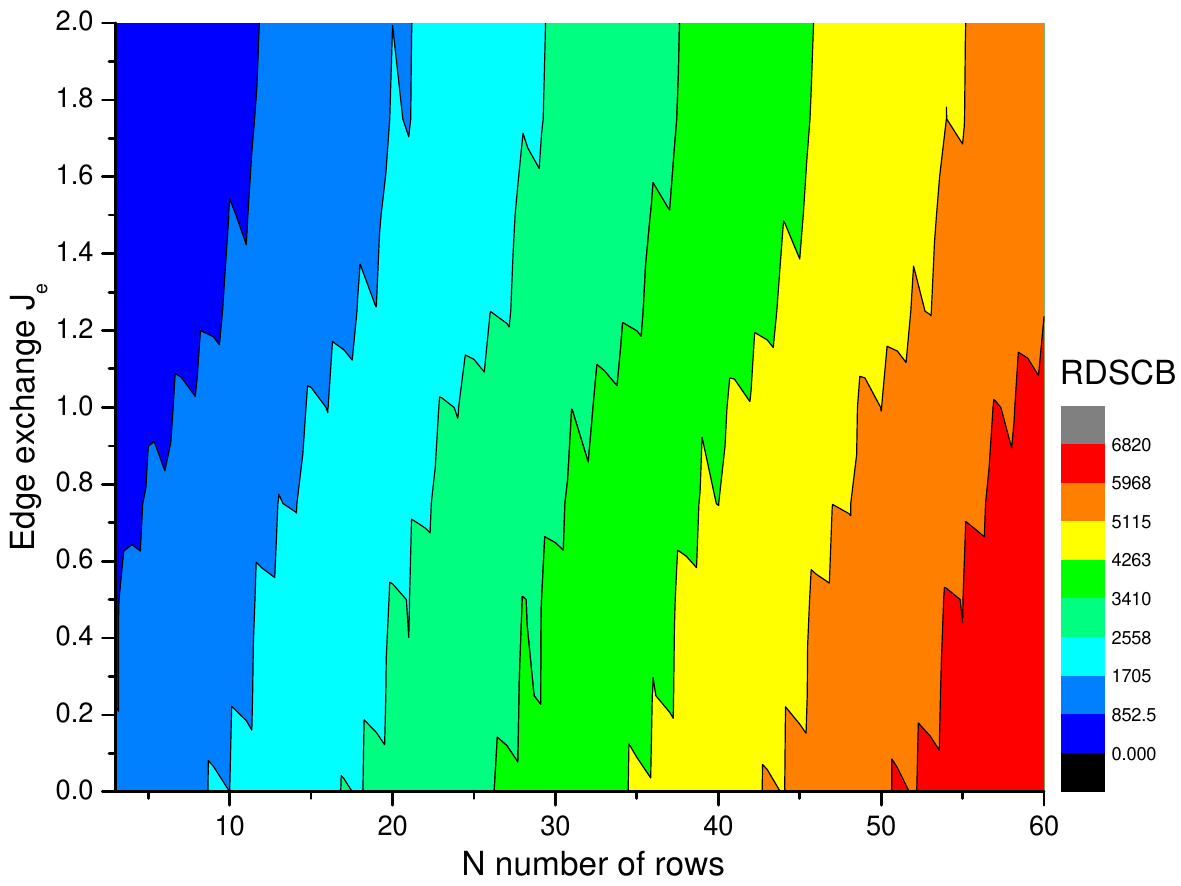}
\caption{The effects of edge exchange and zigzag stripe width on its RDSCB} \label{fig:mgfarmedge}
\end{figure}

Figure \ref{fig:mgfarmedge} shows the color contour plot for the effects of edge exchange and zigzag stripe width on its RDSCB.  It is clear from the figure that RDSCB is decreasing with the increasing of the edge exchange which is shown as a curvature in the colored RDSCB stripes where the zigzags lines at the boundary of the colored stripes come from armchair types depends. The RDSCB is increasing with increasing the stripes width which agree with with result in figure \ref{fig:mgfarmwidth}.

\subsubsection{The effects of impurities on armchair stripe RDSCB}
The study of magnetic impurities effects on armchair stripe is important for expected applications as in case of zigzag stripes. In this section the results are represented for the effects of introducing one and two lines of magnetic impurities on armchair 20, 21 and 22 width stripes on their RDSCB.

As in case of the zigzag case, there are two parameters for the impurities that engineering the energy band for magnetic armchair stripes: The first one is the strength of magnetic interaction represented by line of impurity exchange $J_I$  between the impurities line and the stripe materials \cite{rim1}, which here take the range of values from 0 to 2 in the units of stripe materials magnetic exchange $J$. The second parameter is the impurities line position, which can take the value from second to one line before the stripe end, the line position is alternative between even position number.

Figure \ref{fig:mgfarm1i} shows the effects of one line of impurities position and impurities exchange for $N=20$ and $N=22$ armchair stripes on their RDSCB. The figure shows that the behavior of the RDSCB is nearly the same for both even width stripes, beginning from impurities exchange with value $1.0$ which means no impurities, as impurities exchange increases the RDSCB decreases and it is  independent on the impurity line position. As impurities exchange decreases than 1.0, the RDSCB increases as impurities exchange reach 0.0 the RDSCB has maximum value and it is independent on the impurity line position. This is due to the creation of expanded edge localized states in Fermi level, the existence of one line impurity divide the armchair 20 and 22 width stripes to two interacting substripes one odd and one even with no probability to have symmetry between them, which make this position independent.

\begin{figure}[hp]
\centering
\begin{tabular}{cc}
\includegraphics[scale=.75]{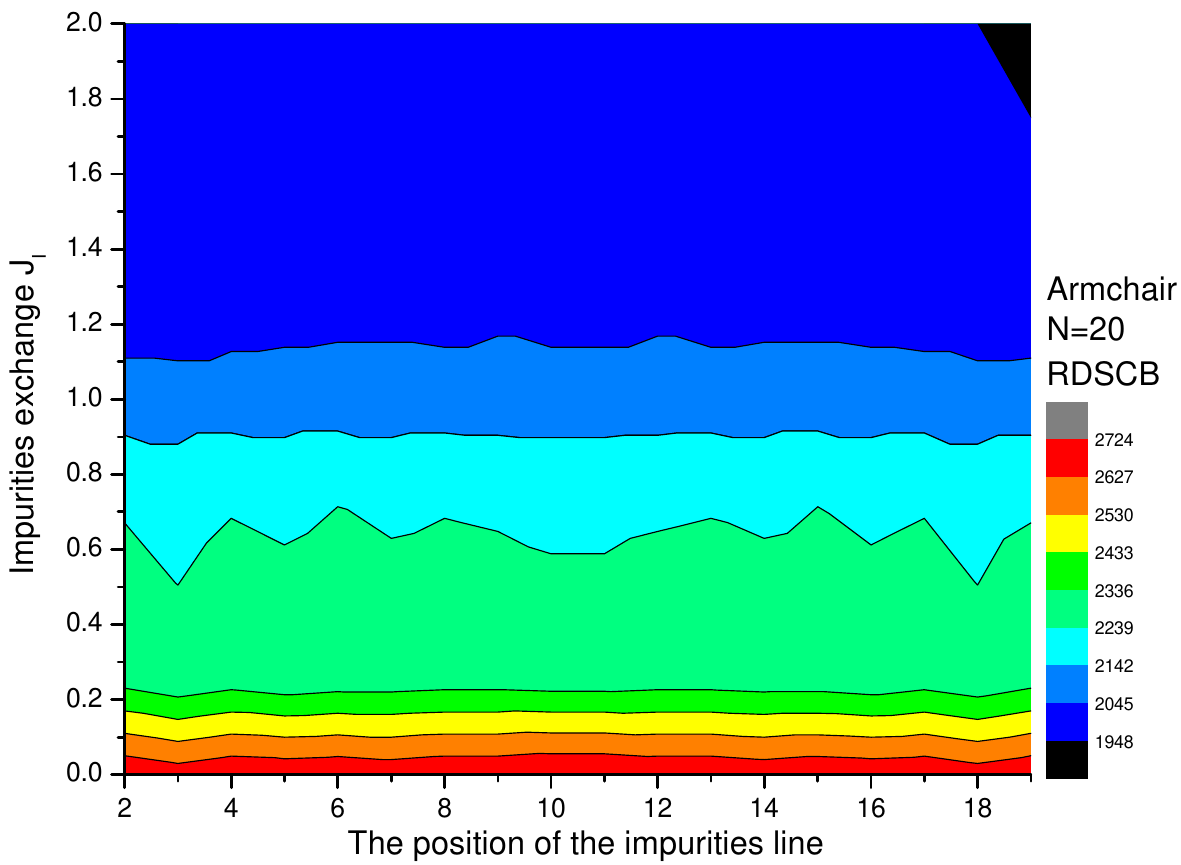}
 \\
\includegraphics[scale=.75]{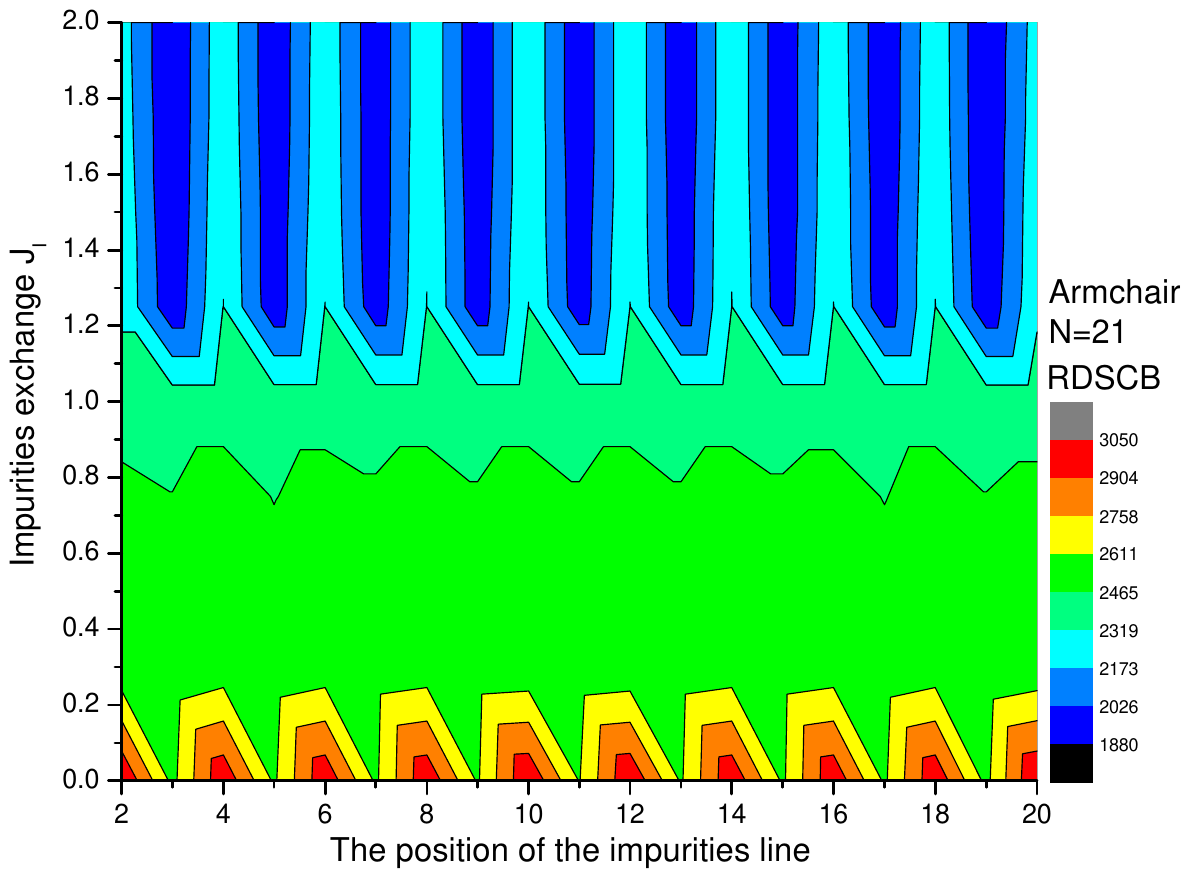} \label{fig:mgfarm2i}\\
\includegraphics[scale=.75]{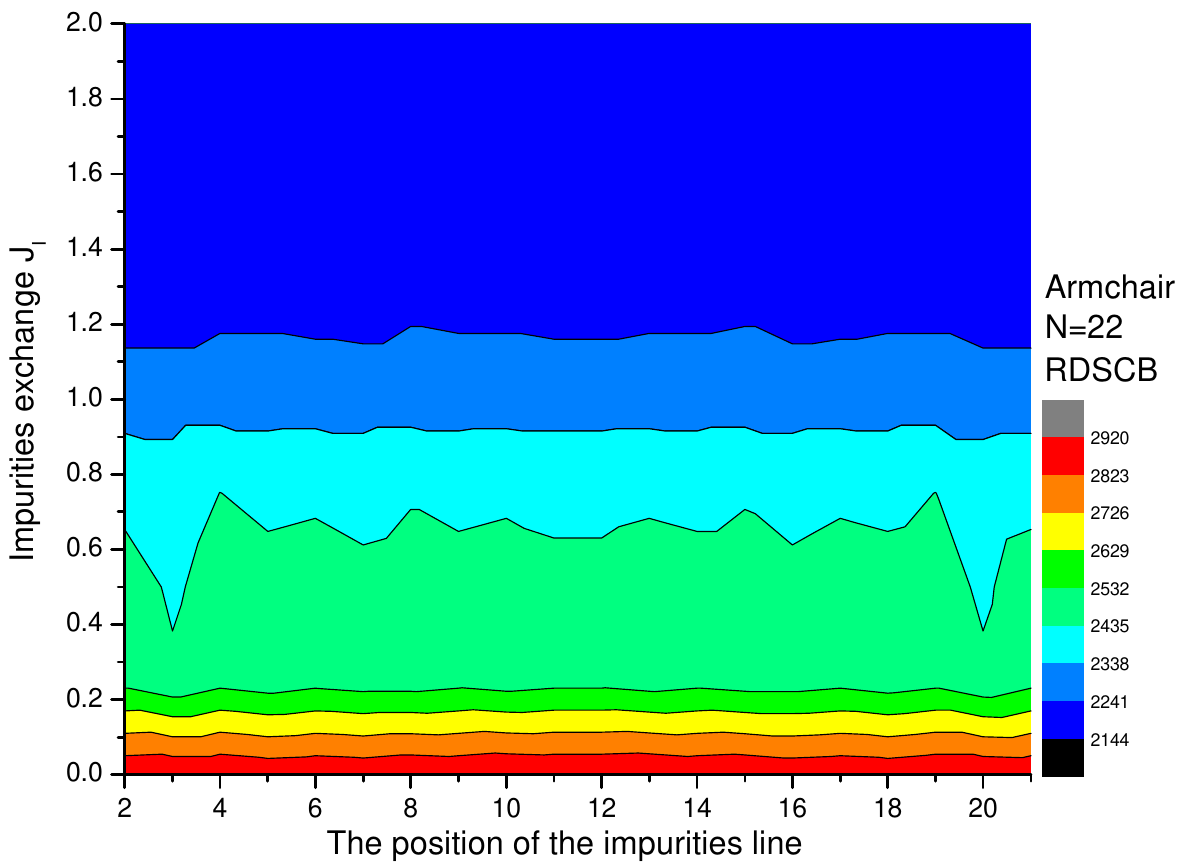}\label{fig:mgfarm3i}
\end{tabular}
\caption{The effect of one line of impurities position and impurities exchange
on armchair stripe RDSCB for stripes width $ N=20,21$ and 22 respectively.} \label{fig:mgfarm1i}
\end{figure}

Figure \ref{fig:mgfarm1i} shows the effects of one line of impurities position and impurities exchange for $N=21$ armchair stripe on its RDSCB, beginning from impurities exchange with value $1.0$ which means no impurities, as impurities exchange increases the RDSCB decreases periodically with maximum decreasing in odd impurity line positions. As impurities exchange decreases than 1.0, the RDSCB increases as impurities exchange reach 0.2 the RDSCB has maximum value in even impurity line positions. The behavior can be explained by the fact that the existence of one line impurity divide the armchair 21 width stripe to two interacting substripes with their width parities depends on the impurity line position, in case of even impurity line position the two substripes are one odd and one even which is same of armchair stripe with 20 width, while in case of odd impurity line position the two substripes are even, which increase the symmetry and degeneracy between the two substripes and reduce the value of RDSCB.

\begin{figure}[hp]
\centering
\begin{tabular}{cc}
\includegraphics[scale=.75]{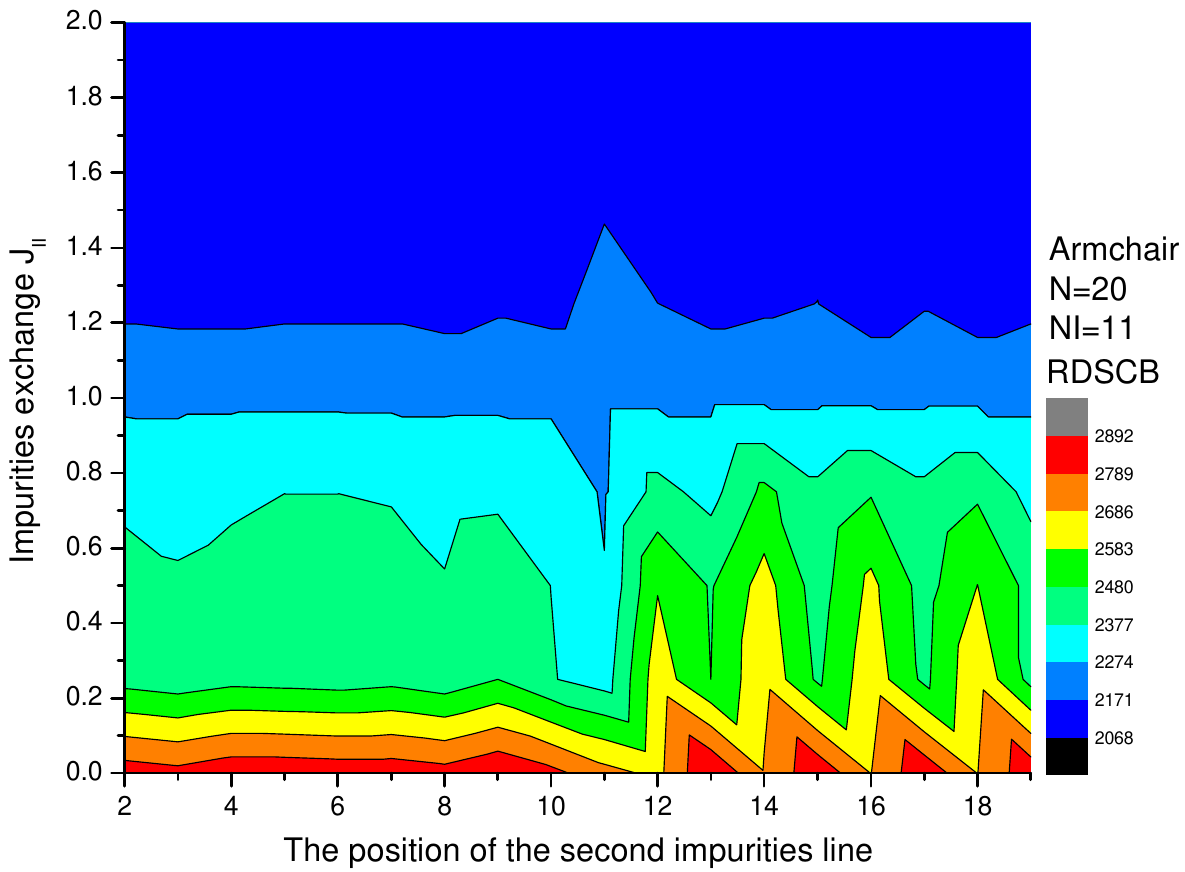}
 \\
\includegraphics[scale=.75]{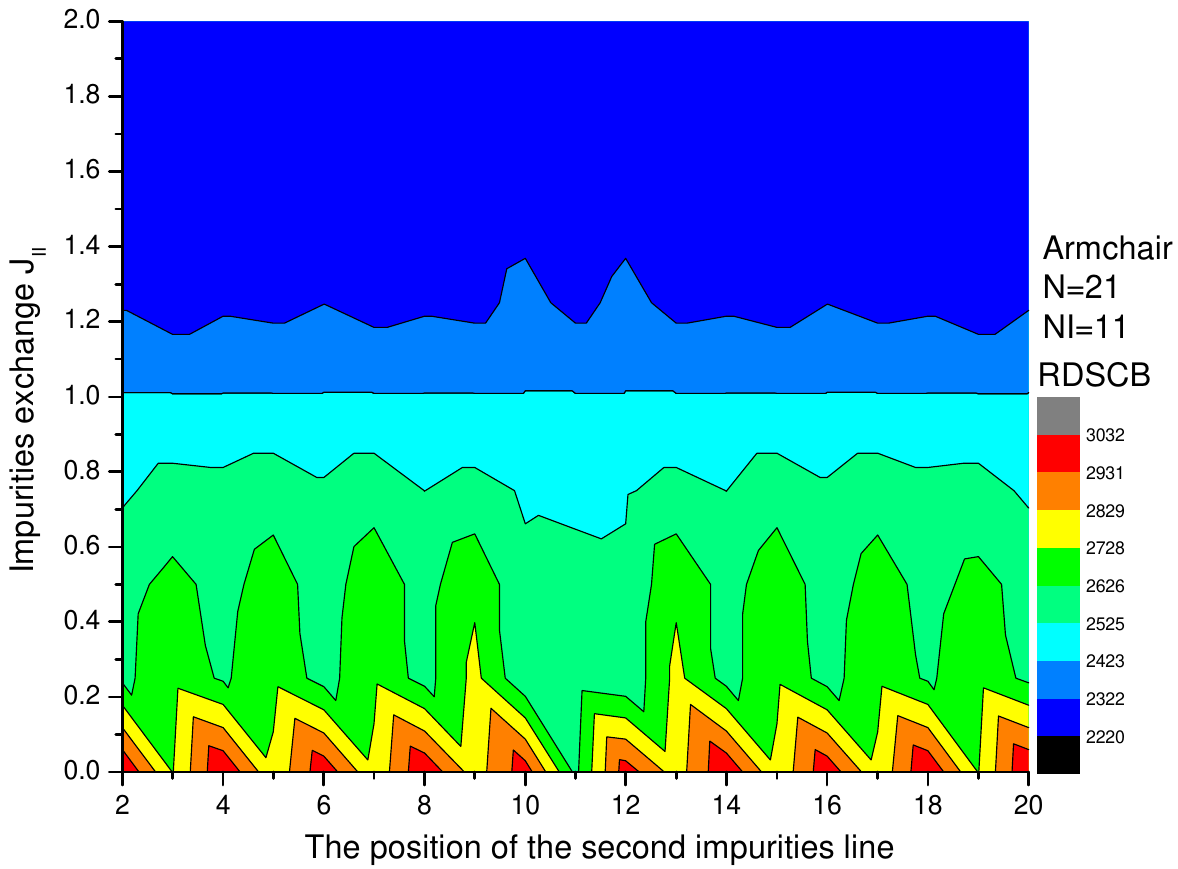} \label{fig:mgfarm2iii}\\
\includegraphics[scale=.75]{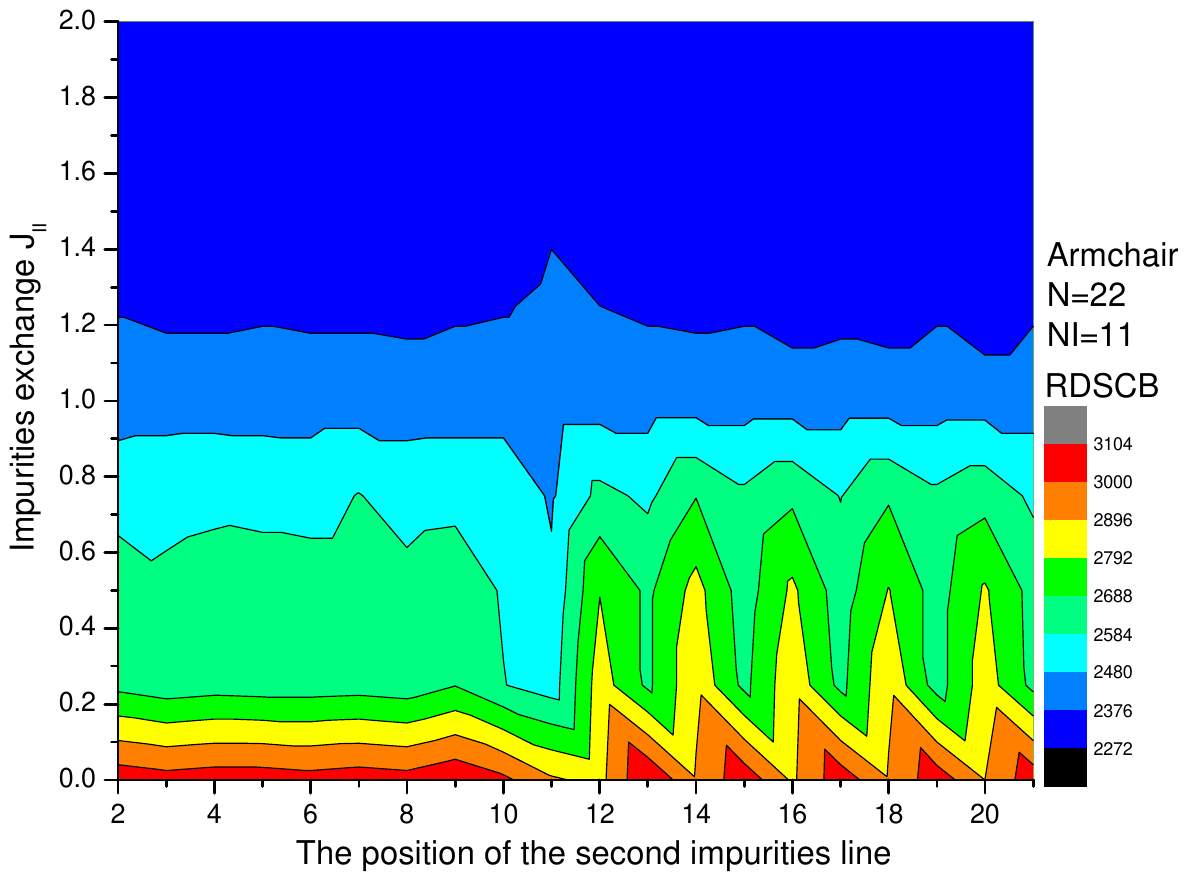}\label{fig:mgfarm3iii}
\end{tabular}
\caption{The effect of second line of impurities position and impurities exchange
on armchair stripe RDSCB with one line of impurities @ N = 11 with impurities
exchange $J_I = 0$ for stripes width $ N=20,21$ and 22 respectively.} \label{fig:mgfarm1iii}
\end{figure}

Figure \ref{fig:mgfarm1iii} shows the effect of second impurities line position with impurities exchange $J_{II}$ for $N=20$ and $N=22$  armchair stripes with one line of impurities at $N=11$ with impurities exchange $J_I=0$ on its RDSCB. The figures show that the behavior of the RDSCB is nearly the same for both even width stripes as the RDSCB is dependent on the position of second impurities line with respect to the position of first impurities line, which is explained by the fact that the first impurities line divide the armchair 20 and 22 stripes to two substripes, the first one is even stripe with 10 width for both 20 and 22 stripes and the second one is odd stripe with 9 and 11 widths for stripes. When the second line change its position in the even substripes, the RDSCB shows the same behavior of one line impurities in even armchair stripes.
At first impurities line position, the second line superimposed on the first line left only the effect of first line on the RDSCB. As the second line change its position in the odd substripes, the RDSCB shows similar behavior of one line impurities in odd armchair stripes with shift for the RDSCB peaks form even second impurities line positions to odd second impurities line positions at second impurities line exchange equal to zero due to the interaction with first even substripe with 10 width.

Figure \ref{fig:mgfarm1iii} shows the effect of second impurities line position with impurities exchange $J_{II}$ for $N=21$  armchair stripes with one line of impurities at $N=11$ with impurities exchange $J_I=0$ on its RDSCB, beginning from impurities exchange with value $1.0$ which means no second impurities, as impurities exchange increases, the RDSCB decreases independently on second impurity line positions. As impurities exchange decreases than 1.0, the RDSCB increases as impurities exchange reach 0.0 the RDSCB has maximum value in even impurity line positions.

\subsubsection{The effect of armchair stripe width on its energy band gap}
The direct energy band gab seen at $q_x=0$ above in armchair stripes is similar to graphene armchair ones and it is very important from applications point of view \cite{Bing,Neto1,rim1}. The relation between the graphene armchair nanoribbons types, its width and its band gab has been studied extensively both experimentally \cite{PhysRevLett.98.206805} and theoretical \cite{PhysRevLett.97.216803,PhysRevB.73.045432,PhysRevB.59.8271,JPSJ.65.1920,PhysRevB.54.17954}. It has been found that the main factor in the armchair nanoribons energy band gab of types $3i$ and $3i+1$ behavior is the quantum confinement which reflected in the inverse dependance of energy band gab $E_g$ on the stripe width $W$  such that $E_g\sim W^{-1}$.

\begin{figure}[h]
\centering
\includegraphics[scale=1]{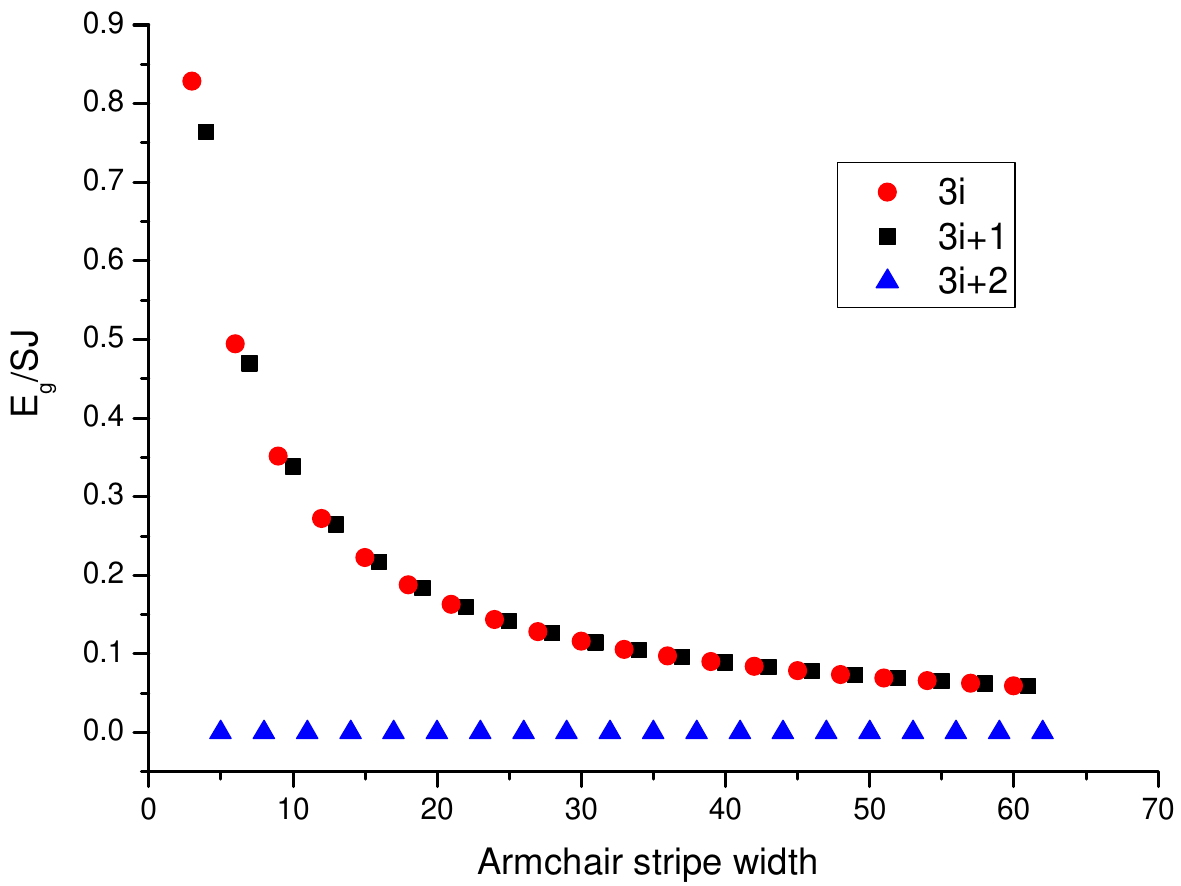}
\caption{The variation of bandgaps of the three types armchair stripes as a function of width (number of rows) $W$} \label{fig:mgfarmbandgaps}
\end{figure}

We used our model to study the variation of bandgaps of the three types magnetic armchair stripes as a function of width (number of rows) the results are shown in Figure \ref{fig:mgfarmbandgaps}. The Figure shows that the energy band gap $E_g$ armchair types $3i$ and $3i+1$ have the same width $W$ dependance seen in graphene armchair nanoribons, i.e. $E_g\sim W^{-1}$, which show the close similarity between graphene nanoribons and magnetic stripes.

\subsubsection{The effects of edge exchange on armchair stripe energy band gap}
\begin{figure}
  \begin{tabular}{cc}
\includegraphics[scale=.6]{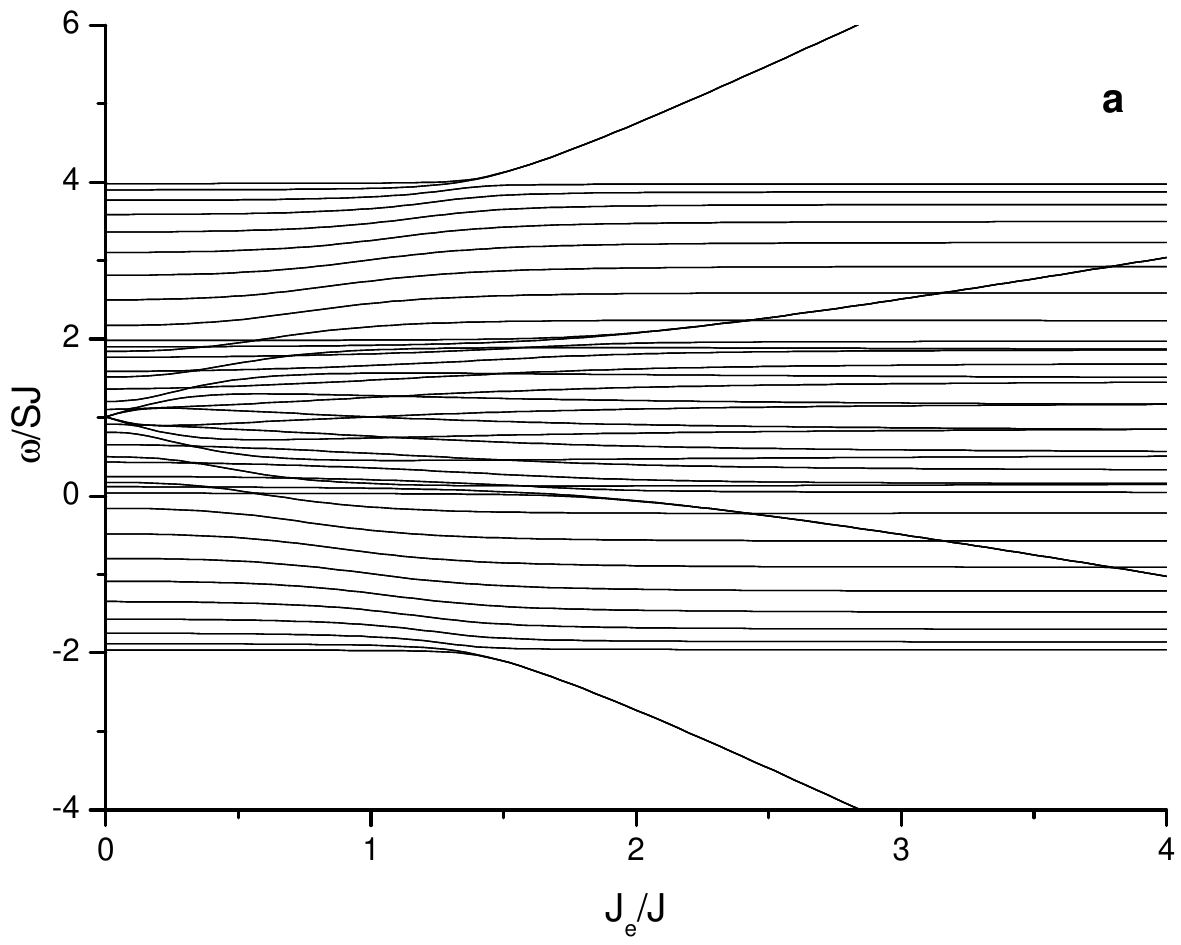}& \includegraphics[scale=.6]{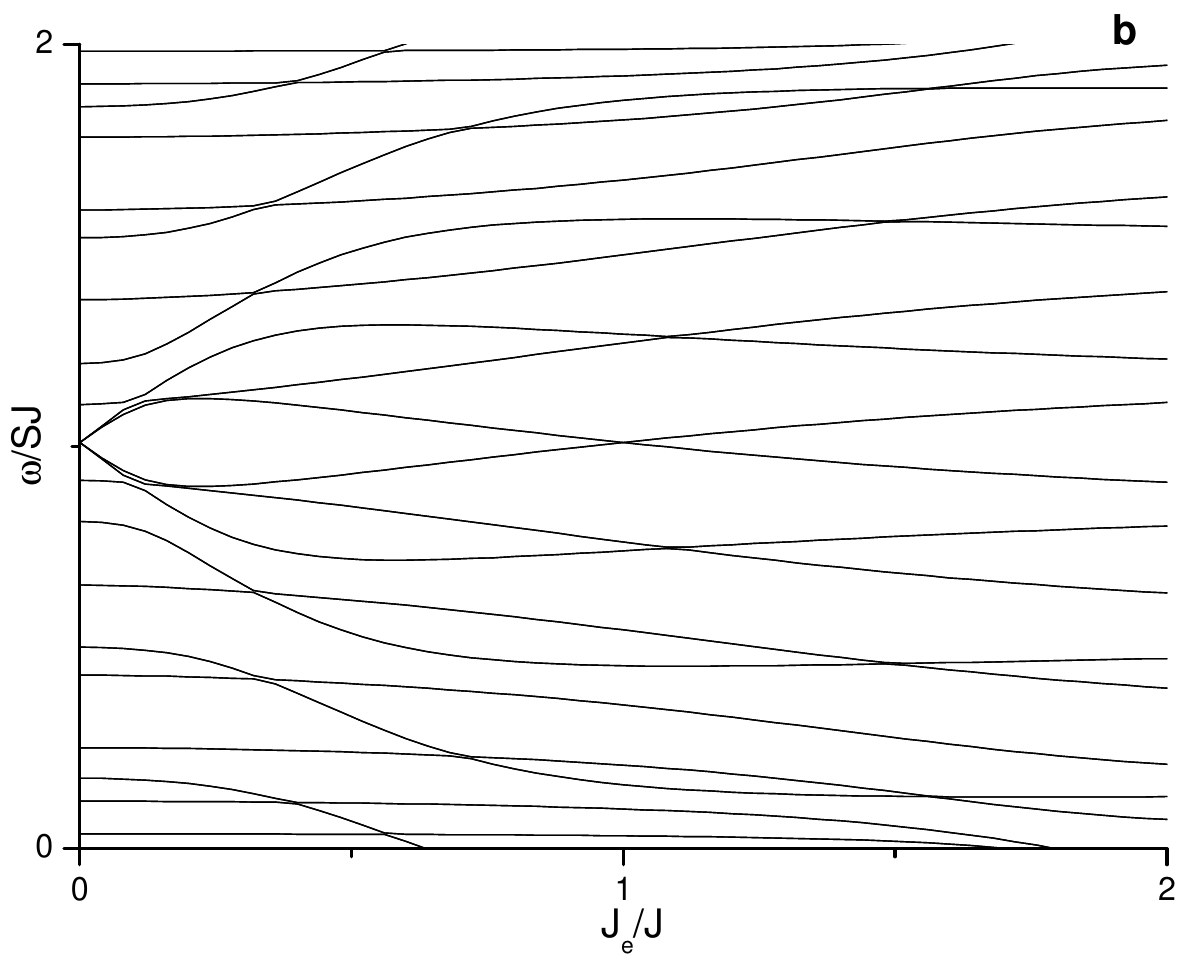}
 \\
\includegraphics[scale=.6]{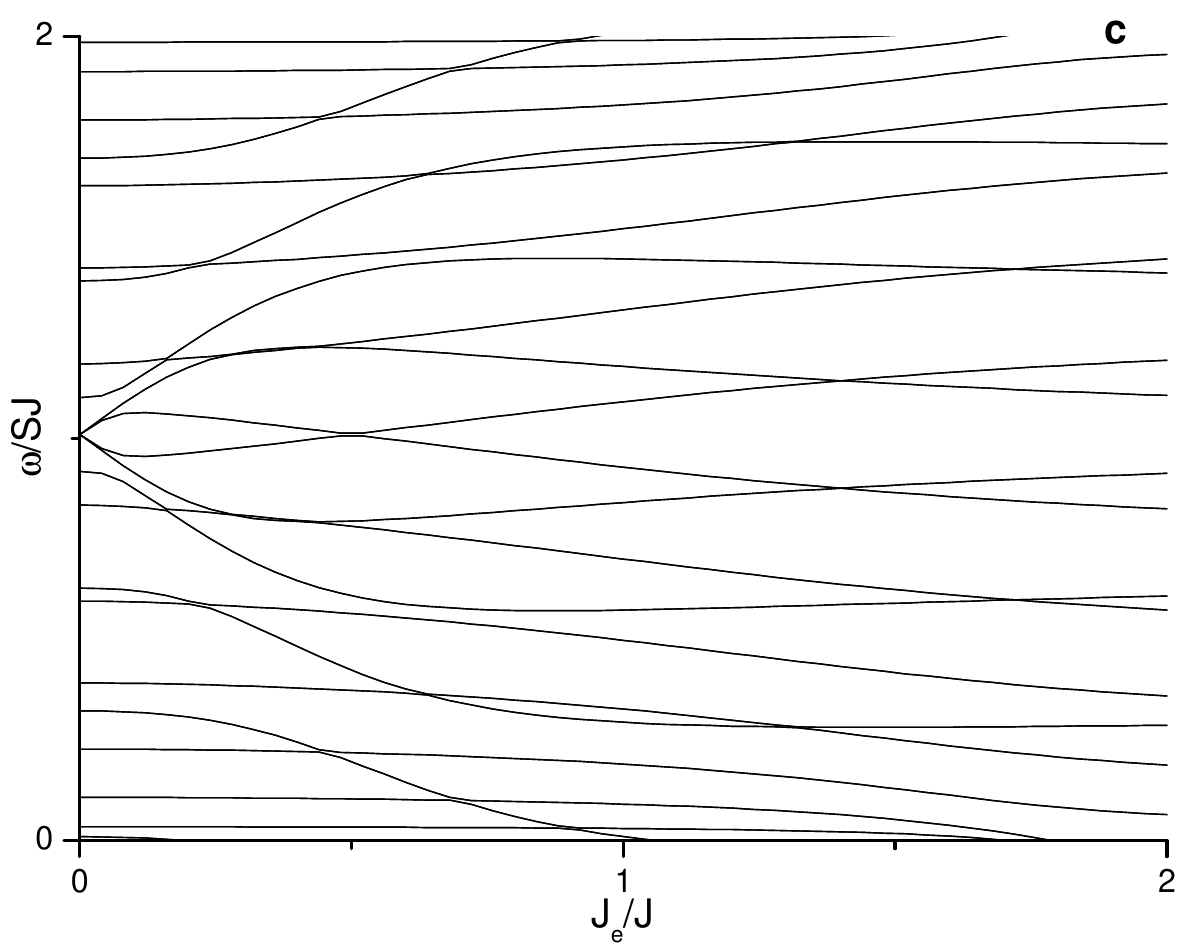}&\includegraphics[scale=.6]{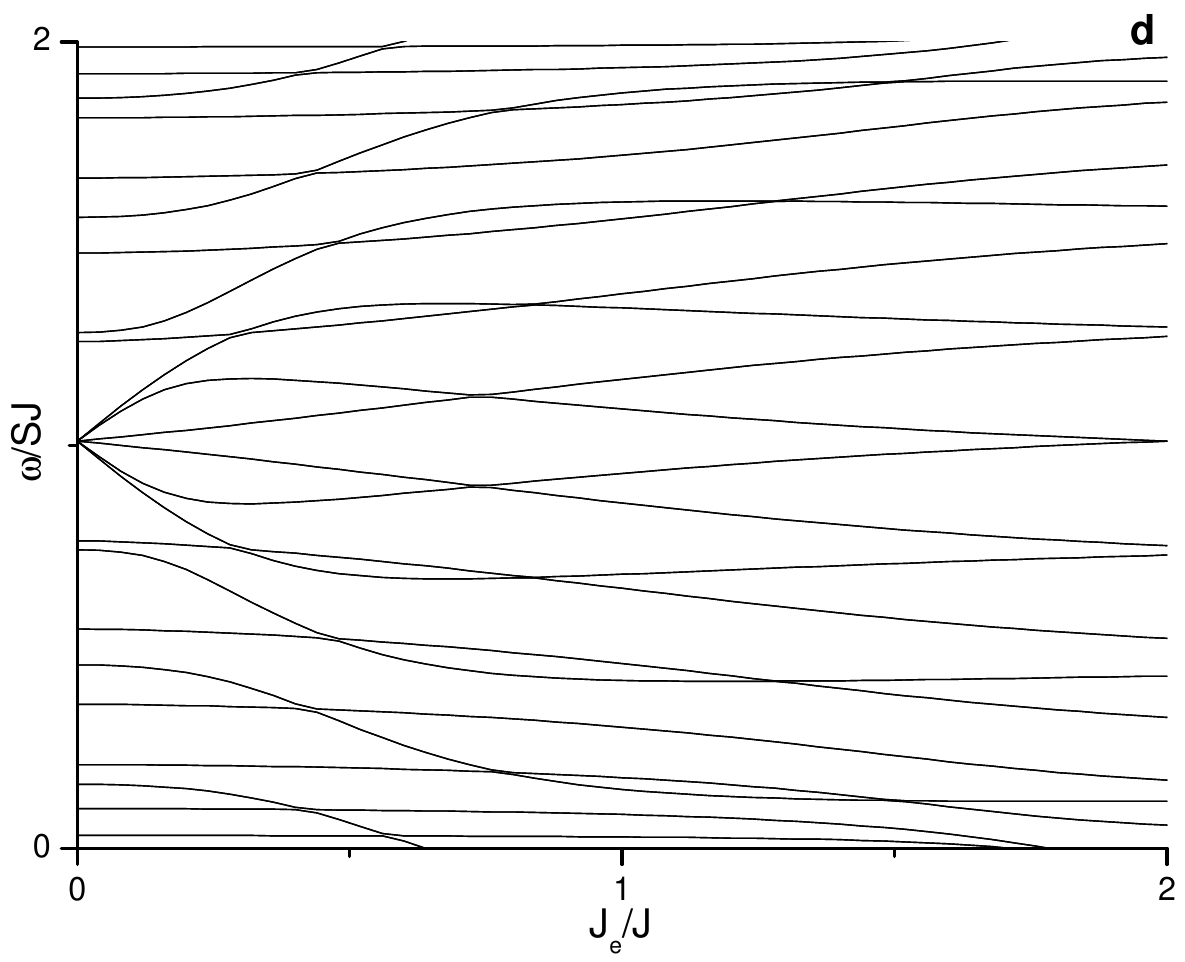}
\end{tabular}
  \caption{$q_x=0$ dispersion of the modes (a) 20-line armchair stripe as a function of the strength of the edge exchange $J_e$ which amplified in the region
around $(\omega/SJ)=\alpha$ in (b). For (c) and (d) the amplified lower energy region for
N=21 and 22 respectively}\label{edgeband}
\end{figure}

In our study of width effect on the energy band gab of magnetic armchair stripe, we assumed that the edges sites have the same exchange as interior sites, i.e. $J_e=J$, but this is not the case, as edges sites have different coordination number consequently their exchange is different from interior, also there is a technical possibility to engineering their magnetic properties and as it is found in armchair graphene nanoribbons case that the edges play important rule in their energy band gap \cite{PhysRevLett.97.216803}, the same is expected for magnetic stripes.   Figure \ref{edgeband} shows the behavior of spin wave energy modes at direct band gap point, i.e. $q_x=0$ with the variation of the edges exchange strength $J_e$.  Figure \ref{edgeband} (a) shows the behavior of all modes for a 20-line armchair stripe with the variation of the edges exchange strength, as edges exchange increases some modes begin to bend and crossing other modes some of them leave all the stripe energy band, the over all behavior is the same for 21 and 22 stripes.

Figure \ref{edgeband} (b), (c), and (d) show the amplified region
around $(\omega/SJ)=\alpha$ for 20, 21 and 22 stripe widths respectively. They show similar behavior beginning at zero edge exchange the minimum  conduction band mode and the maximum valance band meet together and consequently there is no band gap which shown as a localized states at Fermi level $(\omega/SJ)=\alpha$, as the edge exchange increases the two mode splitting producing a band gap for the stripes widths which then crossing each other at certain value of edge exchange depending on the stripe type. In general as the edge exchange approach the value of interior sites exchange, the modes rearranged to show the behavior seen before in Figure \ref{fig:mgfarm1}.

\begin{figure}[h]
\centering
\includegraphics[scale=1]{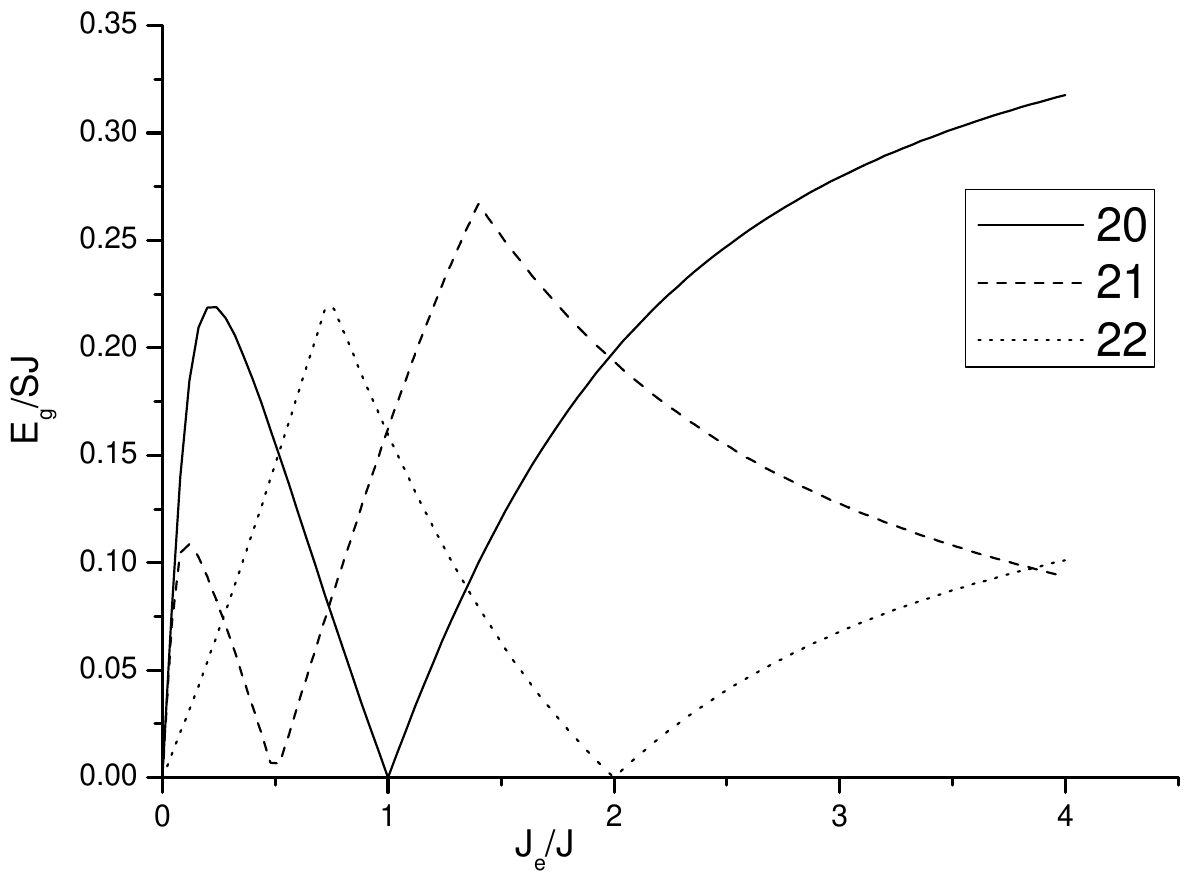}
\caption{Variation of the energy gap against the strength of the
edge exchange for an armchair stripes. Solid line for a
20-line ribbon, dashed line for 21 lines, and dotted line for a
stripe with 22 lines.} \label{edgeband2}
\end{figure}

Figure \ref{edgeband2} shows the variation of the energy gap, which is the difference between the minimum conduction band mode and the maximum valance band mode in the stripe, against the strength of the edge exchange for 20, 21 and 22 armchair stripes. The three energy band gap starts from zero and increases to reach maximum and then decreasing to reach zero again at edge exchange value depending on the stripe width, which are 0.5 for 21 stripe, 1.0 for 20 stripe and 2.0 for 22 stripe. After this minimum each energy band gap starts to increases again, the 20 stripe energy band gap increases to seam constant value, while the 21 stripe energy band gap increase to new maximum value and the decreases, and the 22 stripe energy band gap decreases slowly.

\subsubsection{The effects of impurities on armchair stripe energy band gap}
\begin{figure}
  \begin{tabular}{cc}
\includegraphics[scale=.6]{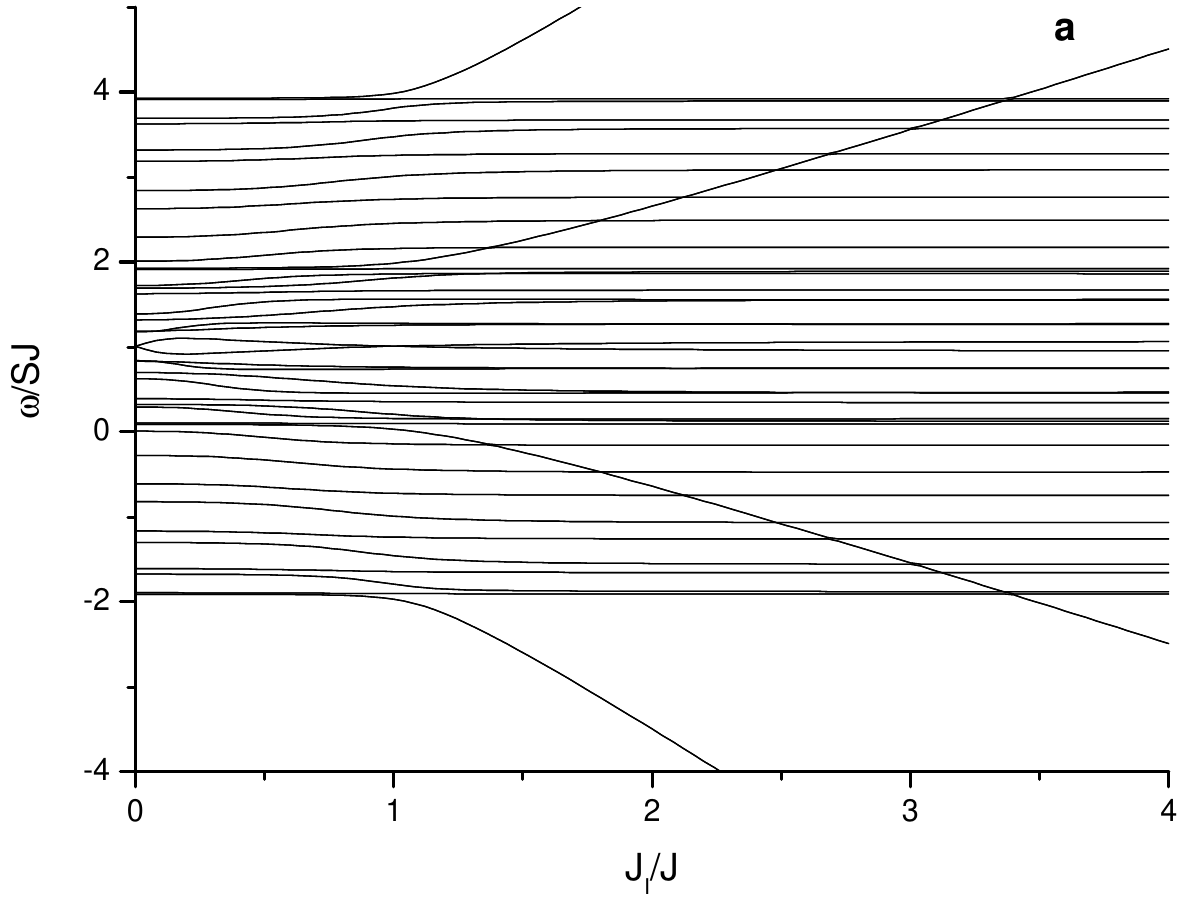}& \includegraphics[scale=.6]{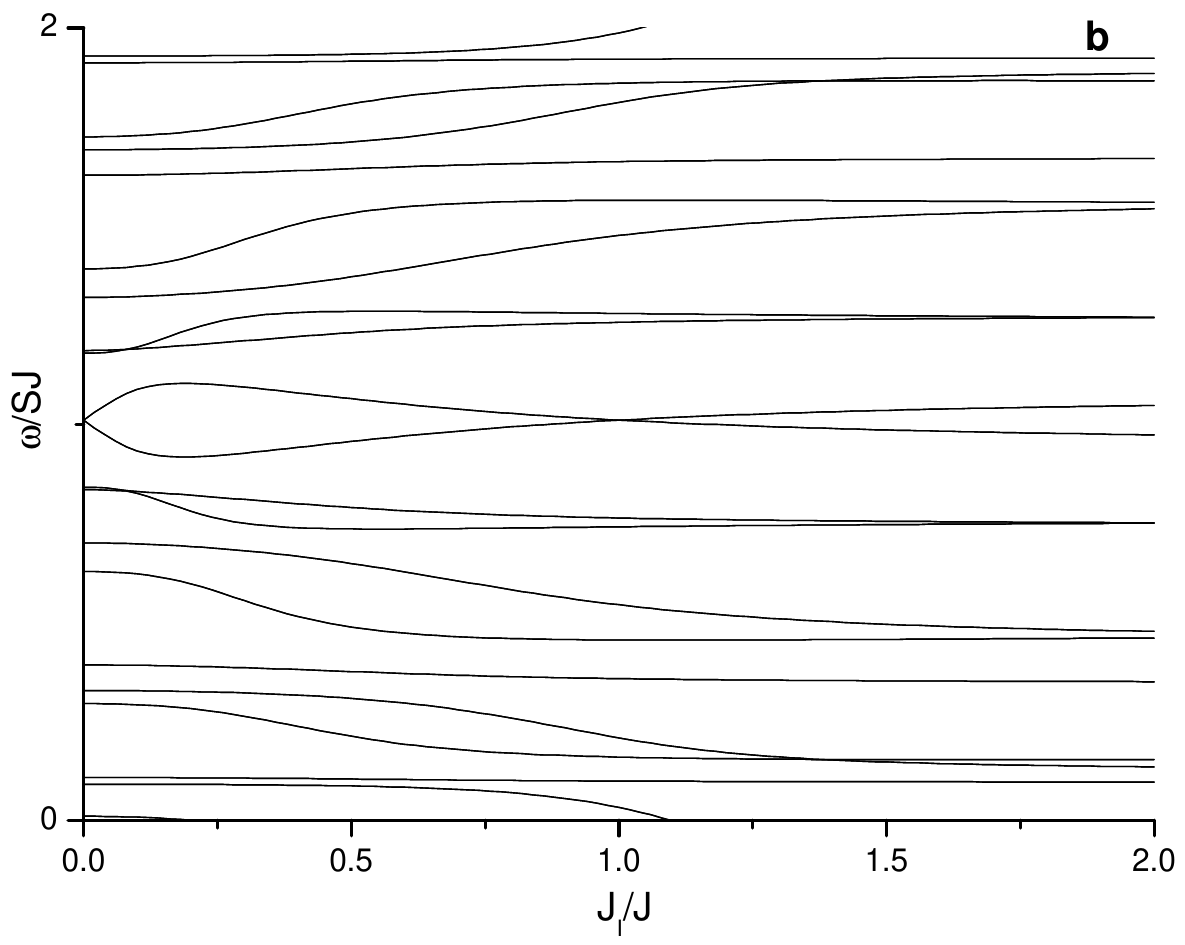}
 \\
\includegraphics[scale=.6]{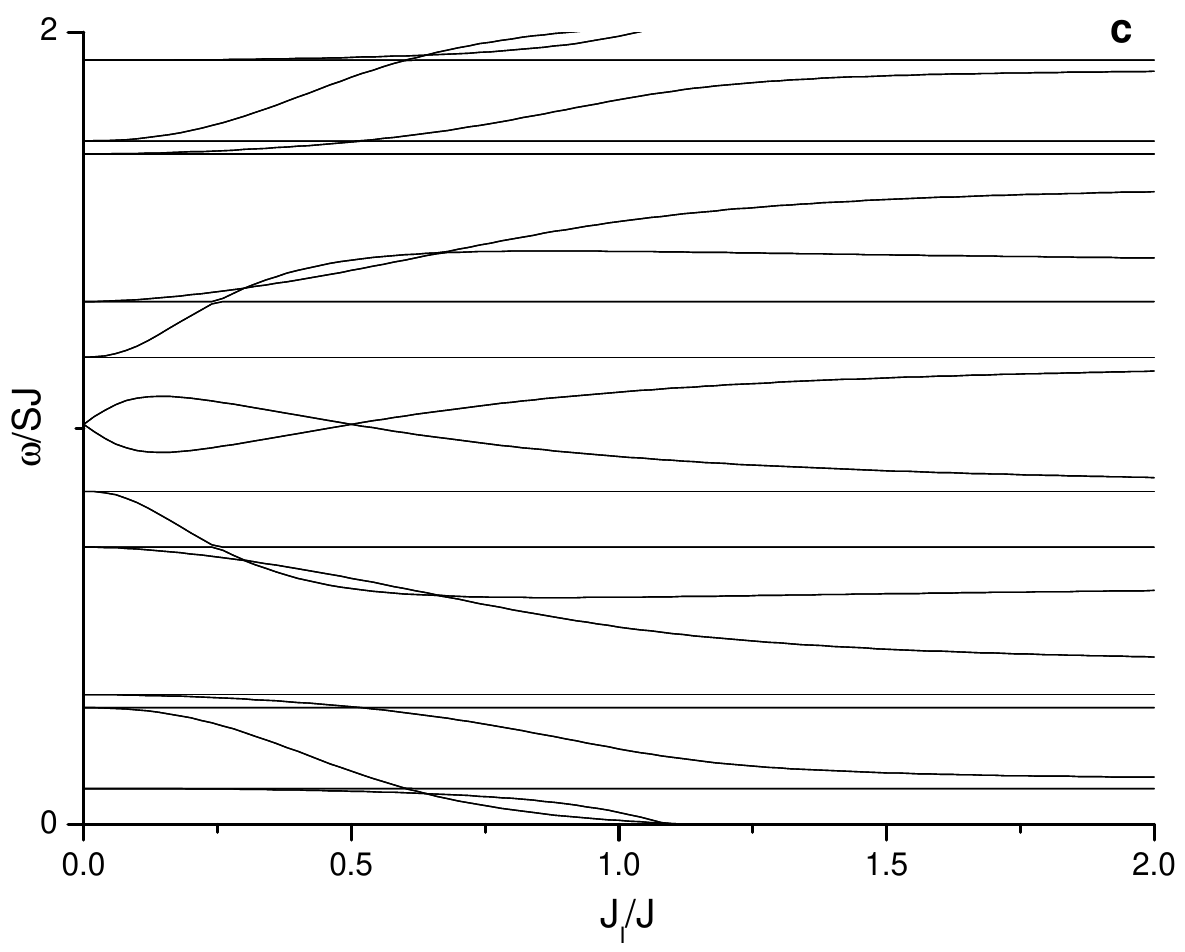}&\includegraphics[scale=.6]{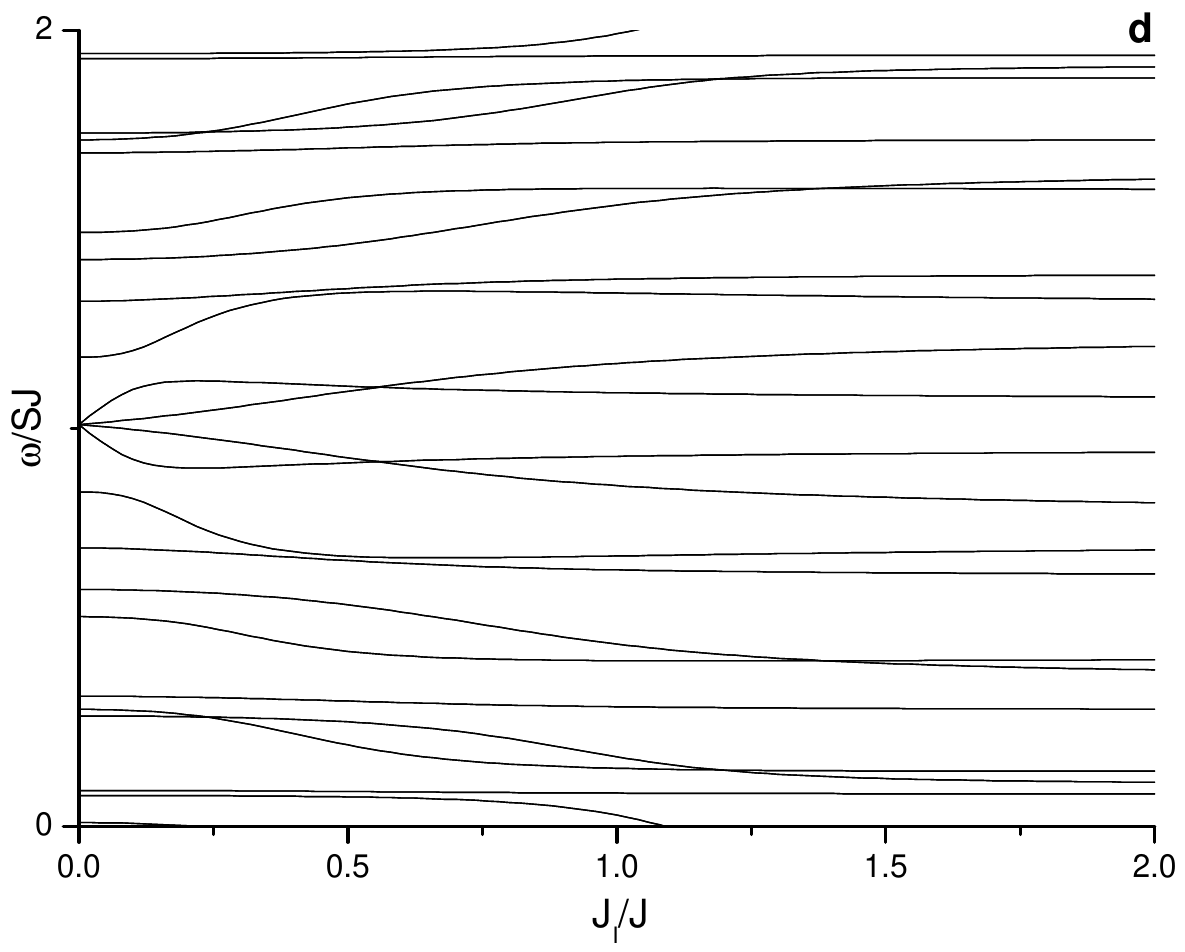}
\end{tabular}
  \caption{$q_x=0$ dispersion of the modes (a) 20-line armchair stripe with
an impurities line in the 11th line as a function of the strength of the impurity exchange $J_I$, which amplified  in the region
around $(\omega/SJ)=\alpha$ in (b). For (c) and (d) the amplified lower energy region for
N=21 and 22, respectively}\label{iedgeband}
\end{figure}

Introducing the impurities in armchair stripes modify their energy bands as seen in Figures \ref{fig:mgfarm1ii} and \ref{fig:mgfarm1iii}. The value of impurities exchange there chosen to be equal to zero, but as impurities exchange increases from zero the band gap behavior should be similar to graphene case \cite{rim1}. Figure \ref{iedgeband} shows the behavior of spin wave energy modes at direct band gap point, i.e. $q_x=0$ with the variation of the impurities exchange strength $J_I$, where the impurities line in the 11th line.  The Figure \ref{iedgeband} (a) shows the behavior of all modes for a 20-line armchair stripe with the variation of the impurities exchange strength. As impurities exchange increases, some modes begin to bend and crossing other modes some of them leave all the stripe energy band, the over all behavior is the same for 21 and 22 stripes, which is the same behavior seen before in edge exchange and graphene case \cite{rim1}.

Figure \ref{iedgeband} (b), (c), and (d) show the amplified region
around $(\omega/SJ)=\alpha$ for 20, 21 and 22 stripe widths respectively. They show similar behavior beginning at zero impurities exchange the minimum  conduction band mode and the maximum valance band meet together. Consequently, there is no band gap which shown as a localized states at Fermi level $(\omega/SJ)=\alpha$, as the impurities exchange increases the two mode splitting producing a band gap for the stripes widths which then crossing each other at certain value of impurities exchange depending on the stripe type. In general, as the impurities exchange approach the value of interior sites exchange, the modes rearranged to show the behavior seen before in edge exchange case.
\begin{figure}[h]
\centering
\includegraphics[scale=1]{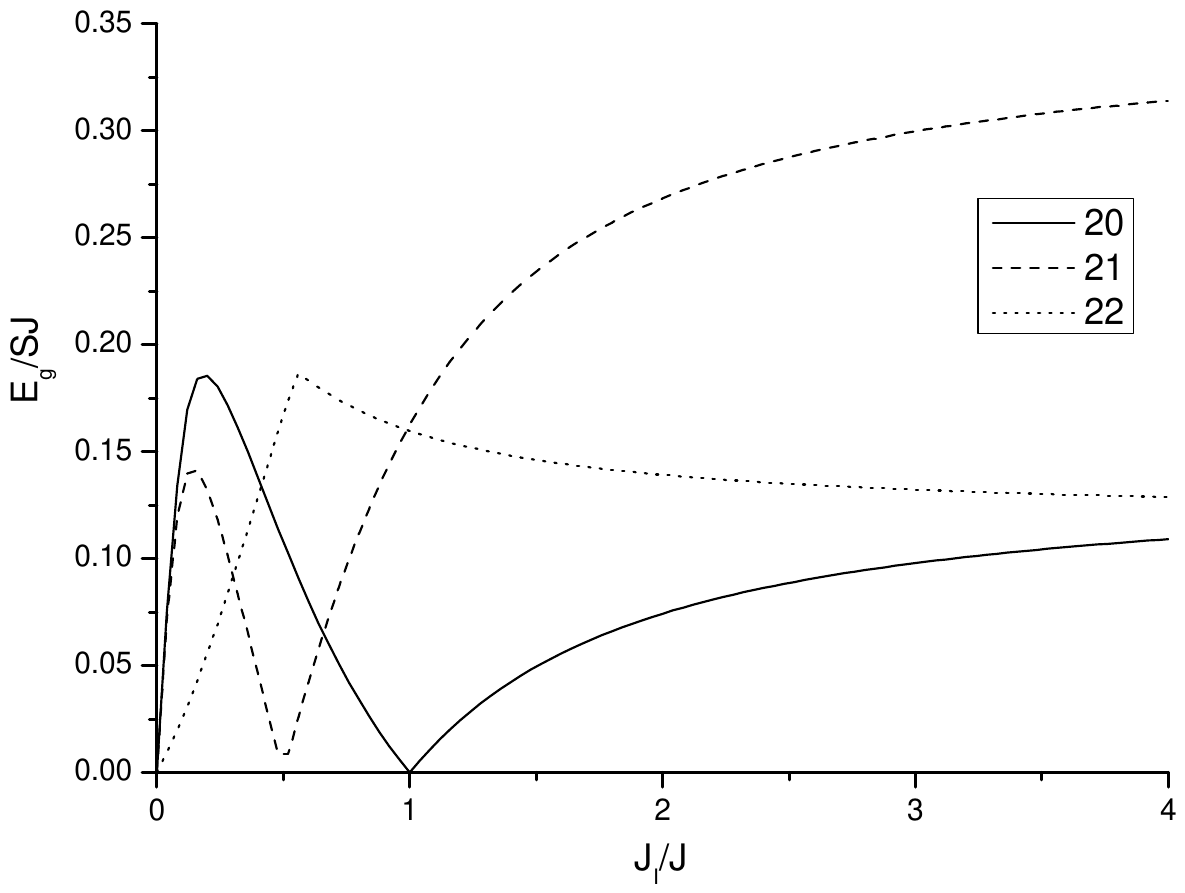}
\caption{Variation of the energy gap against the strength of the
impurity exchange with impurities line in the 11th line of armchair stripes. Solid line for a
20-line stripe, dashed line for 21 lines, and dotted line for a
stripe with 22 lines. The impurities are always in line
number 11.} \label{iedgeband2}
\end{figure}

Figure \ref{iedgeband2} shows, as in the case of edge exchange, the variation of the energy gap against the strength of the impurities exchange for 20, 21 and 22 armchair stripes. The three energy band gap starts from zero and increases to reach maximum, while 22 stripe begin to decreases slowly to nearly constant value, and the 20 and 21 stripe begin to decreasing to reach zero again at impurities exchange value depending on the stripe width, which are 0.5 for 21 stripe, 1.0 for 20 stripe. After this minimum for the two stripes each energy band gap starts to increases again, while the 20 stripe energy band gap increase slowly to seam constant value, the 21 stripe energy band gap increase faster to nearly constant.
\begin{figure}[hp]
  \centering
  \begin{tabular}{cc}
\includegraphics[scale=.6]{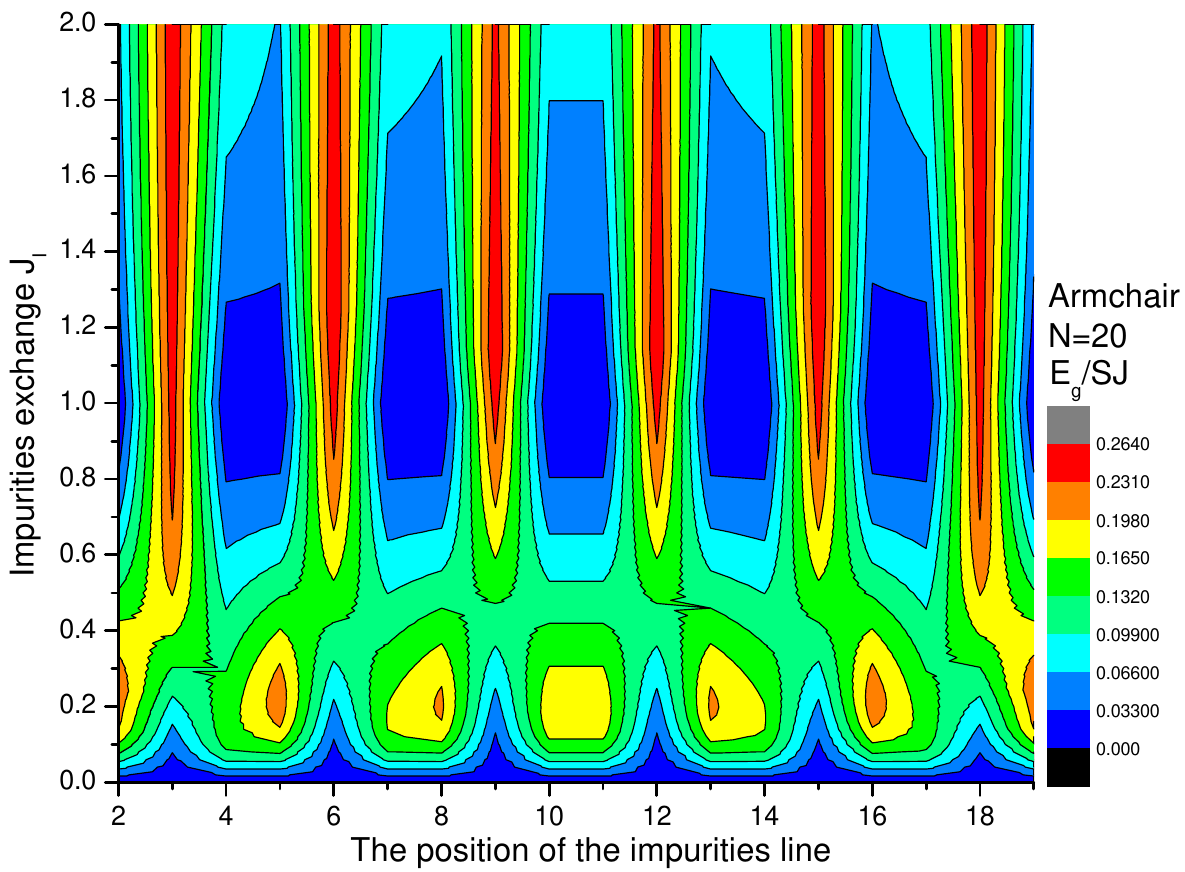}& \includegraphics[scale=.6]{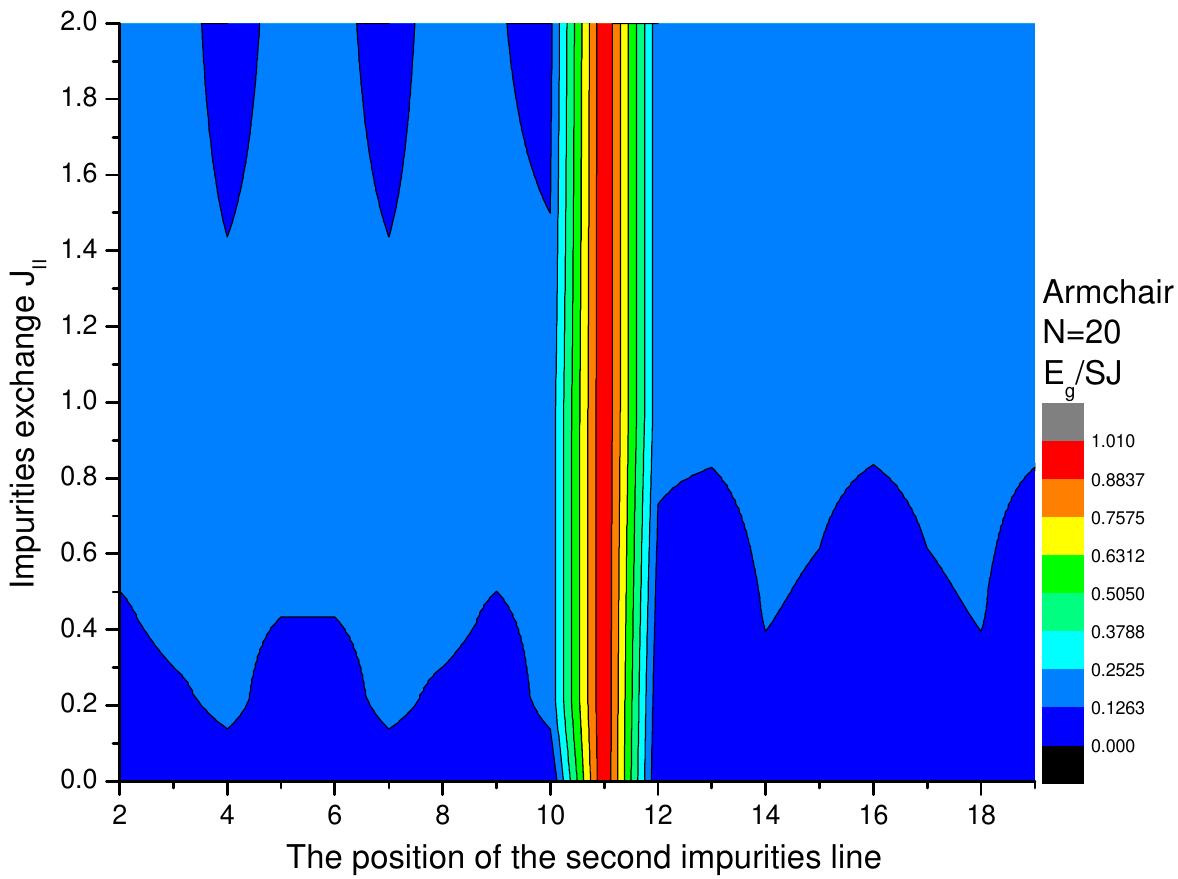}\\
\includegraphics[scale=.6]{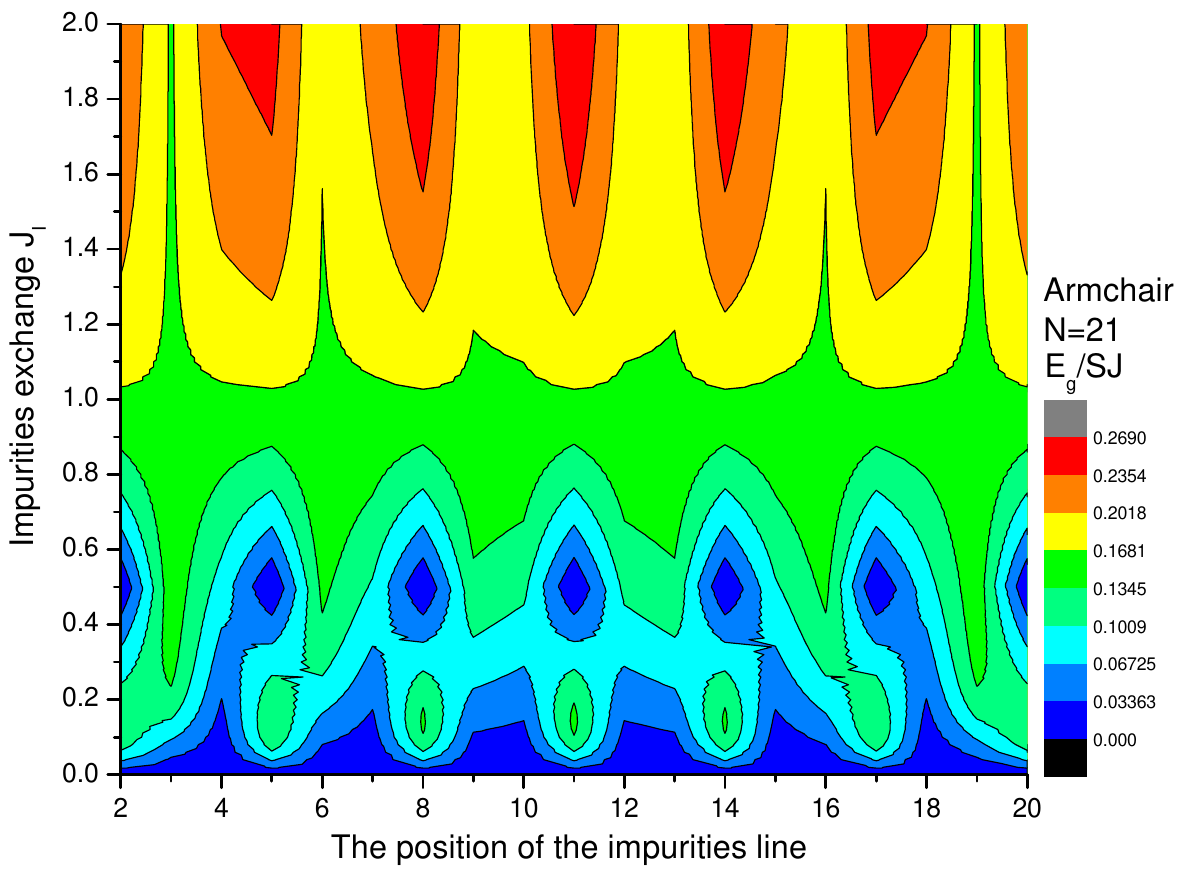}&\includegraphics[scale=.6]{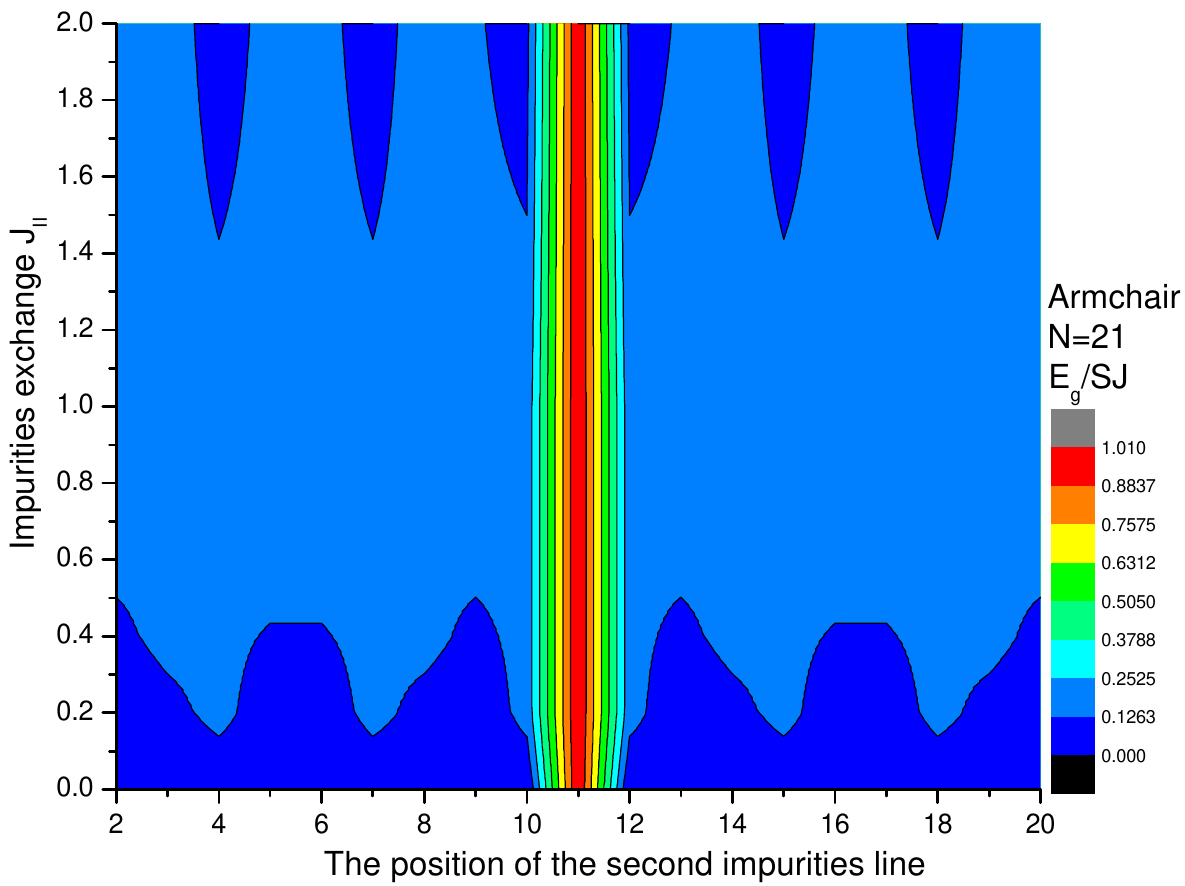}\\
\includegraphics[scale=.6]{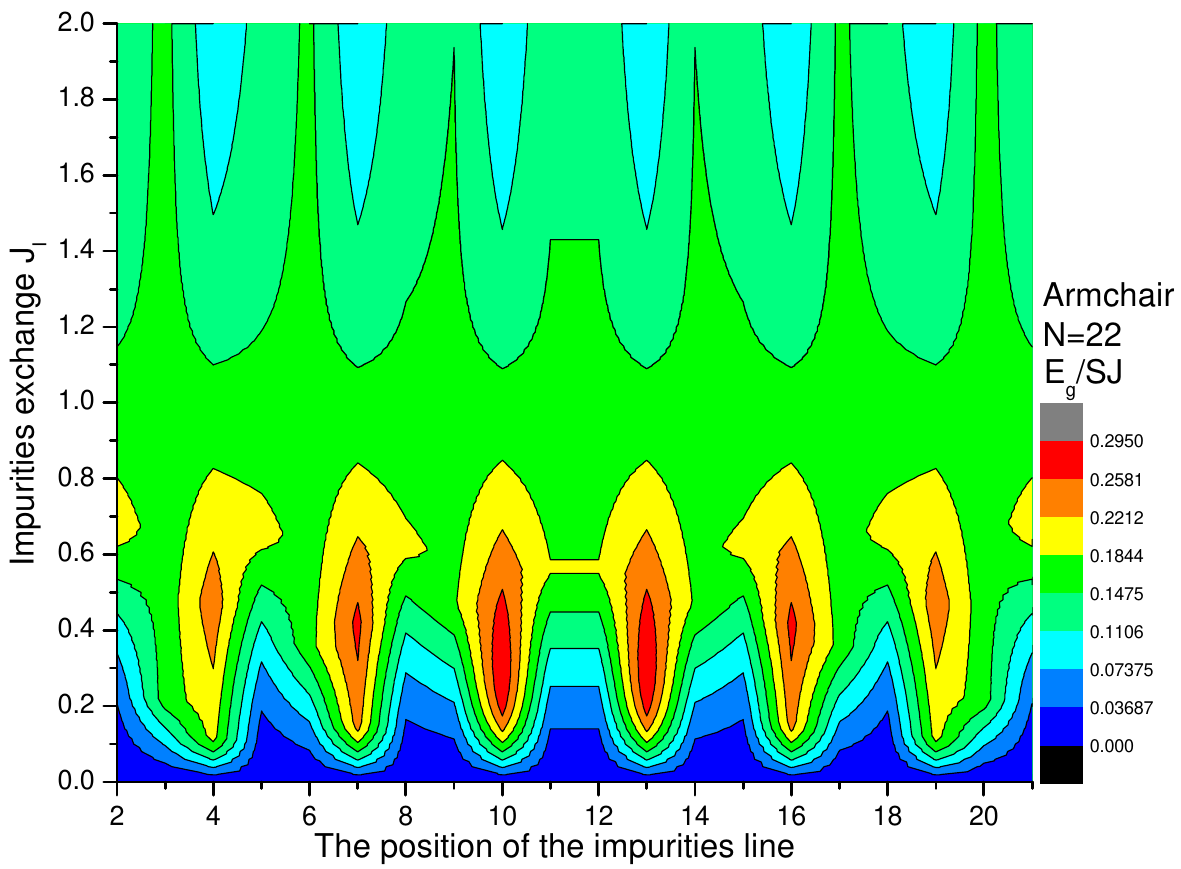}&\includegraphics[scale=.6]{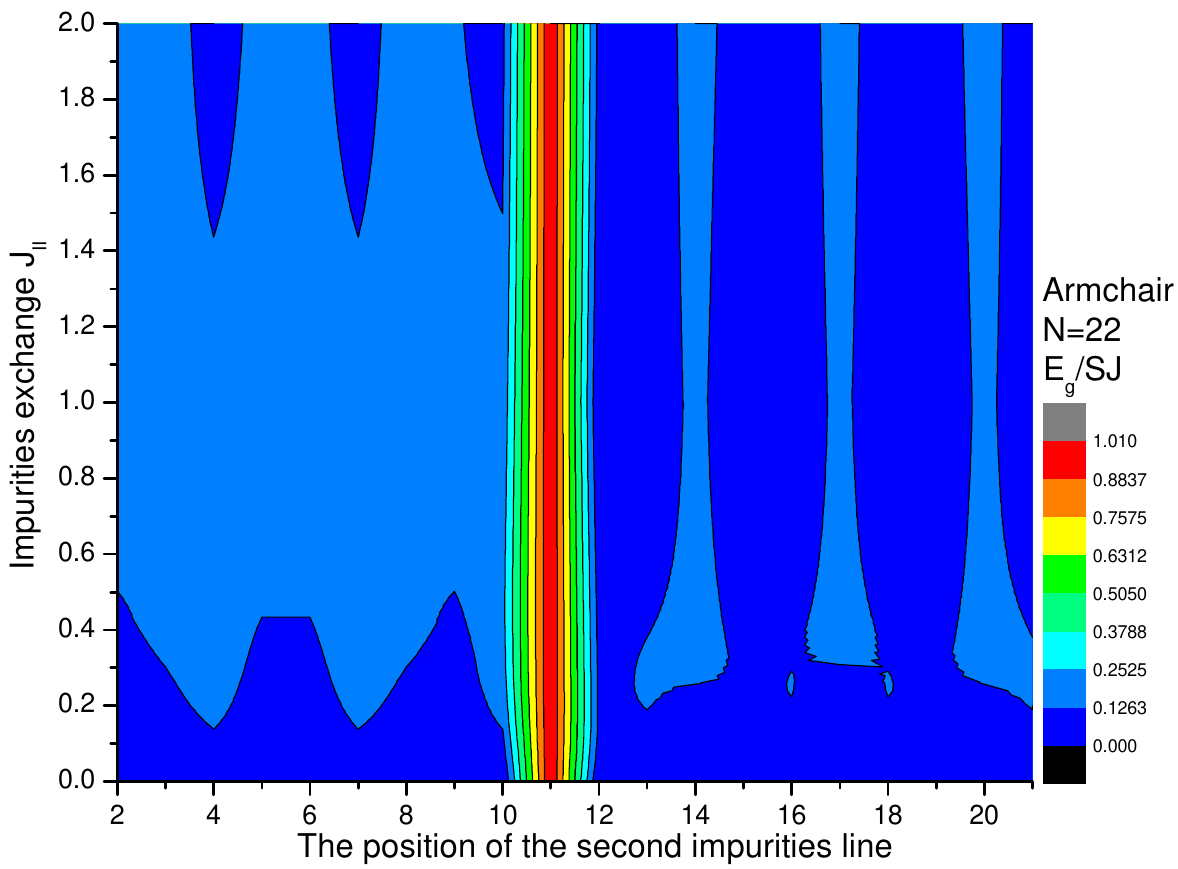}
\end{tabular}
  \caption{Color contour plot of the energy gap for the armchair stripes width 20, 21, and 22
left side showing the energy gap with the position of the first impurity line varying
from line 2 to N-1. Right side showing the energy gap with the position of the second impurity line varying
from line 2 to N-1, with first impurities line at
row number 11 with $J_I=0$.}\label{contourbandgap}
\end{figure}

Figures \ref{contourbandgap} show color contour plot of the behavior of energy band gap for the three stripes as a function in both impurities lines positions and their impurity exchange strength. The right part shows a periodic variation in band gap of the three stripes
with changing both the position and the exchange strength of one impurity line, also the variation is symmetric about the stripe center which is the same behavior found in graphene \cite{rim1}. The left part show the removing of the periodic variation in band gap for the three stripes while its symmetric about center is removed for the 20 and 22 two stripe which is the same behavior for their RDSCB. The left part is due to adding a second impurity line to the three stripes while fixing the first impurity line in row number 11 with zero impurity exchange, which show the great ability to tune the energy band gap of armchair stripes with adding one or more magnetic impurity line.

\section{Discussion and Conclusions}
In this work, the second quantization form of Heisenberg
Hamiltonian for ferromagnetic short range (a geometrical quantity) interaction between nanodots represented by nearest neighbor (NN) exchange $J_{ij}$, is used to study the allowed spin waves modes, i.e. dispersion relations (energy band), for 2D Honeycomb Lattice (a geometrical quantity).

The results of this study for ferromagnetic dots
2D honeycomb lattice stripes show almost coincidence with the results of graphene nanoribbons described by tight binding Hamiltonian for electronic short range (a geometrical quantity) interaction between carbons atoms represented by nearest neighbor (NN) hopping $t_{ij}$ for 2D Honeycomb Lattice (a geometrical quantity).

From technological point of view, those results are very encouraging to fabricate a magnetic counterpart to graphene, which will lead to a new technology especially in the field of spintronic devices and magnetic applications. Also, the results show that many obtained researches results of graphene can be easily applicable to the magnetic case.

\begin{acknowledgments}
This research has been supported by the Egyptian Ministry of Higher Education
and Scientific Research (MZA).
\end{acknowledgments}

\bibliography{xbib2}

\end{document}